# Definitional Functoriality for Dependent (Sub)Types – Extended version

THÉO LAURENT, Inria, France
MEVEN LENNON-BERTRAND, University of Cambridge, United Kingdom
KENJI MAILLARD, Gallinette Project Team, Inria, France

Dependently typed proof assistant rely crucially on definitional equality, which relates types and terms that are automatically identified in the underlying type theory. This paper extends type theory with definitional *functor laws*, equations satisfied propositionally by a large class of container-like type constructors $F$ : Type $\to$ Type, equipped with a $\mathrm{map}_F : (A \to B) \to F\,A \to F\,B$, such as lists or trees. Promoting these equations to definitional ones strengthens the theory, enabling slicker proofs and more automation for functorial type constructors. This extension is used to modularly justify a structural form of coercive subtyping, propagating subtyping through type formers in a map-like fashion. We show that the resulting notion of coercive subtyping, thanks to the extra definitional equations, is equivalent to a natural and implicit form of subsumptive subtyping. The key result of decidability of type-checking in a dependent type system with functor laws for lists has been entirely mechanized in Coq.

This is the extended version of [34].

## 1 INTRODUCTION

Dependent type theory is the foundation of many proof assistants: Coq [56], Lean [45], Agda [5], Idris [14], F* [54]. At its heart lies definitional equality, an equational theory that is automatically decided by the implementation of these proof systems. The more expressive definitional equality is, the less work is requested from users to identify objects. However, there is a fundamental tension at play: making the equational theory too rich leads to both practical and theoretical issues, the most prominent one being the undecidability of definitional equality. This default plagues the otherwise appealing Extensional Type Theory (ETT) [41], a type theory which makes every provable equality definitional, thus making ETT rather impractical as a basis for a proof assistant [15]. As a result, to design usable proof assistants we need to carve out a well-behaved equational theory, that strikes the right balance between expressivity and decidability. In this paper, we show that we can maintain this subtle balance while extending intensional type theory with map operations making the functorial character of type formers explicit, and satisfying *definitional functor laws*. We prove in particular that definitional equality and type-checking remain decidable in this extension, that we dub $\mathrm{MLTT_{map}}$.

The map primitives introduced in $\mathrm{MLTT_{map}}$ have a computational behaviour reminiscent of *structural subtyping*, which propagates existing subtyping structurally through type-formers, and should satisfy reflexivity and transitivity laws similar to the functor laws. Guided by the design of $\mathrm{MLTT_{map}}$, we devise a second system, $\mathrm{MLTT_{coe}}$, with explicit coercions witnessing structural subtyping. To gauge the expressivity of $\mathrm{MLTT_{coe}}$, we relate it to a third system, $\mathrm{MLTT_{sub}}$, where subtyping is implicit, as users of a type system should expect. A simple translation $|\cdot|$ from $\mathrm{MLTT_{coe}}$ to $\mathrm{MLTT_{sub}}$ erases coercions. We show that this erasure can be inverted, elaborating coercions back.

Authors' addresses: Théo Laurent, theo.laurent@inria.fr, Inria, Paris, France; Meven Lennon-Bertrand, Meven.Lennon-Bertrand@cl.cam.ac.uk, University of Cambridge, Cambridge, United Kingdom; Kenji Maillard, kenji.maillard@inria.fr, Gallinette Project Team, Inria, Nantes, France.

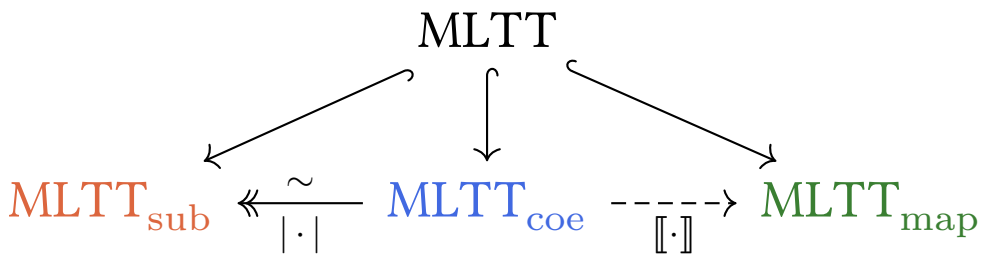

Fig. 1. Relation between MLTT, $\text{MLTT}_{\text{map}}$, $\text{MLTT}_{\text{coe}}$ and $\text{MLTT}_{\text{sub}}$. Arrows denote type-and-conversion-preserving translations between type theories.[1]

For this to be type preserving, it is crucial that $\text{MLTT}_{\text{coe}}$ satisfies our new definitional equalities, which allows us to reflect the equations implicitly satisfied in $\text{MLTT}_{\text{sub}}$ due to coercions being transparent. Fig. 1 synthesizes the three theories that we introduce and their relationships. They all extend Martin-Löf Type Theory (MLTT) [41]. Let us now explore in more detail these three systems.

*Functors and Their Laws.* The notion of functor is pervasive both in mathematics [40] and functional programming [36], capturing the concept of *a parametrized construction applying to objects and their transformations.* Reformulated in type theory, a type former $F : \text{dom}(F) \to \text{Type}$ is a functor when it is equipped with an operation $\text{map}_F\ f : F\,A \to F\,B$ for any morphism $f : \text{hom}_F(A, B)$ between two objects $A$,$B$ in the domain $\text{dom}(F)$ of $F$. Here, $\text{dom}(F)$ must be endowed with the structure of a category[2], with specified composition $\circ^F$ and identities $\text{id}^F$, and $\text{map}_F$ must preserve those:

$$\text{map}_F\ \text{id}^F = \text{id} \qquad \text{(id-eq)}$$

$$(\text{map}_F\ f) \circ (\text{map}_F\ g) = \text{map}_F\,(f \circ^F g) \qquad \text{(comp-eq)}$$

These two equations are known as the *functor laws.* For many container-like functors, such as $\textbf{List}\ A$, lists of elements taken in a type $A$, a map function can be defined in vanilla type theory such that these equations can be shown *propositionally*, *e.g.* by induction. Such propositional equations need however to be used explicitly, putting an extra burden on users and possibly causing coherence issues typical when working with propositional equalities [57]. This is not acceptable: such simple and natural identifications should hold definitionally!

*Example 1.1 (Representation Change).* Consider a dataset of pairs of a number and a boolean, represented as a list. For compatibility purpose, we may need to embed these pairs into a larger dataset using

$$\begin{aligned}\texttt{glue}\,(r\ :\ \{a : \mathbf{N}; b : \mathbf{B}\}) : \{x : \mathbf{B}; y : \mathbf{N}; z : \mathbf{N}\} &\stackrel{\text{def}}{=} \\ \{x := r.b; y := r.a; z := \texttt{if}\ r.b\ \texttt{then}\ r.a\ \texttt{else}\ 42\}.\end{aligned}$$

Going from one dataset to the other amounts to mapping either `glue` or its left inverse `glue_retr`, which forgets the extra field:

$$\begin{aligned}\text{map}_{\textbf{List}}\ \texttt{glue} &: \textbf{List}\ \{a : \mathbf{N}; b : \mathbf{B}\} \to \textbf{List}\ \{x : \mathbf{B}; y : \mathbf{N}; z : \mathbf{N}\}, \\ \text{map}_{\textbf{List}}\ \texttt{glue_retr} &: \textbf{List}\ \{x : \mathbf{B}; y : \mathbf{N}; z : \mathbf{N}\} \to \textbf{List}\ \{a : \mathbf{N}; b : \mathbf{B}\}.\end{aligned}$$

[2]Morphisms $\text{hom}_F(A, B)$ in $\text{dom}(F)$ are not constrained to be type theoretic functions. Accordingly, composition need not to be literally the composition of functions and the specified identities can differ from the identity $\lambda x.x$.

If the functor laws only hold propositionally, each consecutive simplification of back and forth changes of representation needs to be explicitly lifted to lists, and applied. The uncontrolled accumulation of repetitive proof steps, even as simple as these, can quickly burden proof development. In presence of definitional functor laws, instead, any sequence of representation changes will reduce to a single $\text{map}_{\textbf{List}}$: the boilerplate of explicitly manipulating the functor laws is handled automatically by the type theory. Moreover, observe that in this example the retraction $\texttt{glue_retr} \circ \texttt{glue} \cong \text{id}$ is definitional thanks to surjective pairing. Combined with definitional functor laws, the following simplification step is discharged automatically by the type-checker:[3]

$$\text{map}_{\textbf{List}}\ \texttt{glue_retr}\ (\text{map}_{\textbf{List}}\ \texttt{glue}\ l) \cong \text{map}_{\textbf{List}}\ \text{id}\ l \cong l$$

Note that these equations are valid in any context, in particular under binders, whereas for propositional identifications, rewriting under binders is only possible in presence of the additional axiom of function extensionality.

*Example 1.2 (Coherence of Coercions).* Proof assistants may provide the ability for users to declare automatically-inserted functions acting as glue code (coercions in Coq, instance arguments in Agda, has_coe typeclass in Lean). Working with natural ($\mathbf{N}$), integer ($\mathbf{Z}$) and rational ($\mathbf{Q}$) numbers, we want every $\mathbf{N}$ to be automatically coerced to an integer, and so declare a natToZ coercion. Similarly, we can also declare a ZToQ coercion. If we write $0$ (a $\mathbf{N}$) where a $\mathbf{Q}$ is expected, this is accepted, and $0$ is silently transformed to $\texttt{ZToQ}\ (\texttt{natToZ}\ 0)$.

Now, if we want the same mechanism to apply when we pass the list $[0 :: 1 :: 2]$ to a function expecting a $\textbf{List}\,\mathbf{Q}$, we need to provide a way to propagate the coercions on lists. We can expect to solve this problem by declaring $\text{map}_{\textbf{List}}$ as a coercion, too: whenever there is a coercion $f : A \to B$, then $\text{map}_{\textbf{List}}\ f$ should be a coercion from $\textbf{List}\ A$ to $\textbf{List}\ B$. However, by doing so, we would cause more trouble than we solve, as there would be two coercions from $\textbf{List}\,\mathbf{N}$ to $\textbf{List}\,\mathbf{Q}$, $\text{map}_{\textbf{List}}(\texttt{ZToQ} \circ \texttt{natToZ})$ and $(\text{map}_{\textbf{List}}\ \texttt{ZToQ}) \circ (\text{map}_{\textbf{List}}\ \texttt{natToZ})$. In the absence of definitional functor laws for $\text{map}_{\textbf{List}}$, these two are *not* definitionally equal. To add insult to injury, coercions are by default not printed to the user, yielding puzzling error messages like "*$l$ and $l$ are not convertible*" (!), because one is secretly $\text{map}_{\textbf{List}}(\texttt{ZToQ} \circ \texttt{natToZ})\ l$ while the other is $\text{map}_{\textbf{List}}\ \texttt{ZToQ}\ (\text{map}_{\textbf{List}}\ \texttt{natToZ}\ l)$. This makes $\text{map}_{\textbf{List}}$ virtually unusable with coercions.

*Structural Subtyping.* This last example suggests a connection with *subtyping*. Subtyping equips the collection of types with a *subtyping order* $\preccurlyeq$ that allows to seamlessly transport terms from a subtype to a supertype, *i.e.* from $A$ to $A'$ when $A \preccurlyeq A'$. An important aspect of subtyping is *structural subtyping*, *i.e.* how subtyping extends structurally through type formers of the type theory. Typically, we want to have $\textbf{List}\ A \preccurlyeq \textbf{List}\ A'$ whenever $A \preccurlyeq A'$. In the context of the F* program verification platform that heavily uses refinement subtyping, the inability to propagate subtyping on inductive datatypes such as lists has been a long-standing issue that never got solved properly [27]. The absence of structural subtyping also has a history of causing difficulties to Agda [16, 23].

*Definitional Equalities for Subtyping.* From the perspective of users of interactive theorem prover, subtyping should be implicit, transparently providing the expected glue to smoothen the writing of complex statements. From a meta-theoretical perspective, on the other hand, it is useful to explicitly represent all the necessary information of a typing derivation, including where subtyping is used. The first approach is known as *subsumptive subtyping*, on the left, whereas the latter is

[3]We formalize this example, showing that this conversion indeed holds in our system, in file Example_1_1.

embodied by *coercive subtyping*, on the right:

$$\textsc{Sub}\ \frac{\Gamma \vdash_{\mathrm{sub}} t : A \qquad \Gamma \vdash_{\mathrm{sub}} A \preccurlyeq A'}{\Gamma \vdash_{\mathrm{sub}} t : A'} \qquad\qquad \textsc{Coe}\ \frac{\Gamma \vdash_{\mathrm{coe}} t : A \qquad \Gamma \vdash_{\mathrm{coe}} A \preccurlyeq A' \lhd}{\Gamma \vdash_{\mathrm{coe}} \mathrm{coe}_{A,A'}\, t : A'}$$

We want to present subsumptive subtyping to users, but ground the system on the algebraic, better-behaved coercive subtyping. Informally, an application of Sub in the subsumptive type theory $\mathrm{MLTT}_{\mathrm{sub}}$ should correspond to an application of Coe in the coercive type theory $\mathrm{MLTT}_{\mathrm{coe}}$. However, given a derivation $\mathcal{D}$ of $\Gamma \vdash_{\mathrm{sub}} t : A$ we can apply Sub together with a reflexivity proof $\Gamma \vdash_{\mathrm{sub}} A \preccurlyeq A$ to yield a new derivation $\mathcal{D}'$ with the same conclusion $\Gamma \vdash_{\mathrm{sub}} t : A$. $\mathcal{D}$ and $\mathcal{D}'$ should respectively correspond to terms $\Gamma \vdash_{\mathrm{coe}} t' : A$ and $\Gamma \vdash_{\mathrm{coe}} \mathrm{coe}_{A,A}\, t' : A$ in $\mathrm{MLTT}_{\mathrm{coe}}$. Since $t'$ and $\mathrm{coe}_{A,A}\, t'$ both erase to the same $\mathrm{MLTT}_{\mathrm{sub}}$ term $t$, they need to be equated if we want both type theories to be equivalent. Similarly, transitivity of subtyping implies that coercions should compose definitionally, that is $\Gamma \vdash_{\mathrm{coe}} \mathrm{coe}_{B,C}(\mathrm{coe}_{A,B}\, t') \cong \mathrm{coe}_{A,C}\, t' \lhd C$ should always hold in $\mathrm{MLTT}_{\mathrm{coe}}$.

*Functor Laws Meet Structural Subtyping.* Luo and Adams [38] showed that the functorial composition law comp-eq is enough to make structural coercive subtyping compose definitionally, because a structural coercion between lists $\mathrm{coe}_{\mathbf{List}\, A, \mathbf{List}\, B}$ behaves just as the function obtained by mapping $\mathrm{coe}_{A,B}$ on every element of the list. We further investigate this bridge between coercive subtyping and functoriality of type formers, in particular the identity functor law id-eq needed to handle reflexivity of subtyping, and extend Luo and Adams's limited type system to full-blown Martin-Löf Type Theory (MLTT), with universes and large elimination. This understanding leads to a modular design of subtyping: structural subtyping for a type former relies on a functorial structure, and can be considered orthogonally to other type formers of the theory or to the base subtyping. Moreover, definitional functor laws are sufficient to make structural coercive subtyping for any type former expressive and flexible enough to interpret subsumptive subtyping.

*Contributions.* We make the following contributions:

- we design $\mathrm{MLTT}_{\mathrm{map}}$, an extension of Martin-Löf Type Theory (MLTT) exhibiting the functorial nature of standard type formers ($\Pi, \Sigma, \mathbf{List}, \mathbf{W}, \mathbf{Id}, +$), and satisfying definitional functor laws (Section 3);
- we mechanize the metatheory of a substantial fragment of $\mathrm{MLTT}_{\mathrm{map}}$ in Coq, extending a formalization of MLTT [3], proving it is normalizing and has decidable type-checking (Section 4);
- we develop bidirectional presentations for $\mathrm{MLTT}_{\mathrm{sub}}$ and $\mathrm{MLTT}_{\mathrm{coe}}$, which extend MLTT respectively with subsumptive and coercive subtyping;
- we leverage these presentations and the extra functorial equations satisfied by coe in $\mathrm{MLTT}_{\mathrm{coe}}$ to give back and forth, type-preserving translations between the two systems (Section 5).

The necessary technical background, notations and definitions for MLTT are introduced in Section 2, while Section 6 details the related and future work.

This is an extended version of [34]. It mainly extends Section 2, adds some figures to the text, the discussion on eliminator fusion in Section 3.3, provides a small instance of base subtyping using records in Section 5.1 and has complete proofs and type systems in appendices. The appendix also contains a detailed description of the formalisation presented in Section 4, which was originally provided as an accompanying note alongside [33] for the artefact evaluation process.

$\boxed{\Gamma \vdash T}$ Type $T$ is well-formed under context $\Gamma$

$\boxed{\Gamma \vdash t : T}$ Term $t$ has type $T$ under context $\Gamma$

$$\textsc{Var}\ \frac{\vdash \Gamma \qquad (x : A) \in \Gamma}{\Gamma \vdash x : A} \qquad \textsc{Sort}\ \frac{\vdash \Gamma}{\Gamma \vdash \mathrm{Type}_i : \mathrm{Type}_{i+1}} \qquad \textsc{El}\ \frac{\Gamma \vdash A : \mathrm{Type}_i}{\Gamma \vdash A}$$

$$\textsc{Fun}\ \frac{\Gamma \vdash A : \mathrm{Type}_i \quad \Gamma, x : A \vdash B : \mathrm{Type}_i}{\Gamma \vdash \Pi\, x : A.B : \mathrm{Type}_i} \qquad \textsc{Abs}\ \frac{\Gamma \vdash A \quad \Gamma, x : A \vdash t : B}{\Gamma \vdash \lambda\, x : A.t : \Pi\, x : A.B} \qquad \textsc{App}\ \frac{\Gamma \vdash t : \Pi\, x : A.B \quad \Gamma \vdash u : A}{\Gamma \vdash t\ u : B[u]}$$

$$\textsc{List}\ \frac{\Gamma \vdash A : \mathrm{Type}_i}{\Gamma \vdash \mathbf{List}\, A : \mathrm{Type}_i} \qquad \textsc{Nil}\ \frac{\Gamma \vdash A}{\Gamma \vdash \varepsilon_A : \mathbf{List}\, A} \qquad \textsc{Cons}\ \frac{\Gamma \vdash A \qquad \Gamma \vdash a : A \quad \Gamma \vdash l : \mathbf{List}\, A}{\Gamma \vdash a ::_A l : \mathbf{List}\, A}$$

$$\textsc{ListInd}\ \frac{\Gamma \vdash A \qquad \Gamma \vdash s : \mathbf{List}\, A \qquad \Gamma, z : \mathbf{List}\, A \vdash P \quad \Gamma \vdash b_\varepsilon : P[\varepsilon_A] \qquad \Gamma, x : A, y : \mathbf{List}\, A, z : P[y] \vdash b_{::} : P[x ::_A y]}{\Gamma \vdash \mathrm{ind}_{\mathbf{List}\, A}(s; z.P; b_\varepsilon, x.y.z.b_{::}) : P[s]} \qquad \textsc{Conv}\ \frac{\Gamma \vdash t : A \quad \Gamma \vdash A \cong B}{\Gamma \vdash t : B}$$

Fig. 2. Declarative typing for MLTT (complete rules: Appendix C.1)

$\boxed{\Gamma \vdash t \cong u : A}$ Terms $u$ and $v$ are convertible at type $A$ under context $\Gamma$

$\boxed{\Gamma \vdash A \cong B}$ Types $A$ and $B$ are convertible under context $\Gamma$

$$\beta\textsc{Fun}\ \frac{\Gamma \vdash A \qquad \Gamma, x : A \vdash t : B \qquad \Gamma \vdash u : A}{\Gamma \vdash (\lambda\, x : A.t)\ u \cong t[u] : B[u]} \qquad \eta\textsc{Fun}\ \frac{\Gamma, x : A \vdash f\, x \cong g\, x : B}{\Gamma \vdash f \cong g : \Pi\, x : A.B}$$

$$\beta\textsc{Nil}\ \frac{\Gamma \vdash A \quad \Gamma, z : \mathbf{List}\, A \vdash P \qquad \Gamma \vdash b_\varepsilon : P[\varepsilon_A] \qquad \Gamma, x : A, y : \mathbf{List}\, A, z : P[y] \vdash b_{::} : P[x ::_A y]}{\Gamma \vdash \mathrm{ind}_{\mathbf{List}\, A}(\varepsilon_A; l.P; b_\varepsilon, x.y.z.b_{::}) \cong b_\varepsilon : P[\varepsilon_A]}$$

$$\beta\textsc{Cons}\ \frac{\Gamma \vdash A \qquad \Gamma \vdash a : A \qquad \Gamma \vdash l : \mathbf{List}\, A \quad \Gamma, z : \mathbf{List}\, A \vdash P \qquad \Gamma \vdash b_\varepsilon : P[\varepsilon_A] \qquad \Gamma, x : A, y : \mathbf{List}\, A, z : P[y] \vdash b_{::} : P[x ::_A y]}{\Gamma \vdash \mathrm{ind}_{\mathbf{List}\, A}(a ::_A l; z.P; b_\varepsilon, x.y.z.b_{::}) \cong b_{::}[\mathrm{id}, a, l, \mathrm{ind}_{\mathbf{List}\, A}(l; z.P; b_\varepsilon, x.y.z.b_{::})] : P[a ::_A l]}$$

$$\textsc{ConvConv}\ \frac{\Gamma \vdash t \cong t' : A \qquad \Gamma \vdash A \cong A'}{\Gamma \vdash t \cong t' : A'} \qquad \textsc{ElConv}\ \frac{\Gamma \vdash A \cong A' : \mathrm{Type}_i}{\Gamma \vdash A \cong A'}$$

$$\textsc{Refl}\ \frac{\Gamma \vdash t : A}{\Gamma \vdash t \cong t : A} \qquad \textsc{Sym}\ \frac{\Gamma \vdash t \cong u : A}{\Gamma \vdash u \cong t : A} \qquad \textsc{Trans}\ \frac{\Gamma \vdash t \cong u : A \qquad \Gamma \vdash u \cong v : A}{\Gamma \vdash t \cong v : A}$$

Fig. 3. Declarative conversion for MLTT (complete rules: Appendix C.1)

## 2 TYPE THEORY AND ITS METATHEORY

We work in the setting of dependent type theories *à la* Martin-Löf (MLTT) [41], an ideal abstraction of the type theories underlying existing proof assistants such as Agda, Coq, F$^\star$ or Lean. MLTT employs five categories of judgements, characterizing the well-formed contexts ($\vdash \Gamma$), types ($\Gamma \vdash T$) and terms ($\Gamma \vdash t : T$) (Figure 2), and providing the equational theory on types ($\Gamma \vdash A \cong B$) and terms ($\Gamma \vdash t \cong u : A$) (Figure 3). Two terms related by this equational theory are said to be *definitionally equal* or *convertible*, to stress the fact that these terms will be identified by proof assistants implementing this theory, without any need for manual equational reasoning.

*Variables and substitution.* Throughout the paper, we use named variables ($x, y$ ...) for readability purposes, but we think of them as de Bruijn indices, which is what we use in the Coq formalization. In particular, we do not consider freshness conditions. A substitution $\sigma$ consists of a list of terms, and we write $t[\sigma]$ for its parallel substitution in the term $t$. The substitution $(\mathrm{id}, u)$ replaces the 0th de Bruijn index by the term $u$, and applies the identity substitution to all other variables, leaving them intact. We will write $t[(\mathrm{id}, u)]$ simply as $t[u]$, which would be written $t[x := u]$ in more verbose notation, if $x$ correspond to the 0th de Bruijn index in $t$. Typing in all systems is extended pointwise to substitutions in a standard fashion, see Appendix C.1.

*Negative Types: Dependent Products and Sums.* Dependent function types, noted $\Pi\, x : A.B$, are introduced using $\lambda$-abstraction $\lambda\, x : A.t$ and eliminated with application $t\, u$. We also include dependent sum types $\Sigma\, x : A.B$, introduced with pairs $(t, u)_{x.B}$ and eliminated through projections $\pi_1\, p$ and $\pi_2\, p$. Both come with an η-law.

*Universes of Types.* Our type theories feature a countable hierarchy of universes $\mathrm{Type}_i$, which are types for types. Any inhabitant of a universe is a well-formed type, and, in order to make the presentation compact, we do not repeat rules applying both for universes and types: all rules given for terms of some universe $\mathrm{Type}_i$ have a counterpart as a type judgement whenever it makes sense.

For instance, in addition to Fun, we have the type-level equivalent

$$\textsc{FunTy}\ \frac{\Gamma \vdash A \qquad \Gamma, x : A \vdash B}{\Gamma \vdash \Pi\, x : A.B}$$

*Positive Types: Inductives.* As we study the functorial status of type formers, *parametrized* inductive types are our main focus. Our running example is the type of lists $\mathbf{List}\, A$, parametrized by a type $A$, and inhabited by the empty list $\varepsilon_A$ and the consing $hd ::_A tl$ of a head $hd : A$ onto a tail $tl : \mathbf{List}\, A$. Lists are eliminated using the dependent eliminator $\mathrm{ind}_{\mathbf{List}\, A}(s; l.P; b_\varepsilon, x.y.z.b_{::})$, which performs induction on the scrutinee $s$, returning a value in $P[s]$, using the two branches $b_{::}$ and $b_\varepsilon$, corresponding to the two constructors $::$ and $\varepsilon$. $b_{::}$ binds three variables corresponding to the head, tail and induction hypothesis on the tail. More generally, strictly positive recursive datatypes are often presented in MLTT via $\mathbf{W}\, x : A.B$, the type of well-founded trees with nodes labelled by $a : A$ of arity $B\, a$. Finally, Martin-Löf's identity type $\mathbf{Id}\ A\, x\, y$ represents equalities between two elements $x, y : A$, and is introduced with the reflexivity proof $\mathrm{refl}_{A,a} : \mathbf{Id}\ A\, a\, a$. A general inductive type scheme is outside the scope of this paper, but the specific types we treat ($\mathbf{List}$, $\mathbf{W}$, $\mathbf{Id}$ and $+$) cover all aspects of inductive types: recursion, branching, parameters, and indices. Moreover, they can emulate all indexed inductive types [1, 8, 28], although we will see in Section 3.1 that this encoding interacts poorly with functor laws. As $\mathbf{0}$ and $\mathbf{1}$ are not parametrized, their presentation in our setting is entirely standard.

*Rules in the Paper.* Due to space constraints, we focus in the text on the most interesting rules, and on two types: dependent functions and lists. Together, they cover the interesting points of our work: dependent product types have a binder and come with an η-law; lists are a parametrized datatype, for which definitional functor laws are challenging. Complete rules are given in Appendix C.

### 2.1 Metatheoretical Properties

In order to show that the extensions of MLTT from Figure 1 are well-behaved, we establish the following meta-theoretical properties.

*Consistency and canonicity.* In order to be logically sound, a type theory should have no closed term of the empty type, *i.e.* there should be no $t$ such that $\vdash t : \mathbf{0}$. This *consistency* property is an easy consequence of *canonicity*, which characterizes the inhabitants of inductive types in the empty context as those obtained by repeated applications of constructors, up to conversion. Consistency follows, as $\mathbf{0}$ has no constructor.

*Decidability of type-checking and conversion.* A proof assistant should also be able to check whether a proof is valid, *i.e.* whether a typing judgement is derivable. In a dependent type system where terms essentially encode the structure of derivations, the main obstacle to decidability of typing is that of conversion.

*Normal forms for terms and derivations.* In order to establish both consistency and decidability, we exhibit a function computing *normal forms* of terms. Inspecting the possible normal forms in the empty context entails canonicity. Moreover, conversion of normal forms is easily decided, and so we can build on normalization to decide conversion. Finally, we can go further, and use normalization to build canonical representatives of typing and conversion derivations, which we rely on to relate our different systems.

*Injectivity of type constructors.* A more technical, but equally important property is injectivity of type constructors, for instance that whenever $\Pi\, x : A.B \cong \Pi\, x : A'.B'$, then $A \cong A'$ and $B \cong B'$. This property fails in extensional type theory, where the equational theory is too rich. For dependent type theories, injectivity of type constructors is the main stepping stone towards subject reduction, the fact that reduction is type-preserving, and thus included in conversion.

### 2.2 Neutrals, Normals, and Reduction

Before getting to how we establish these properties, we must introduce a last element: computation. Indeed, most conversion rules can be seen not just as equalities but be oriented as computations to be performed. This leads to the definition of weak-head reduction $\leadsto^\star$ in Figure 4, an evaluation strategy for open terms which reduces just as much as needed in order to uncover the head constructor of a term. This means reducing not just at top level: if our term is an application, we might need to reduce the function in order to expose a $\lambda$-abstraction and subsequently β-reduce the term with the (call-by-name) rule βRed. However, we do *not* allow reduction in the argument of an application, so that reduction remains deterministic: there is at most one possible reduct for any term. Weak-head reduction is the only reduction that is used throughout this article.

The normal forms (nf) for weak-head reduction, *i.e.* the terms that cannot reduce, are inductively characterized at the bottom of Figure 4, together with the companion notion of neutral forms (ne). Normal forms can be either a canonical term, starting with a head constructor (for instance, a $\lambda$-abstraction or $\varepsilon$), or a neutral term. Neutrals are stuck computations, blocked by a variable, *e.g.* $x\ u$ is stuck on $x$ and cannot reduce further.

### 2.3 Proof techniques

We can now go through the techniques we use to establish the properties of Section 2.1.

$\boxed{t \rightsquigarrow^1 t'}$ Term $t$ weak-head reduces in one step to term $t'$

$$\beta\textsc{Red}\ \frac{}{(\lambda\, x : A.t)\ u \rightsquigarrow^1 t[u]} \qquad \beta\textsc{RedNil}\ \frac{}{\mathrm{ind}_{\mathbf{List}\,A}(\varepsilon_A; x.P; b_\varepsilon, x.y.z.b_{::}) \rightsquigarrow^1 b_\varepsilon}$$

$$\beta\textsc{RedCons}\ \frac{}{\mathrm{ind}_{\mathbf{List}\,A}(a ::_A l; x.P; b_\varepsilon, x.y.z.b_{::}) \rightsquigarrow^1 b_{::}[a, l, \mathrm{ind}_{\mathbf{List}\,A}(l; x.P; b_\varepsilon, x.y.z.b_{::})]}$$

$$\textsc{RedApp}\ \frac{t \rightsquigarrow^1 t'}{t\ u \rightsquigarrow^1 t'\ u} \qquad \textsc{RedInd}\ \frac{t \rightsquigarrow^1 t'}{\mathrm{ind}_{\mathbf{List}\,A}(t; x.P; b_\varepsilon, x.y.z.b_{::}) \rightsquigarrow^1 \mathrm{ind}_{\mathbf{List}\,A}(t'; x.P; b_\varepsilon, x.y.z.b_{::})}$$

$\boxed{t \rightsquigarrow^\star t'}$ Term $t$ weak-head reduces in multiple steps to term $t'$

$$\textsc{RedBase}\ \frac{}{t \rightsquigarrow^\star t} \qquad \textsc{RedStep}\ \frac{t \rightsquigarrow^1 t' \qquad t' \rightsquigarrow^\star t''}{t \rightsquigarrow^\star t''}$$

$$\boxed{\mathrm{nf}\ f} \quad \stackrel{\text{def}}{=} \quad n \mid \Pi\, x : t.t \mid \mathrm{Type}_i \mid \mathbf{List}\, t \mid \lambda\, x : A.t \mid \varepsilon_A \mid t ::_A t \qquad \text{weak-head normal forms}$$

$$\boxed{\mathrm{ne}\ n} \quad \stackrel{\text{def}}{=} \quad x \mid n\ t \mid \mathrm{ind}_{\mathbf{List}\,A}(n; t; t, t) \qquad \text{weak-head neutrals}$$

Fig. 4. Weak-head reduction and normal forms ($t$ stands for an arbitrary term)

*Logical Relations.* Logical relations are our main tool to obtain normalization and canonicity results. At a high-level, we follow the approach of Abel, Öhman, and Vezzosi [2], where the logical relation is based on reducibility, a complex predicate on types and terms, which in particular entails the existence of a weak-head normal form. The key property is the fundamental lemma, stating that every well-typed term is reducible, *i.e.* that the logical relation is a model of MLTT. The existence of (deep) normal forms is obtained through the inspection of reducibility derivations for a term, since they contain iterated reduction steps to a normal form.

We use the logical relation not only to characterize the normal forms of terms but also the conversion between them, showing that a proof of convertibility between two terms can be transformed to a canonical shape interleaving weak-head reduction sequences and congruence steps between weak-head normal forms. We detail in Section 4 the novel challenges we encountered when adapting the approach of Abel, Öhman, and Vezzosi [2] to parametrized inductive types.

*Bidirectional Typing and Algorithmic Conversion.* Our second tool is a presentation of conversion and typing that, while still inductively defined, is as close as possible to an actual implementation. Typing is bidirectional [32, 46], *i.e.* decomposed into type inference and type checking, and essentially follows Lennon-Bertrand [32].[4] We use bidirectional typing for its rigid, canonical derivation structure, rather than for its ability to cut down type annotations on terms. Thus, although we use bidirectional judgements, all our terms infer a type, in contrast to what is common in the bidirectional literature [21, 43], where some terms can only be checked.

Algorithmic conversion, presented in Figure 5 combines ideas from both bidirectional typing and the presentation of Abel, Öhman, and Vezzosi [2]. Crucially, it gets rid entirely of the generic transitivity rule for conversion, and instead uses term-directed reduction, intertwined with comparison of the heads of weak-head normal forms. Algorithmic conversion is mutually defined with a second relation, dedicated to comparing weak-head neutral forms, called when encountering

[4]In line with Lennon-Bertrand [32], we pick $\rhd$ as the symbol for inference, and $\lhd$ as the one for checking, to avoid clashes with Coq's => in the formalization.

$\boxed{\Gamma \vdash n \approx n' \rhd T}$ Neutrals $n$ and $n'$ are comparable, inferring the type $T$

$$\textsc{NVar}\ \frac{(x : T) \in \Gamma}{\Gamma \vdash x \approx x \rhd T} \qquad \textsc{NApp}\ \frac{\Gamma \vdash n \approx_{\mathrm{h}} n' \rhd \Pi\, x : A.B \qquad \Gamma \vdash u \cong u' \lhd A}{\Gamma \vdash n\ u \approx n'\ u' \rhd B[u]}$$

$\boxed{\Gamma \vdash t \cong_{\mathrm{h}} t' \lhd T}$ Reduced terms $t$ and $t'$ are convertible at type $T$

$$\textsc{CList}\ \frac{\Gamma \vdash A \cong A' \lhd \mathrm{Type}_i}{\Gamma \vdash \mathbf{List}\ A \cong_{\mathrm{h}} \mathbf{List}\ A' \lhd \mathrm{Type}_i} \qquad \textsc{CProd}\ \frac{\Gamma \vdash A \cong A' \lhd \mathrm{Type}_i \qquad \Gamma, x : A' \vdash B \cong B' \lhd \mathrm{Type}_i}{\Gamma \vdash \Pi\, x : A.B \cong_{\mathrm{h}} \Pi\, x : A'.B' \lhd \mathrm{Type}_i}$$

$$\textsc{CFun}\ \frac{\Gamma, x : A \vdash f\ x \cong f'\ x \lhd B}{\Gamma \vdash f \cong_{\mathrm{h}} f' \lhd \Pi\, x : A.B} \qquad \textsc{CCons}\ \frac{\Gamma \vdash a \cong a' \lhd A'' \qquad \Gamma \vdash l \cong l' \lhd \mathbf{List}\ A''}{\Gamma \vdash a ::_A l \cong_{\mathrm{h}} a' ::_{A'} l' \lhd \mathbf{List}\ A''}$$

$$\textsc{NeuList}\ \frac{\Gamma \vdash n \approx_{\mathrm{h}} n' \rhd S}{\Gamma \vdash n \cong_{\mathrm{h}} n' \lhd \mathbf{List}\ A} \qquad \textsc{NeuNeu}\ \frac{\mathrm{ne}\ M \qquad \Gamma \vdash n \approx n' \rhd N}{\Gamma \vdash n \cong_{\mathrm{h}} n' \lhd M}$$

$\boxed{\Gamma \vdash t \cong t' \lhd T}$ Terms $t$ and $t'$ are convertible at type $T$

$\boxed{\Gamma \vdash n \approx_{\mathrm{h}} n' \rhd T}$ Neutrals $n$ and $n'$ are comparable, inferring reduced type $T$

$$\textsc{TmRed}\ \frac{t \rightsquigarrow^\star u \qquad t' \rightsquigarrow^\star u' \qquad T \rightsquigarrow^\star U \qquad \Gamma \vdash u \cong_{\mathrm{h}} u' \lhd U}{\Gamma \vdash t \cong t' \lhd T} \qquad \textsc{NRed}\ \frac{\Gamma \vdash n \approx n' \rhd T \qquad T \rightsquigarrow^\star S \qquad \mathrm{nf}\, S}{\Gamma \vdash n \approx_{\mathrm{h}} n' \rhd S}$$

Fig. 5. Algorithmic conversion (complete rules: Appendix C.2)

| Judgement | Input(s) | Inputs are well-formed |
|---|---|---|
| $\Gamma \vdash t \rhd T$ | $\Gamma, t$ | $\vdash \Gamma$ |
| $\Gamma \vdash t \lhd T$ | $\Gamma, T, t$ | $\vdash \Gamma$ and $\Gamma \vdash T$ |
| $\Gamma \vdash T \cong T' \lhd$ | $\Gamma$, $T$ and $T'$ | $\vdash \Gamma, \Gamma \vdash T$ and $\Gamma \vdash T'$ |
| $\Gamma \vdash t \cong t' \lhd T$ | $\Gamma, t, t'$ and $T$ | $\vdash \Gamma, \Gamma \vdash T, \Gamma \vdash t : T$ and $\Gamma \vdash t' : T$ |
| $\Gamma \vdash t \approx t' \rhd T$ | $\Gamma$, $t$ and $t'$ | $\vdash \Gamma$, ne $t$, ne $t'$, and $\exists A, A'$ s.t. $\Gamma \vdash t : A, \Gamma \vdash t' : A'$ |

Fig. 6. Well-formed inputs (for $\cong_{\mathrm{h}}$, $\approx_{\mathrm{h}}$, $\rhd_{\mathrm{h}}$, similar to their non-reduced variants)

neutrals at positive types. We think of general conversion as "checking", *i.e.* taking a type as input, while neutral comparison is "inferring", *i.e.* the type is an output. In turn, conversion is used in the following typing rule to compare the inferred type for $t$ with the one it should check against.

$$\textsc{Check}\ \frac{\Gamma \vdash t \rhd T' \qquad \Gamma \vdash T' \cong T \lhd}{\Gamma \vdash t \lhd T}$$

Using the consequences of the logical relation, we can show that this algorithmic presentation has many desirable properties. For instance, transitivity is admissible, even though there is no dedicated rule. Collecting the properties derived from the logical relation, we can obtain our second main objective: equivalence between the algorithmic and declarative presentations.

Property 2.1 (Equivalence of the Presentations). *If* $\Gamma \vdash t : T$, *then* $\Gamma \vdash t \lhd T$. *Conversely, if* $\vdash \Gamma$, $\Gamma \vdash T$ *and* $\Gamma \vdash t \lhd T$, *then* $\Gamma \vdash t : T$.

Note that the implication from the bidirectional judgement to the declarative one only holds if the context and type are well-formed. In general, our algorithmic presentations are "garbage-in, garbage-out": they maintain well-formation of types and contexts, but do not enforce them. Thus, most properties of the algorithmic derivations only hold if their inputs are well-formed, in the sense of Figure 6. Note that in checking and inference modes, while the term is an input, it is of course not assumed to be well-formed in advance, since this is what the judgement itself asserts. This algorithmic, syntax-directed presentation is well suited for implementations and to establish relationships between type systems.

## 3 A FUNCTORIAL TYPE THEORY

We develop an extension $\mathrm{MLTT}_{\mathrm{map}}$ of MLTT with primitive $\mathrm{map}_F$ operations for each parametrized type former $F$ of MLTT, that is $\Pi, \Sigma, +$, $\mathbf{List}$, $\mathbf{W}$, and $\mathbf{Id}$. These map operations internalize the functorial character of the type formers,[5] and by design *definitionally* satisfy the functor laws for each type former $F$:

$$\mathrm{map}_F\ \mathrm{id} \cong \mathrm{id} \qquad \text{(id-eq)}$$

$$\mathrm{map}_F\ f \circ \mathrm{map}_F\ g \cong \mathrm{map}_F\ (f \circ g) \qquad \text{(comp-eq)}$$

Section 3.1 describes the structure needed on type formers to state their functoriality in $\mathrm{MLTT}_{\mathrm{map}}$. In Section 3.2 we show how definitionally functorial $\mathrm{map}_F$ are definable in vanilla MLTT for type formers with an η-law. Section 3.3 introduces the main content of this paper, required to enforce the functor laws on inductive type formers: the extension of the equational theory on neutral terms. We explain the technical design choices needed to define and use the logical relations for $\mathrm{MLTT}_{\mathrm{map}}$ and obtain as a consequence that the theory enjoys consistency, canonicity, and decidable conversion and type-checking. We implement these design choices in Coq for a simplified but representative version of $\mathrm{MLTT}_{\mathrm{map}}$, with one universe and the $\Pi$, $\Sigma$, $\mathbf{List}$ and $\mathbf{N}$ type formers, with their respective map operators. This formalization is detailed in Section 4.

### 3.1 Functorial Structure on Type Formers

In order to state the functor laws for a type former $F$, such as $\Pi, \Sigma, \mathbf{List}, \mathbf{W}, \mathbf{Id}$, we must specify the categorical structures involved. A type former $F$ is parametrized by a telescope of parameters that we collectively refer to as $\mathrm{dom}(F)$, and produces a type. We will always equip the codomain Type of a type former $F$ with the category structure of functions between types, with the standard identity and composition. Note that composition is associative and unital up to conversion, thanks to η-laws on function types.

The domain $\mathrm{dom}(F)$ of a type former must also be equipped with the structure of a category. We introduce the judgement $\Delta \vdash_{\mathrm{map}} X : \mathrm{dom}(F)$ to stand for a substitution in context $\Delta$ of the telescope of parameters of $F$. Then, given two such instances $X_1$ and $X_2$ of parameters for $F$, morphisms between $X_1$ and $X_2$ are classified by the judgement $\Delta \vdash_{\mathrm{map}} \varphi : \mathrm{hom}_F(X_1, X_2)$. We require $\mathrm{dom}(F)$ to be also equipped with identities and a definitionally associative and unital composition:

[5] These equations are all propositionally true in MLTT, proven by induction for datatypes.

| Type former $F$ | Domain $\Delta \vdash_{\text{map}} X : \text{dom}(F)$ | Morphisms $\Delta \vdash_{\text{map}} \varphi : \text{hom}_F(\cdot_1, \cdot_2)$ |
|---|---|---|
| **List** | $X = (A) \wedge \Delta \vdash_{\text{map}} A$ | $\varphi = (f) \wedge \Delta \vdash_{\text{map}} f : A_1 \to A_2$ |
| $\Pi$ | $X = (A, B) \wedge \Delta \vdash_{\text{map}} A$ <br> $\wedge \quad \Delta, a : A \vdash_{\text{map}} B$ | $\varphi = (f, g) \wedge \Delta \vdash_{\text{map}} f : A_2 \to A_1$ <br> $\wedge\, \Delta, a : A_2 \vdash_{\text{map}} g : B_1[f\,a] \to B_2$ |
| $\Sigma$ | idem | $\varphi = (f, g) \wedge \Delta \vdash_{\text{map}} f : A_1 \to A_2$ <br> $\wedge\, \Delta, a : A_1 \vdash_{\text{map}} g : B_1 \to B_2[f\,a]$ |
| **W** | idem | $\varphi = (f, g) \wedge \Delta \vdash_{\text{map}} f : A_1 \to A_2$ <br> $\wedge\, \Delta, a : A_1 \vdash_{\text{map}} g : B_2[f\,a] \to B_1$ |
| **Id** | $X = (A, x, y) \wedge \Delta \vdash_{\text{map}} A$ <br> $\wedge \quad \Delta \vdash_{\text{map}} x : A$ <br> $\wedge \quad \Delta \vdash_{\text{map}} y : A$ | $\varphi = (f) \wedge \Delta \vdash_{\text{map}} f : A_1 \to A_2$ <br> $\wedge \quad \Delta \vdash_{\text{map}} f\,x_1 \cong x_2 : A_2$ <br> $\wedge \quad \Delta \vdash_{\text{map}} f\,y_1 \cong y_2 : A_2$ |
| $+$ | $X = (A, B) \wedge \Delta \vdash_{\text{map}} A$ <br> $\wedge \quad \Delta \vdash_{\text{map}} B$ | $\varphi = (f, g) \wedge \Delta \vdash_{\text{map}} f : A_1 \to A_2$ <br> $\wedge \quad \Delta \vdash_{\text{map}} g : B_1 \to B_2$ |

Fig. 7. Domain and categorical structure on type formers

$$\frac{\Delta \vdash_{\text{map}} X : \text{dom}(F)}{\Delta \vdash_{\text{map}} \text{id}^F_X : \text{hom}_F(X, X)} \qquad \frac{\Delta \vdash_{\text{map}} \varphi : \text{hom}_F(X, Y) \qquad \Delta \vdash_{\text{map}} \psi : \text{hom}_F(Y, Z)}{\Delta \vdash_{\text{map}} \psi \circ^F \varphi : \text{hom}_F(X, Z)}$$

For instance, for dependent products, $\text{dom}(\Pi)$ and $\text{hom}_\Pi$ are given by

$$\begin{array}{rcl} \Delta \vdash_{\text{map}} (A, B) : \text{dom}(\Pi) & \iff & \Delta \vdash_{\text{map}} A \wedge \Delta, a : A \vdash_{\text{map}} B \\ \Delta \vdash_{\text{map}} (f, g) : \text{hom}_\Pi((A_1, B_1), (A_2, B_2)) & \iff & \Delta \vdash_{\text{map}} f : A_2 \to A_1 \quad \wedge \\ & & \quad \Delta, a : A_2 \vdash_{\text{map}} g : B_1[f\,a] \to B_2 \end{array}$$

with identity $\text{id}^\Pi_{(A,B)} \stackrel{\text{def}}{=} (\text{id}_A, \text{id}_B)$ and composition $(f, g) \circ^\Pi (f', g') \stackrel{\text{def}}{=} (f' \circ f, g \circ g')$.

The domain and morphism for each type former are described in Figure 7. Identities and compositions are given by the categorical structure on Type for **List** and **Id**, and are defined componentwise, for $\Sigma$, **W** and $+$, similarly to $\Pi$. Figure 8 presents the conversion rules of $\text{MLTT}_{\text{map}}$, extending those of MLTT with general functoriality rules and specific rules for each type former. For each type former $F$, $\text{map}_F$ is introduced using Map and witnesses the functorial nature of $F$, that is $F$ maps morphisms $\varphi$ in its domain between two instances of its parameters $X, Y$ (left implicit) to functions between types

$$\Delta \vdash_{\text{map}} \varphi : \text{hom}_F(X, Y) \qquad \Longrightarrow \qquad \Delta \vdash_{\text{map}} \text{map}_F\ \varphi : F\,X \to F\,Y$$

These mapping operations obey the two functor laws, as stated by MapId and MapComp.

The computational behaviour of maps, as defined by weak-head reduction, depends on the type former. On $\Pi$ and $\Sigma$, map is defined by its observation, namely application for $\Pi$ and first and second projections for $\Sigma$. On inductive types such as **List**, **W**, **Id** and $+$, map traverses constructors, applying the provided morphism on elements of the parameter type(s), and itself to recursive arguments. This corresponds to the usual notion of map on lists. On **W**-types, the map operation relabels the nodes of the trees using its first component, and reorganizes the subtrees according to its second component. On identity types, the reflexivity proof $\text{refl}_{A_1,a}$ at a point $a : A_1$ is mapped to the reflexivity proof at $f\,a : A_2$ for $f : A_1 \to A_2$. On sum types $A + B$, either the first or second

For each type former $F$ ($\Pi$, $\Sigma$, **List**, **W**, **Id**, $+$)

$$\textsc{Map}\ \frac{\Gamma \vdash_{\mathrm{map}} X, Y : \mathrm{dom}(F) \qquad \Gamma \vdash_{\mathrm{map}} f : \mathrm{hom}_F(X,Y)}{\Gamma \vdash_{\mathrm{map}} \mathrm{map}_F\, f : F\,X \to F\,Y} \qquad \textsc{MapId}\ \frac{\Gamma \vdash_{\mathrm{map}} X : \mathrm{dom}(F) \qquad \Gamma \vdash_{\mathrm{map}} t : F\,X}{\Gamma \vdash_{\mathrm{map}} \mathrm{map}_F\, \mathrm{id}^F_X\ t \cong t : F\,X}$$

$$\textsc{MapComp}\ \frac{\Gamma \vdash_{\mathrm{map}} X, Y, Z : \mathrm{dom}(F) \qquad \Gamma \vdash_{\mathrm{map}} g : \mathrm{hom}_F(X,Y) \qquad \Gamma \vdash_{\mathrm{map}} f : \mathrm{hom}_F(Y,Z) \qquad \Gamma \vdash_{\mathrm{map}} t : F\,X}{\Gamma \vdash_{\mathrm{map}} \mathrm{map}_F\ f\ (\mathrm{map}_F\, g\, t) \cong \mathrm{map}_F (f \circ^F g)\, t : F\,Z}$$

Specific rules

$$\mathrm{map}_{\mathbf{List}}\ f\ (hd :: tl) \rightsquigarrow^1 f\ hd :: \mathrm{map}_{\mathbf{List}}\ f\ tl \qquad \mathrm{map}_{\mathbf{List}}\ f\ \varepsilon \rightsquigarrow^1 \varepsilon$$

$$\pi_1\ (\mathrm{map}_\Sigma\ f\, p) \rightsquigarrow^1 (\pi_1\, f)\ (\pi_1\, p) \qquad \pi_2\ (\mathrm{map}_\Sigma\ f\, p) \rightsquigarrow^1 (\pi_2\, f)\ (\pi_2\, p)$$

$$\mathrm{map}_\Pi\ f\, h\, t \rightsquigarrow^1 (\pi_2\, f)\ (h\ (\pi_1\ f\ t)) \qquad \mathrm{map}_{\mathbf{Id}}\ f\ \mathrm{refl}_{A_1,a} \rightsquigarrow^1 \mathrm{refl}_{A_2,f\,a}$$

$$\mathrm{map}_+\ (f,g)\ (\mathrm{inj}^l\ a) \rightsquigarrow^1 \mathrm{inj}^l\ (f\,a) \qquad \mathrm{map}_+\ (f,g)\ (\mathrm{inj}^r\ b) \rightsquigarrow^1 \mathrm{inj}^r\ (g\,b)$$

$$\mathrm{map}_{\mathbf{W}}\{T_1\}\{T_2\} f\ (\sup a\ k) \rightsquigarrow^1$$
$$\sup_{x.\pi_2\,T_2} (\pi_1\ f\ a)\ (\lambda\, x : (\pi_2\, T_2\ (\pi_1\ f\ a)).\ \mathrm{map}_{\mathbf{W}}\ f\ (k\ (\pi_2\, g\, x)))$$

$$\textsc{RedMapComp}\ \frac{\mathrm{ne}\ n \qquad F \in \{\mathbf{List}, \mathbf{Id}, +, \mathbf{W}\}}{\mathrm{map}_F\ f\ (\mathrm{map}_F\, g\, n) \rightsquigarrow^1 \mathrm{map}_F (f \circ^F g)\ n}$$

Fig. 8. $\mathsf{MLTT}_{\mathrm{map}}$ (extends Figures 2 to 5, complete rules: Appendix C.3)

component of the morphism $(f, g)$ is employed depending on the constructor $\mathrm{inj}^l$ or $\mathrm{inj}^r$. Each reduction rule has a corresponding conversion rule that can be found in Appendix C.3.

*Functorial Maps and Type Former Encodings.* Positive sum types $A + B$ can be simulated in MLTT by the type $\Sigma\, b : \mathbf{B}\,.\delta(b, A, B)$, using the branching operation $\delta(b, A, B) \stackrel{\text{def}}{=} \mathrm{ind}_{\mathbf{B}}(b; z.\,\mathrm{Type}_i; A, B)$. This encoding admits the adequate introduction and elimination rules. It induces a mapping from $\mathrm{dom}(+)$ to $\mathrm{dom}(\Sigma)$, sending a morphism $\Delta \vdash_{\mathrm{map}} (f, g) : \mathrm{hom}_+((A_1, B_1), (A_2, B_2))$ to the morphism $\Delta \vdash_{\mathrm{map}} (\mathrm{id}_{\mathbf{B}}, f \oplus g) : \mathrm{hom}_\Sigma((\mathbf{B}, \delta(b, A, B)), (\mathbf{B}, \delta(b, A', B')))$ where $f \oplus g$ is

$$\Delta, b : \mathbf{B} \vdash_{\mathrm{map}} \mathrm{ind}_{\mathbf{B}}(b; z.\delta(z, A, B) \to \delta(z, A', B'); f, g) : \delta(b, A, B) \to \delta(b, A', B').$$

We can show by case analysis on $\mathbf{B}$ that this mapping satisfies the propositional functor laws. However, it falls short from satisfying the definitional ones.[6] It is thus not enough to compose $\mathrm{map}_\Sigma$ with this mapping to obtain a functorial action on sum types $A + B$, and explains why we add $+$ primitively.

This obstruction to inductive encodings would motivate a general definition of functorial map for a scheme of indexed inductive types. However, it seems already non-trivial to specify the categorical structure on the domain of an arbitrary inductive type, let alone generate the type and equations for the corresponding map operation. Thus, we rather concentrate on understanding the theory on quintessential examples, leaving out a general treatment to future work.

[6] This would amount to an instance of the η-law for $\mathbf{B}$.

### 3.2 Extensional Types and Map

A type $A$ is extensional when its elements are characterized by their observation, *i.e.* any element is convertible to its η-expansion, an elimination followed by an introduction – an equation usually called η-law. For extensional type formers, it is possible to define a map operation satisfying the functor laws. In MLTT and $\mathrm{MLTT}_{\mathrm{map}}$, both (strong) dependent sums Σ and dependent products Π have such extensionality laws, and so their map operations are definable.

$$\begin{aligned}\mathrm{map}_{\Pi}\ ((g,f) : \mathrm{hom}_{\Pi}((A,B),(A',B')))\ (h : \Pi(x:A)B) &\stackrel{\text{def}}{=} \lambda\, x : A'.f\ (h\ (g\ x))\\ \mathrm{map}_{\Sigma}\ ((g,f) : \mathrm{hom}_{\Sigma}((A,B),(A',B')))\ (p : \Sigma(x:A)B) &\stackrel{\text{def}}{=} (g\ (\pi_1\, p), f\ (\pi_2\, p))\end{aligned}$$

Lemma 3.1. $\mathrm{map}_{\Pi}$ *and* $\mathrm{map}_{\Sigma}$ *satisfy the definitional functor laws* MapId *and* MapComp*.*

Appendix D.1 gives a direct proof, and the accompanying artifact also shows that the functor laws hold for Coq's Π and Σ types.[7] The specific rules of Figure 8 hold by β-reduction.

### 3.3 New Equations for Neutral Terms in Dependent Type Theory

Inductive types in MLTT do not satisfy a definitional η-law. For identity types, the η-law is equivalent to the equality reflection principle of extensional MLTT, whose equational theory is undecidable [15, 29]. Extensionality principles for inductive types with recursive occurrences as **List** or **W** are also likely to break the decidability of the equational theory, by adapting an argument for streams [42]. The result of the previous section hence does not apply, and it is instructive to look at the actual obstruction. Consider the case of **List**, and the equation for preservation of identities:

$$\Gamma \vdash_{\mathrm{map}} \mathrm{map}_{\mathbf{List}}\ \mathrm{id}_A^{\mathbf{List}}\ l \cong l : \mathbf{List}\ A. \qquad (\star)$$

If we were to define $\mathrm{map}_{\mathbf{List}}$ by induction on lists as is standard, we would get

$$\mathrm{map}_{\mathbf{List}}(f : A \to B)\,(l : \mathbf{List}\ A) \stackrel{\text{def}}{=} \mathrm{ind}_{\mathbf{List}}(\mathbf{List}\ B; l; \varepsilon_B, hd.tl.ih_{tl}.(f\, hd) ::_B ih_{tl})$$

We can observe that Eq. ($\star$) is validated on closed canonical terms of type **List**:

$$\begin{aligned}\mathrm{map}_{\mathbf{List}}\ \mathrm{id}_A\ \varepsilon_A \cong \varepsilon_A\ &\mathrm{map}_{\mathbf{List}}\ \mathrm{id}_A\ (hd ::_A tl)\\ &\cong (\mathrm{id}_A\ hd) ::_A \mathrm{map}_{\mathbf{List}}\ \mathrm{id}_A\ tl \stackrel{\text{ind. hyp.}}{\cong} hd ::_A tl\end{aligned}$$

However, on neutral terms, typically variables, we are stuck as long as we stay within the equational theory of MLTT:

$$A : \mathrm{Type}, x : \mathbf{List}\ A \nvdash \mathrm{map}_{\mathbf{List}}\ \mathrm{id}_A\ x \cong x : \mathbf{List}\ A.$$

In order to validate Eq. ($\star$), $\mathrm{MLTT}_{\mathrm{map}}$ must thus at the very least extend the equational theory on neutral terms. Allais, McBride, and Boutillier [6] show in the simply-typed case that these equations between neutral terms are actually the only obstruction to functor laws, and in the remainder of this section we discuss how to adapt MLTT to this idea.

[7] In file mapPiSigmaFunctorLaws.

| | | | |
|---|---|---|---|
| $\boxed{\text{nf } f}$ | $\stackrel{\text{def}}{=}$ | $\cdots \mid c$ | weak-head normal forms |
| $\boxed{\text{ne } n}$ | $\stackrel{\text{def}}{=}$ | $\cdots \mid \text{ind}_{\textbf{List}\,A}(c;t;t)$ | weak-head neutrals |
| $\boxed{\text{cne } c}$ | $\stackrel{\text{def}}{=}$ | $n \mid \text{map}_{\textbf{List}}\, f\, n$ | compacted neutrals |

Fig. 9. Weak-head normal and neutrals for $\text{MLTT}_{\text{map}}$ (extends Figure 4)

*Map Composition and Compacted Neutrals.* The first step in order to validate the functor laws is to get as close as possible to a canonical representation during reduction. In order to deal with composition of maps, we extend reduction with RedMapComp, merging consecutive stuck maps. In order to preserve the deterministic nature of weak-head reduction, map compaction should only apply when no other rule does. To achieve this, the type former $F$ should not be extensional, because $\text{map}_{\Pi}$ is already handled through the η-expansion of CFun, and similarly for $\text{map}_{\Sigma}$. Moreover, the mapped term should be neither a canonical form where map already has a computational behaviour, nor a map itself that could fire the same rule. To control this, we separate neutrals, which cannot contain a map as their head, and *compacted neutrals*, which can start with at most one map, as shown in Figure 9 alongside normal forms. Allais, McBride, and Boutillier [6] also features a similar decomposition of normal forms into three different classes, although their normal forms for lists are more complex than ours as they validate more definitional equations than functor laws.

*Map on Identities.* For identities, using a similar reduction-based approach is difficult: turning the equation $\Gamma \vdash_{\text{map}} \text{map}_{\textbf{List}}\ \text{id}_A\ l \cong l : \textbf{List}\ A$ into a reduction raises issues similar to those encountered with η-laws. Orienting it as an expansion $l \rightsquigarrow^{\star} \text{map}_{\textbf{List}}\ \text{id}_A\ l$ requires knowledge of the type to ensure the expansion only applies to lists, and is potentially non-terminating. Accommodating type-directed reduction would require a deep reworking of our setting.

As a result, just like for η on functions in rule CFun, we implement this rule as part of conversion, rather than as a reduction. We also incorporate it carefully in the notion of reducible conversion in the logical relation, where we have access to enough properties of the type theories. Since the equation is always validated by canonical forms, we only need to enforce it on compacted neutrals. The logical relation for an inductive type $I$ ($\textbf{List}$, $\textbf{W}$, $\textbf{Id}$, $+$) thus specifies that a neutral $n$ is reducibly convertible to a compacted neutral $\text{map}_I\ f\, m$, whenever the neutrals $n$ and $m$ are convertible and $f$ agrees with the identity of $\text{dom}(I)$ on any neutral term. See MapNeConvRedL in the next section for the exact rule.

*Eliminators: fusion or no fusion?* When considering the interaction between map and the eliminator $\text{ind}_{\textbf{List}}$, a design choice arises: should we also fuse them, *i.e.* implement the following reduction rule, which pushes the map from the scrutinee into the branches?

$$\begin{aligned}&\text{ind}_I(\text{map}_{\textbf{List}}\ f\, n; l.P; b_{\varepsilon}, a.l.h.b_{::}) \rightsquigarrow^1\\ &\qquad \text{ind}_I(n; l.P[\text{map}_{\textbf{List}}\ f\, l]; b_{\varepsilon}, a.l.h.b_{::}[\text{id}, f\, a, \text{map}_{\textbf{List}}\ f\, l, h])\end{aligned}$$

From the point of view of functorial equations, this is not necessary. Thus, in Figure 9 and the rest of this paper we take the most conservative approach, and do not add this rule.

However, in the setting of subsumptive bidirectional subtyping, this fusion is necessary if we wish to infer the parameters of the inductive types from the scrutinee (as in Fus below), rather than store them in the induction node (as in NoFus).

$$\textsc{Fus}\ \frac{\Gamma \vdash_{\text{sub}} s \rhd_{\text{h}} \textbf{List}\, A \qquad \Gamma, l : \textbf{List}\, A \vdash_{\text{sub}} P \rhd_{\text{h}} \text{Type} \qquad \ldots}{\Gamma \vdash_{\text{sub}} \text{ind}_{\textbf{List}}(s; l.P; \ldots) \rhd P[s]}$$

$$\textsc{NoFus}\ \frac{\Gamma \vdash_{\text{sub}} A \lhd \qquad \Gamma \vdash_{\text{sub}} s \lhd \textbf{List}\, A \qquad \Gamma, l : \textbf{List}\, A \vdash_{\text{sub}} P \rhd_{\text{h}} \text{Type} \qquad \ldots}{\Gamma \vdash_{\text{sub}} \text{ind}_{\textbf{List}\, A}(s; l.P; \ldots) \rhd P[s]}$$

Rule Fus is more appealing, as it removes an unnecessary conversion test between the type of $s$ and that stored in the node. Yet, elaborating it to a coercive system requires this target to have the extra fusion law above. Intuitively, this is because rule Fus does not fix the parameter type at which the eliminator is typed, and so this parameter can change, which in a coercive system corresponds to pushing coercions into the branches, as in the fusion equation above.

*Experimenting $\text{MLTT}_{\text{map}}$ through rewrite rules.* Even though we have not attempted a justification of the metatheory of $\text{MLTT}_{\text{map}}$ with a presentation purely based on rewriting, it is still possible to use oriented version of the functor laws to experiment with this theory: Agda experimentally supports rewrite rules [17] while ongoing implementation work exists for Coq [25]. As an illustration, we implemented Example 1.1 in Agda.[8] Concretely, we postulate a new constant $\text{map}_F$ and add the following rules:

$$\text{map}_F\ B'\, C\ f\,(\text{map}_F\ A\, B\, g\, x) \rightsquigarrow^1 \text{map}_F\ A\, C\,(\lambda z : A.f\,(g\, z))\, x \qquad \text{(comp-rew)}$$

$$\text{map}_F\ A\, A'\,(\lambda z : A''.z)\, x \rightsquigarrow^1 x \qquad \text{(id-rew)}$$

together with the usual definition of $\text{map}_F$ on the constructors of the type former $F$. We rely on typing information to enforce that redundant data coincide, for instance that $A, A'$ and $A''$ are convertible in id-rew.

## 4 FORMALIZING NEW EQUATIONS FOR NEUTRAL LISTS

In this section we expose the main components of the accompanying Coq formalization, which covers normalization, equivalence of declarative and algorithmic typing, decidability of type-checking, and canonicity for a subset of $\text{MLTT}_{\text{map}}$ with $\mathbf{0}, \mathbf{N}, \Pi, \Sigma,$ **List** and a single universe. The formalization extends a port to Coq [3] of a previous Agda formalization [2], which has already been extended multiple times [24, 47, 48]. We focus on the challenges to establish the functor laws on lists, and direct the reader either to the Coq code, or to Abel, Öhman, and Vezzosi and Adjedj et al. for other details. The formalization spans ~26k lines of code, approximately 9k of which are specific to our extension with lists and definitionally functorial maps and are new compared to Adjedj et al. Text in blue refer to files in the companion artifact.

### 4.1 A Logical Relation with Functor Laws on List

The Coq development defines both declarative and algorithmic presentations of $\text{MLTT}_{\text{map}}$ and proves their equivalence through a logical relation parametrized by a generic typing interface[9] instantiated by both presentations. Beyond generic variants of the typing and conversion judgement, the interface uses two extra judgements: $\Gamma \vdash_{\text{map}} t \rightsquigarrow^\star t' : A$ stating that $t$ reduces to $t'$ and that they are both well typed at type $A$ in context $\Gamma$; and $\Gamma \vdash_{\text{map}} n \approx n' : A$ stating that $n$ and $n'$ are convertible neutral terms.

[8]See file `map.agda` in the companion artifact.
[9]Defined in GenericTyping

*Definition of the Logical Relation.* In presence of dependent types, the standard strategy of reducibility proofs defining reducibility of terms by induction on their types fails. Rather, reducibility of types and of terms are define mutually mutually, the latter defined out of a witness of the former, and the former reusing the latter for the universe. Following Abel, Öhman, and Vezzosi [2], we thus first define for each type former $F$ what it means to be a type reducible as $F$, and then what it means to be a reducible term and reducibly convertible terms at such a type reducible as $F$. A type is then reducible if it is reducible as $F$ for some type former $F$. As we extend the logical relation to handle $\mathbf{List}$ and $\mathrm{map}_{\mathbf{List}}$, we focus on a high level description of the reducibility of types as lists and the reducible convertibility of terms of type $\mathbf{List}$, the most challenging elements in the definition.[10] Two points required specific attention with respect to prior work. First, to handle the fact that constructors contain their parameters, we need to impose reducible conversions between these and the parameters coming from the type. Second, in order to validate composition of map on neutrals that may contain a map, we need to equip neutrals with additional reducibility data, rather than pure typing information.

A type $X$ is reducible as a list in context $\Gamma$, written $\Gamma \Vdash_{\mathbf{List}} X$, if it weak-head reduces to $\mathbf{List}\ A$ for some parameter type $A$ reducible in any context $\Delta$ extending $\Gamma$ via a weakening $\rho : \mathsf{Wk}(\Delta, \Gamma)$. If $\mathfrak{R} : \Gamma \Vdash_{\mathbf{List}} X$ is a witness that $X$ is reducible as a list, then $\mathbb{P}(\mathfrak{R})$ stands for the parameter type $A$ of this witness, and $\mathbb{P}_{\Vdash}(\mathfrak{R}) : \Pi\{\rho : \mathsf{Wk}(\Delta, \Gamma)\}.\Delta \Vdash \mathbb{P}(\mathfrak{R})[\rho]$ is its witness of reducibility.

Reducible conversion of terms as lists $\Gamma \Vdash t \cong t' : A \mid \mathfrak{R}$ is defined in Figure 10. Two terms $t$ and $t'$ are reducibly convertible as lists with respect to the witness of reducibility $\mathfrak{R} : \Gamma \Vdash_{\mathbf{List}} X$ if they reduce to normal forms $v, v'$ that are reducibly convertible as normal forms of type list $\Gamma \Vdash_{\mathrm{nf}} v \cong v' : A \mid \mathfrak{R}$ (ListRed). Straightforwardly, two canonical forms are convertible if they are both $\varepsilon$ (NilRed) or both $- :: -$ (ConsRed) with reducibly convertible heads and tails.

For compacted neutral forms, we need to consider four cases according to whether each of the left or the right hand-side term is a $\mathrm{map}_{\mathbf{List}}$. NeRed provides the easy case where both terms are actually neutral, with a single premise requiring that these are convertible as neutrals for the generic typing interface. MapMapConvRed gives the congruence rule for stuck $\mathrm{map}_{\mathbf{List}}$, relating $\mathrm{map}_{\mathbf{List}}\ f\, n$ and $\mathrm{map}_{\mathbf{List}}\ f'\, n'$ when the mapped lists $n$ and $n'$ are convertible as neutrals and the bodies $f\,x$ and $f'\,x$ of the functions are reducibly convertible. Note that at this point of the logical relation, we do not know that the domain of the functions $f$ and $f'$ is reducible, only that their codomain is, as provided by $\mathbb{P}_{\Vdash}(\mathfrak{R})$. This constraint motivates both the η-expansion of the functions on the fly before comparing them, and the necessity of a Kripke-style quantification on larger contexts for the reducibility of the parameter type $\mathbb{P}_{\Vdash}(\mathfrak{R})$, together ensuring that the recursive reducible conversion happens at a reducible type, namely an adequate instance of $\mathbb{P}(\mathfrak{R})$. Finally, the symmetric rules NeMapConvRedR and MapNeConvRedL deal with the comparison of a $\mathrm{map}_{\mathbf{List}}$ against a neutral $n$, that can be morally thought as $\mathrm{map}_{\mathbf{List}}\ \mathrm{id}\ n$, and indeed the premises correspond to what one would obtain with MapMapConvRed in that case, up to an inlined β-reduction step.

*Validity of the Functor Laws.* All the expected properties extend to this new logical relation: reflexivity, symmetry, transitivity, irrelevance with respect to reducible conversion, stability by weakening and anti-reduction.[11] These properties are essential in order to show that the logical relation validates the functor laws on any reducible term. The proof proceeds through an usual argument for logical relations: on canonical forms, the functor laws hold as observed already in Section 3.3; on compacted neutrals and neutral forms, we need to show that any compositions of $\mathrm{map}_{\mathbf{List}}$ reduce to a single map of a function with a reducible body, which amounts to show that

[10] Available in file LogicalRelation.
[11] Available in the directory LogicalRelation.

$$\textsc{ListRed}\ \frac{\Gamma \vdash_{\mathrm{map}} t \rightsquigarrow^{\star} v : \mathbf{List}\ \mathbb{P}(\mathfrak{R}) \qquad \Gamma \vdash_{\mathrm{map}} t' \rightsquigarrow^{\star} v' : \mathbf{List}\ \mathbb{P}(\mathfrak{R}) \qquad \Gamma \Vdash_{\mathrm{nf}} v \cong v' : X \mid \mathfrak{R}}{\Gamma \Vdash t \cong t' : X \mid \mathfrak{R}}$$

$$\textsc{ConsRed}\ \frac{\begin{array}{c}\Gamma \Vdash \mathbb{P}(\mathfrak{R}) \cong P \mid \mathbb{P}_{\Vdash}(\mathfrak{R}) \qquad \Gamma \Vdash \mathbb{P}(\mathfrak{R}) \cong P' \mid \mathbb{P}_{\Vdash}(\mathfrak{R}) \\ \Gamma \Vdash hd \cong hd' : \mathbb{P}(\mathfrak{R}) \mid \mathbb{P}_{\Vdash}(\mathfrak{R}) \qquad \Gamma \Vdash tl \cong tl' : X \mid \mathbb{P}_{\Vdash}(\mathfrak{R})\end{array}}{\Gamma \Vdash_{\mathrm{nf}} hd ::_{P} tl \cong hd' ::_{P'} tl' : X \mid \mathfrak{R}}$$

$$\textsc{NilRed}\ \frac{\begin{array}{c}\Gamma \Vdash \mathbb{P}(\mathfrak{R}) \cong P \mid \mathbb{P}_{\Vdash}(\mathfrak{R}) \\ \Gamma \Vdash \mathbb{P}(\mathfrak{R}) \cong P' \mid \mathbb{P}_{\Vdash}(\mathfrak{R})\end{array}}{\Gamma \Vdash_{\mathrm{nf}} \varepsilon_{P} \cong \varepsilon_{P'} : X \mid \mathfrak{R}} \qquad \textsc{NeRed}\ \frac{\Gamma \vdash_{\mathrm{map}} n \approx n' : \mathbf{List}\ \mathbb{P}(\mathfrak{R})}{\Gamma \Vdash_{\mathrm{nf}} n \cong n' : X \mid \mathfrak{R}}$$

$$\textsc{MapNeConvRedL}\ \frac{\begin{array}{c}\Gamma \vdash_{\mathrm{map}} n \approx n' : \mathbf{List}\ \mathbb{P}(\mathfrak{R}) \\ \Gamma, x : \mathbb{P}(\mathfrak{R}) \Vdash f\, x \cong x : \mathbb{P}(\mathfrak{R}) \mid \mathbb{P}_{\Vdash}(\mathfrak{R})\end{array}}{\Gamma \Vdash_{\mathrm{nf}} \mathrm{map}_{\mathbf{List}}\ f\, n \cong n' : X \mid \mathfrak{R}} \qquad \textsc{NeMapConvRedR} \ldots$$

$$\textsc{MapMapConvRed}\ \frac{\Gamma \vdash_{\mathrm{map}} n \approx n' : \mathbf{List}\ A \qquad \Gamma, x : A \Vdash f\, x \cong f'\, x : \mathbb{P}(\mathfrak{R}) \mid \mathbb{P}_{\Vdash}(\mathfrak{R})}{\Gamma \Vdash_{\mathrm{nf}} \mathrm{map}_{\mathbf{List}}\ f\, n \cong \mathrm{map}_{\mathbf{List}}\ f'\, n' : X \mid \mathfrak{R}}$$

Fig. 10. Reducible convertibility of lists (where $\mathfrak{R}$ is a proof of $\Gamma \Vdash_{\mathbf{List}} X$)

composing reducible functions produces reducible outputs on reducible inputs. This last step in the proof reflect our assumption that the categorical structure equipping domains of type formers, here dom($\mathbf{List}$), should be definitionally associative and unital.

## 4.2 Deciding Conversion and Typechecking for $\mathsf{MLTT}_{\mathrm{map}}$

Instantiating the generic typing interface of the logical relation with declarative typing provides metatheoretic consequences of the existence of normal forms, among which normalization, injectivity of type constructors and subject reduction. Using those, we can show that algorithmic typing is sound directly by induction, and also that it fits the generic typing interface of the logical relation, which lets us derive that it is complete with respect to declarative typing.

This part of the proof is close to Abel, Öhman, and Vezzosi [2] and Adjedj et al. [3]. The main change is that we adapt algorithmic conversion to reflect the addition of compacted neutrals in our definition of normal forms, by introducing a third mutually defined relation to compare these compacted neutrals. The main idea is summed up in rules ListNeConv and ListNeMap below: when comparing compacted neutrals, we use the new relation $\approx_{\mathrm{map}}$, which simulates the behaviour of the logical relation from Figure 10 on compacted neutrals.

$$\textsc{ListNeConv}\ \frac{\Gamma \vdash_{\mathrm{map}} c \approx_{\mathrm{map}} c' \lhd \mathbf{List}\ A}{\Gamma \vdash_{\mathrm{map}} c \cong_{\mathrm{h}} c' \lhd \mathbf{List}\ A}$$

$$\textsc{ListNeMap}\ \frac{\Gamma \vdash_{\mathrm{map}} n \approx_{\mathrm{h}} n' \rhd \mathbf{List}\ A \qquad \Gamma, x : A \vdash_{\mathrm{map}} f\, x \cong x \lhd B}{\Gamma \vdash_{\mathrm{map}} \mathrm{map}_{\mathbf{List}}\ f\, n \approx_{\mathrm{map}} n' \lhd \mathbf{List}\ B}$$

Using this second, algorithmic, instance as a specification, we can show the soundness and completeness of a conversion-checking function extending that of Adjedj et al. [3] with lists and $\boxed{\Gamma \vdash_{\text{sub}} T \preccurlyeq_{\text{h}} T' \lhd}$ Reduced type $T$ is a subtype of reduced type $T'$

$$\textsc{UniSub}\ \frac{}{\Gamma \vdash_{\text{sub}} \text{Type}_i \preccurlyeq_{\text{h}} \text{Type}_i \lhd} \qquad \textsc{ProdSub}\ \frac{\Gamma \vdash_{\text{sub}} A' \preccurlyeq A \lhd \quad \Gamma, x : A' \vdash_{\text{sub}} B \preccurlyeq B' \lhd}{\Gamma \vdash_{\text{sub}} \Pi x : A.B \preccurlyeq_{\text{h}} \Pi x : A'.B' \lhd}$$

$$\textsc{ListSub}\ \frac{\Gamma \vdash_{\text{sub}} A \preccurlyeq A' \lhd}{\Gamma \vdash_{\text{sub}} \textbf{List}\ A \preccurlyeq_{\text{h}} \textbf{List}\ A' \lhd} \qquad \textsc{NeuSub}\ \frac{\Gamma \vdash_{\text{sub}} n \approx_{\text{h}} n' \rhd T}{\Gamma \vdash_{\text{sub}} n \preccurlyeq_{\text{h}} n' \lhd}$$

Fig. 11. Algorithmic subtyping between reduced types (extends Figure 5, complete rules: Appendix C.7)

neutral compaction. Thus, via the equivalence of declarative and algorithmic conversion, we obtain decidability of the rich equational theory of (declarative) $\text{MLTT}_{\text{map}}$. The details of the function's internals are given in Appendix B.5.

## 5 SUBTYPING, COERCIVE AND SUBSUMPTIVE

The main application we develop for our definitional functor laws is structural subtyping. More precisely, we describe two extensions of MLTT. The first, $\text{MLTT}_{\text{sub}}$, has subsumptive subtyping: whenever $\vdash_{\text{sub}} t : A \preccurlyeq A'$, then also $\vdash_{\text{sub}} t : A'$, leaving subtyping implicit. The second, $\text{MLTT}_{\text{coe}}$, features coercive subtyping, witnessed by an operator $\text{coe}_{A,A'}\ t$ explicitly marking where subtyping is used and well-typed whenever $\vdash_{\text{coe}} t : A \preccurlyeq A'$. The computational behaviour of coe on each type former is informed by the corresponding map in $\text{MLTT}_{\text{map}}$. Structural coercions can hence be studied modularly in $\text{MLTT}_{\text{map}}$ and tied together in $\text{MLTT}_{\text{coe}}$.

In Section 5.1, we give algorithmic presentations of $\text{MLTT}_{\text{coe}}$ and $\text{MLTT}_{\text{sub}}$. In the context of a proof assistant or dependently typed programming language, $\text{MLTT}_{\text{sub}}$ would be the flexible, user-facing system, and $\text{MLTT}_{\text{coe}}$ its well-behaved specification. We do not develop the equivalence between this algorithmic presentation of $\text{MLTT}_{\text{coe}}$ and its declarative variant, as its proof is similar to the one for $\text{MLTT}_{\text{map}}$.

Section 5.3 relates $\text{MLTT}_{\text{coe}}$ and $\text{MLTT}_{\text{sub}}$: there is a simple erasure $|\cdot|$ from the former to the latter which removes coercions, and we show it is type-preserving; conversely, we show that any well-typed $\text{MLTT}_{\text{sub}}$ term can be elaborated to a well-typed $\text{MLTT}_{\text{coe}}$ term. The extra definitional functor laws are essential at this stage, to ensure that all equalities valid in $\text{MLTT}_{\text{sub}}$ still hold in $\text{MLTT}_{\text{coe}}$. Since we are in a dependently typed system, if equations valid in $\text{MLTT}_{\text{sub}}$ failed to hold in $\text{MLTT}_{\text{coe}}$, elaboration could not be type-preserving. Finally, Section 5.4 discusses the implications of this equivalence for coherence.

### 5.1 The Type Systems $\text{MLTT}_{\text{sub}}$ and $\text{MLTT}_{\text{coe}}$

We focus on the structural aspect of subtyping, and a base case is needed to have a non-trivial subtyping relation, i.e. to relate more types than conversion. We use record types with width and depth subtyping as an illustrative and typical instance for such a base case, but other forms of subtyping would work as well, for instance refinement types with subtyping induced by the implication order on predicates.

$$\textsc{RecTy}\ \frac{\mathcal{L} \in \mathcal{P}_{\mathrm{f}}(\mathrm{Lbl}) \qquad \forall l \in \mathcal{L}.\quad \Gamma \vdash_{\mathrm{sub}} A_l \lhd}{\Gamma \vdash_{\mathrm{sub}} \{l : A_l\}_{l \in \mathcal{L}} \lhd} \qquad \textsc{RecSub}\ \frac{\mathcal{K} \subseteq \mathcal{L} \qquad \forall k \in \mathcal{K}.\quad \Gamma \vdash_{\mathrm{sub}} A_k \preccurlyeq B_k \lhd}{\Gamma \vdash_{\mathrm{sub}} \{l : A_l\}_{l \in \mathcal{L}} \preccurlyeq_{\mathrm{h}} \{k : B_k\}_{k \in \mathcal{K}} \lhd}$$

Fig. 12. Records, typing and subtyping (extends Figures 5 and 11, complete rules: Appendices C.5 and C.6)

*Algorithmic $MLTT_{\mathrm{sub}}$.* This system replaces Check with the following rule, which uses subtyping $\preccurlyeq$ instead of conversion:

$$\textsc{CheckSub}\ \frac{\Gamma \vdash_{\mathrm{sub}} t \rhd T' \qquad \Gamma \vdash_{\mathrm{sub}} T' \preccurlyeq T \lhd}{\Gamma \vdash_{\mathrm{sub}} t \lhd T}$$

Subtyping, defined in Figure 11, orients type-level conversion from Figure 5, taking into account co- and contravariance. It relies on neutral comparison and term-level conversion, both of which are *not* altered with respect to Figure 5: subtyping is a type-level concept only.

*A type of records for a non-trivial instance of subtyping.* While the rules of Figure 11 let us propagate subtyping structurally through type formers, for the resulting system to be any different from MLTT, we need some base non-trivial subtyping. Its exact choice is largely orthogonal to the focus of this paper on the structural aspect of subtyping, and indeed the development of this section is relatively independent of it. Still, for our subtyping not to be degenerate, we must fix something.

We pick records as a simple example, presented in Figure 12. We fix a countable set of labels Lbl, and for each finite subset $\mathcal{L} \subseteq \mathrm{Lbl}$ and $\mathcal{L}$-indexed family of types $A_l$ we introduce a (non-dependent) record type $\{l : A_l\}_{l \in \mathcal{L}}$.[12] To each record type corresponds a record constructor $\{l := a_l\}_{l \in \mathcal{L}}$, as well as projections $t.l$ corresponding to labels $l \in \mathcal{L}$. Subtyping between record types is defined as inclusion of the set of labels, and pairwise subtyping of types at the same label, *i.e.* both depth and width subtyping. Full rules for record constructors and projections are given in Appendix C.5.

*Algorithmic $MLTT_{\mathrm{coe}}$.* In contrast with $\mathrm{MLTT}_{\mathrm{sub}}$, rule Check in $\mathrm{MLTT}_{\mathrm{coe}}$ is *not* altered. Instead, subtyping is only allowed when *explicitly* marked by coe, as follows:

$$\textsc{Coe}\ \frac{\Gamma \vdash_{\mathrm{coe}} A \lhd \qquad \Gamma \vdash_{\mathrm{coe}} A' \lhd \qquad \Gamma \vdash_{\mathrm{coe}} t \lhd A \qquad \Gamma \vdash_{\mathrm{coe}} A \preccurlyeq A' \lhd}{\Gamma \vdash_{\mathrm{coe}} \mathrm{coe}_{A,A'}\ t \rhd A'}$$

Reduction must of course be extended to give an operational behaviour to coe, and is given in Figure 13, together with normal forms. Operationally, $\mathrm{coe}_{A,A'}\ t$ reduces the types $A$ and $A'$ to head normal forms, then behaves like the relevant map, propagating coe recursively. Since $\mathrm{coe}_{A,A'}\ t$ is well-typed only when $A$ is a subtype of $A'$, the type formers of their head normal forms have to agree, ensuring that we can always rely on this behaviour to enact structural subtyping. As for map, rule CoeCoe lets us compact a succession of stuck coe. This only applies to positive types (characterized by $\mathrm{nf}^{\oplus}$): we do not compact coercions between negative/extensional types, but wait for the term to be observed to trigger further reduction.

Neutral conversion is described at the top of Figure 14 and features an additional comparison between compacted neutrals similar to $\mathrm{MLTT}_{\mathrm{map}}$ (ListNeConv). Rule NCoe is a congruence for coercions, where the source and target types necessarily agree by typing invariants, and are thus not compared. Rules NCoeL and NCoeR handle identity coercions. Accordingly, $\approx_{\mathrm{coe}}$ is carefully used whenever normal forms can be compacted neutrals, *e.g.* at neutral and positive types, as

[12] We choose to avoid dependency mainly for the sake of simplicity, but see no difficulty to have dependent records instead.

$\boxed{t \rightsquigarrow^1 t'}$

$$\frac{\mathsf{nf}\ f}{(\mathsf{coe}_{\Pi\, x:A'.B', \Pi\, x:A.B}\ f)\ a \rightsquigarrow^1 \mathsf{coe}_{B'[\mathsf{coe}_{A,A'}\ a], B[a]}(f\ (\mathsf{coe}_{A,A'}\ a))} \qquad \mathsf{coe}_{\mathsf{Type}_i, \mathsf{Type}_i}\ t \rightsquigarrow^1 t$$

$$\mathsf{coe}_{\mathbf{List}\ A, \mathbf{List}\ A'}\ \varepsilon \rightsquigarrow^1 \varepsilon \qquad \mathsf{coe}_{\mathbf{List}\ A, \mathbf{List}\ A'}(h :: t) \rightsquigarrow^1 \mathsf{coe}_{A,A'}\ h :: \mathsf{coe}_{\mathbf{List}\ A, \mathbf{List}\ A'}\ t$$

$$\textsc{CoeL}\ \frac{A \rightsquigarrow^1 A'}{\mathsf{coe}_{A,B}\ t \rightsquigarrow^1 \mathsf{coe}_{A',B}\ t} \qquad \textsc{CoeR}\ \frac{\mathsf{nf}^{\oplus} \text{ or ne } A \qquad B \rightsquigarrow^1 B'}{\mathsf{coe}_{A,B}\ t \rightsquigarrow^1 \mathsf{coe}_{A,B'}\ t}$$

$$\textsc{CoeTm}\ \frac{\mathsf{nf}^{\oplus} \text{ or ne } A, B \qquad t \rightsquigarrow^1 t'}{\mathsf{coe}_{A,B}\ t \rightsquigarrow^1 \mathsf{coe}_{A,B}\ t'} \qquad \textsc{CoeCoe}\ \frac{\mathsf{nf}^{\oplus} \text{ or ne } U, U', T, T' \qquad \mathsf{ne}\ n}{\mathsf{coe}_{U,U'}\ \mathsf{coe}_{T,T'}\ n \rightsquigarrow^1 \mathsf{coe}_{T,U'}\ n}$$

| | | | |
|---|---|---|---|
| $\boxed{\mathsf{nf}\ f}$ | $\stackrel{\text{def}}{=}$ | $n \mid P \mid N \mid \lambda\, x : t.t \mid \varepsilon_t \mid t ::_t t \mid \mathsf{coe}_{N,N}\ f \mid \ldots$ | weak-head normal forms |
| $\boxed{\mathsf{nf}^{\ominus}\ N}$ | $\stackrel{\text{def}}{=}$ | $\Pi\, x : t.t \mid \Sigma\, x : t.t$ | negative whnf types |
| $\boxed{\mathsf{nf}^{\oplus}\ P}$ | $\stackrel{\text{def}}{=}$ | $\mathsf{Type}_i \mid \mathbf{List}\ t \mid \ldots$ | positive whnf types |
| $\boxed{\mathsf{ne}\ n}$ | $\stackrel{\text{def}}{=}$ | $x \mid n\ t \mid n.l \mid \mathsf{ind}_P(c; t; t) \mid \ldots$ | weak-head neutrals |
| $\boxed{\mathsf{cne}\ c}$ | $\stackrel{\text{def}}{=}$ | $n \mid \mathsf{coe}_{P,P}\ n \mid \mathsf{coe}_{n,n}\ n$ | compacted neutrals |

Fig. 13. Weak-head reduction rules for coercion (extends Figure 4, complete rules: Appendix C.8)

$\boxed{\Gamma \vdash_{\mathsf{coe}} t \approx_{\mathsf{coe}} t' \lhd T}$ Compacted neutrals $t$ and $t'$ are comparable at type $T$

$$\textsc{NCoe}\ \frac{\Gamma \vdash_{\mathsf{coe}} n \approx n' \rhd S''}{\Gamma \vdash_{\mathsf{coe}} \mathsf{coe}_{S,T}\ n \approx_{\mathsf{coe}} \mathsf{coe}_{S',T'}\ n' \lhd T''} \qquad \textsc{NCoeL}\ \frac{\Gamma \vdash_{\mathsf{coe}} n \approx n' \rhd S''}{\Gamma \vdash_{\mathsf{coe}} \mathsf{coe}_{S,T}\ n \approx_{\mathsf{coe}} n' \lhd T''}$$

$$\textsc{NCoeR}\ \frac{\Gamma \vdash_{\mathsf{coe}} n \approx n' \rhd S''}{\Gamma \vdash_{\mathsf{coe}} n \approx_{\mathsf{coe}} \mathsf{coe}_{S',T'}\ n' \lhd T''} \qquad \textsc{NNoCoe}\ \frac{\Gamma \vdash_{\mathsf{coe}} n \approx n' \rhd S''}{\Gamma \vdash_{\mathsf{coe}} n \approx_{\mathsf{coe}} n' \lhd T''}$$

$\boxed{\Gamma \vdash_{\mathsf{coe}} t \cong_{\mathsf{h}} t' \lhd T}$

$$\textsc{NeuList}\ \frac{\Gamma \vdash_{\mathsf{coe}} n \approx_{\mathsf{coe}} n' \lhd \mathbf{List}\ A}{\Gamma \vdash_{\mathsf{coe}} n \cong_{\mathsf{h}} n' \lhd \mathbf{List}\ A} \qquad \textsc{NeuNeu}\ \frac{\Gamma \vdash_{\mathsf{coe}} n \approx_{\mathsf{coe}} n' \lhd M \qquad \mathsf{ne}\ M}{\Gamma \vdash_{\mathsf{coe}} n \cong_{\mathsf{h}} n' \lhd M}$$

Fig. 14. Algorithmic comparison of neutrals in $\mathrm{MLTT}_{\mathrm{coe}}$ (extends Figure 5, complete rules: Appendix C.9)

shown at the bottom of Figure 14. Apart from this change, conversion at the term and type level and subtyping are similar to those of $\mathrm{MLTT}_{\mathrm{sub}}$.

## 5.2 Metatheoretic properties of $\mathrm{MLTT}_{\mathrm{coe}}$

All the metatheoretic properties of $\mathrm{MLTT}_{\mathrm{map}}$ mentioned in Section 4 carry over to $\mathrm{MLTT}_{\mathrm{coe}}$. In particular, $\mathrm{MLTT}_{\mathrm{coe}}$ admits weak-head normal forms, a key property in order to properly describe the relationship between subsumptive and coercive subtyping. Following the approach for $\mathrm{MLTT}_{\mathrm{map}}$, we establish this property by first introducing a declarative version $\mathrm{MLTT}_{\mathrm{coe}}$ of $\mathrm{MLTT}_{\mathrm{coe}}$, whose normalisation is obtained through a logical relation, and then, using a further instance of the logical relation, show that the algorithmic and declarative systems with coercive subtyping are equivalent. As a by-product, the equivalence also implies that inferred types in the algorithmic system are principal.

*Declarative $\mathrm{MLTT}_{\mathrm{coe}}$.* The declarative presentation of $\mathrm{MLTT}_{\mathrm{coe}}$, noted $\vdash_{\mathrm{coe}}$, straightforwardly extends MLTT (Figures 2 and 3) with typing and conversion rules for records and coe similar to the ones of the algorithmic presentation. Most importantly, it contains the following two rules for definitional identity and composition of coercions.

$$\textsc{CoeId}\ \frac{\Gamma \vdash_{\mathrm{coe}} t : A}{\Gamma \vdash_{\mathrm{coe}} \mathrm{coe}_{A,A}\, t \cong t : A} \qquad \textsc{CoeTrans}\ \frac{\Gamma \vdash_{\mathrm{coe}} t : A \qquad \Gamma \vdash_{\mathrm{coe}} A \preccurlyeq A' \qquad \Gamma \vdash_{\mathrm{coe}} A' \preccurlyeq A''}{\Gamma \vdash_{\mathrm{coe}} \mathrm{coe}_{A',A''}\, \mathrm{coe}_{A,A'}\, t \cong \mathrm{coe}_{A,A''}\, t : A''}$$

The complete presentation can be found in Appendix C.10. $\mathrm{MLTT}_{\mathrm{coe}}$ serves as a baseline to establish the metatheoretical properties of $\mathrm{MLTT}_{\mathrm{coe}}$. In particular, it normalizes.

**Theorem 5.1 (Weak-head normalization).** *If $\Gamma \vdash_{\mathrm{coe}} t : T$, then there exists a weak-head normal form $t'$ such that $t \rightsquigarrow^\star t'$.*

Proof ideas from $\mathrm{MLTT}_{\mathrm{map}}$ carry over to $\mathrm{MLTT}_{\mathrm{coe}}$, and we did not mechanize this part of the paper, focusing our formalization effort on the most challenging aspect of the theory. We sketch how to extend the logical relation for $\mathrm{MLTT}_{\mathrm{map}}$ to $\mathrm{MLTT}_{\mathrm{coe}}$ – the proofs of equivalence between the declarative and algorithmic systems from the logical relation then remain mostly unchanged.

Proof sketch (extending the logical relation to $\mathrm{MLTT}_{\mathrm{coe}}$). $\mathrm{MLTT}_{\mathrm{coe}}$ has three main differences compared to $\mathrm{MLTT}_{\mathrm{map}}$: record types, subtyping and coercions.

First, we need to define the logical relation at record types and show the validity of introduction and elimination forms. Since, records behave as iterated Cartesian products, the reducibility proof carries over. Thus, reducibility at record types is defined as reducibility of each projection.

Second, we need to extend reducible type-level conversion to handle subtyping. As the structure of the two judgements is exactly the same, apart from the base subtyping case, we can parametrize reducible conversion by a *conversion problem*,[13] a three-valued variant indicating conversion, subtyping, or supertyping, the latter being needed to handle contravariance and the left bias of reducible conversion, which is defined on a proof of reducibility of its left type.

Finally, we need to show that $\mathrm{coe}_{A,A'}\, t$ is reducible whenever $A$ is a reducible subtype of $A'$, and $t$ is reducible at $A$. Because of the former, both must have normal forms which are either constructed with the same type former $F$, both record types, or both neutrals. In the first case, $\mathrm{coe}_{A,A'}\, t$ behaves like $\mathrm{map}_F$, and the proofs from Section 4 carry over. If $A$ and $A'$ are both neutral, $\mathrm{coe}_{A,A'}\, t$ might compact if $t$ is a coercion, but this is also similar to the case of a neutral map for lists in $\mathrm{MLTT}_{\mathrm{map}}$, and so the proof from Section 4 carries over again.

We are left with the case of record types. We need to show that if $t$ is reducible at $\{l : A_l\}_{l \in \mathcal{L}}$ which is a reducible subtype of $\{k : B_k\}_{k \in \mathcal{K}}$ then $\mathrm{coe}_{\{l : A_l\}_{l \in \mathcal{L}}, \{k : B_k\}_{k \in \mathcal{K}}}\, t$ is reducible. By definition of reducibility at record types as reducibility of all projections, and closure of reducibility by anti-reduction, it is enough to show that each $\mathrm{coe}_{A_k,B_k}\, t.k$ is reducible at $B_k$ for $k \in \mathcal{K}$. Combining the reducibility of $t.k$ obtained from that of $t$, together with the induction hypothesis on the reducible subtyping $A_k \preccurlyeq B_k$ completes this step. □

[13] This technique is borrowed from the way cumulativity is handled in MetaCoq [51].

*Equivalence of algorithmic and declarative typing.* In order to transfer the weak-head normalization result of $\mathrm{MLTT}_{\mathrm{coe}}$ to $\mathrm{MLTT}_{\mathrm{coe}}$, we use a second instantiation the logical relation sketched above and obtain soundess and completeness of $\mathrm{MLTT}_{\mathrm{coe}}$ with respect to $\mathrm{MLTT}_{\mathrm{coe}}$.

Theorem 5.2 (Soundness and completeness of algorithmic typing). *If* $\vdash_{\mathrm{coe}} \Gamma$ *and* $\Gamma \vdash_{\mathrm{coe}} t \rhd T$ *then* $\Gamma \vdash_{\mathrm{coe}} t : T$*, and similarly for the other judgements. Conversely, if* $\Gamma \vdash_{\mathrm{coe}} t : T$*, then* $\Gamma \vdash_{\mathrm{coe}} t \lhd T$*, and similarly for the other judgements.*

As a corollary, if $\Gamma \vdash_{\mathrm{coe}} t : T$ then there exists $T_p$ such that $\Gamma \vdash_{\mathrm{coe}} t \rhd T'$ and $\Gamma \vdash_{\mathrm{coe}} T_p \preccurlyeq T \lhd$, and by uniqueness of the inferred type, $T_p$ is actually the principal type of $t$.

## 5.3 Elaboration and Erasure

We can now turn to the correspondence between $\mathrm{MLTT}_{\mathrm{sub}}$ and $\mathrm{MLTT}_{\mathrm{coe}}$. The translation in the forward direction, *erasure* $|\cdot|$, removes coercions $\left|\mathrm{coe}_{A,A'}\ t\right| = t$ and is otherwise a congruence. It is lifted pointwise to contexts. We first show that erasure is sound, meaning that it preserves typing and conversion, and then that it is also invertible, *i.e.* that any well-typed $\mathrm{MLTT}_{\mathrm{sub}}$ term $t'$ elaborates to a well-typed $\mathrm{MLTT}_{\mathrm{coe}}$ term $t$ whose erasure is $t' = |t|$.

*Soundness of Erasure.* Erasure translates from a constrained system to a more liberal one. Establishing its soundness, *e.g.*that conversion and typing are preserved, is relatively easy, as long as the reduction rules of Figure 13 are designed so that erasure preserves them. Indeed, the key point is that reduction rules for coe do *not* change the structure of the erased term, and so erase to exactly zero steps of reduction. In contrast, the rule below is inadequate, as it would η-expand terms at function types more in $\mathrm{MLTT}_{\mathrm{coe}}$ than in $\mathrm{MLTT}_{\mathrm{sub}}$:

$$\mathrm{coe}_{\Pi\, x:A'.B', \Pi\, x:A.B}\ f \rightsquigarrow^1 \lambda\, x : A.\ \mathrm{coe}_{B'\left[\mathrm{coe}_{A,A'}\ x\right], B}(f\ \ \mathrm{coe}_{A,A'}\ x).$$

The two terms remain nonetheless convertible. By induction on $\mathrm{MLTT}_{\mathrm{coe}}$'s typing derivation, one can then show that erasure preserves conversion and subtyping, and finally typing.

Theorem 5.3 (Erasure Preserves Typing). *If* $\Gamma \vdash_{\mathrm{coe}} t \lhd T$ *holds and its inputs are well-formed, then* $|\Gamma| \vdash_{\mathrm{sub}} |t| \lhd |T|$.

*Elaboration.* Elaborating back from $\mathrm{MLTT}_{\mathrm{sub}}$ to $\mathrm{MLTT}_{\mathrm{coe}}$ is more challenging: as we add annotations, we must ensure that these do not hinder conversion. We follow the proof strategy of a similar proof of elaboration soundness in Lennon-Bertrand et al. [35]. The core of the argument are so-called "catch-up lemmas", which ensure that annotations never block redexes. As an example, here is the one for function types.

Lemma 5.4 (Catch up, Function Type). *If* $\Gamma \vdash_{\mathrm{coe}} f\ a \lhd B$ *and* $|f| = \lambda\, x : A'.\ t'$*, then there exists* $t$ *such that* $|t| = t'$ *and* $f\ a \rightsquigarrow^\star t[a]$.

From these catch-up lemmas it follows that erasure is a backward simulation, therefore that it preserves subtyping, and finally that it is type-preserving. Proofs are all by induction, and given in Appendix D.2.

Lemma 5.5 (Erasure is a Backward Simulation). *Assume that* $\Gamma \vdash_{\mathrm{coe}} t : T$*. If* $|t| \rightsquigarrow^\star u'$*, with* $u'$ *a weak-head normal form, then* $t \rightsquigarrow^\star u$*, with* $u$ *a weak-head normal form such that* $|u| = u'$.

Lemma 5.6 (Elaboration Preserves Subtyping). *The following implications hold whenever the inputs of the conclusions are well-formed:*

(1) *if* $|\Gamma| \vdash_{\mathrm{sub}} |T| \preccurlyeq |U| \lhd$*, then* $\Gamma \vdash_{\mathrm{coe}} T \preccurlyeq U \lhd$;
(2) *if* $|\Gamma| \vdash_{\mathrm{sub}} |t| \cong |u| \lhd |T|$*, then* $\Gamma \vdash_{\mathrm{coe}} t \cong u \lhd T$;

(3) *if* $|\Gamma| \vdash_{\mathrm{sub}} |t| \approx |u| \rhd T$*, then* $\Gamma \vdash_{\mathrm{coe}} t \approx u \rhd T$*;*
(4) *and similarly for the other judgements.*

Finally, the main theorem states that we can elaborate terms using implicit subtyping to explicit coercions, in a type-preserving way.

Corollary 5.7 (Elaboration). *If* $\vdash_{\mathrm{coe}} \Gamma$, $\Gamma \vdash_{\mathrm{coe}} T \lhd$ *and* $|\Gamma| \vdash_{\mathrm{sub}} t' \lhd |T|$*, then there exists* $t$ *such that* $\Gamma \vdash_{\mathrm{coe}} t \lhd T$*, and* $|t| = t'$*.*

Importantly, to establish this equivalence we do *not* need to develop any meta-theory for $\mathrm{MLTT}_{\mathrm{sub}}$: having the meta-theory of $\mathrm{MLTT}_{\mathrm{coe}}$ is enough!

Nonetheless, now that the equivalence between the two systems has been established, we can use it to transport meta-theoretic properties, such as normalization, from $\mathrm{MLTT}_{\mathrm{coe}}$ to $\mathrm{MLTT}_{\mathrm{sub}}$.

### 5.4 Coherence

An important property of elaboration is *coherence*, stating that the elaboration of a well-typed term does not depend on its typing derivation. In our algorithmic setting, a term has at most one typing derivation and so at most one elaboration. However, multiple well-typed terms in $\mathrm{MLTT}_{\mathrm{coe}}$ can still erase to the same $\mathrm{MLTT}_{\mathrm{sub}}$ term. While only one of them is the result of elaboration as defined in Corollary 5.7, all these distinct terms should still behave similarly. The following is a direct consequence of Lemma 5.6, and shows that the equations imposed on coe are enough to give us a very strong form of coherence: it holds up to definitional equality, rather than in a weaker, semantic way. Another way to look at this is that the scenario of Example 1.2 cannot happen, thanks to our new equations: if two terms erase to the same coercion-free one in $\mathrm{MLTT}_{\mathrm{sub}}$, then they *must* be convertible in $\mathrm{MLTT}_{\mathrm{coe}}$. Hidden coercions cannot be responsible for failures of conversion.

Theorem 5.8 (Coherence). *If* $t$, $u$ *are such that* $\Gamma \vdash_{\mathrm{coe}} t \lhd T$ *and* $\Gamma \vdash_{\mathrm{coe}} u \lhd T$*, with* $\vdash_{\mathrm{coe}} \Gamma$ *and* $\Gamma \vdash_{\mathrm{coe}} T \lhd$*, and moreover* $|t| = |u|$ *(i.e.* $t$ *and* $u$ *correspond to the same* $\mathit{MLTT}_{\mathrm{sub}}$ *term), then* $\Gamma \vdash_{\mathrm{coe}} t \cong u \lhd T$*.*

Proof. By reflexivity, (obtained through the equivalence with the declarative system), $\Gamma \vdash_{\mathrm{coe}} t \cong t \lhd T$. Using Theorem 5.3 (soundness of erasure), we get $|\Gamma| \vdash_{\mathrm{sub}} |t| \cong |t| \lhd |T|$, and so also $|\Gamma| \vdash_{\mathrm{sub}} |t| \cong |u| \lhd |T|$. But then by Lemma 5.6 (elaboration preserving conversion), we can come back, and obtain $\Gamma \vdash_{\mathrm{coe}} t \cong u \lhd T$. □

As particular cases of this coherence theorem, we can now exhibit the necessity of the functor laws, sharpening the informal argument in the introduction. For the identity law, any well-typed $\mathrm{MLTT}_{\mathrm{coe}}$ term $\mathrm{coe}_{A,A}\, t$ erases to $|\mathrm{coe}_{A,A}\, t| = |t|$ in $\mathrm{MLTT}_{\mathrm{sub}}$, and by coherence we obtain that the conversion $\Gamma \vdash_{\mathrm{coe}} \mathrm{coe}_{A,A}\, t \cong t \lhd A$ is required in $\mathrm{MLTT}_{\mathrm{coe}}$. For the composition law, we have for adequately well-typed terms that $|\mathrm{coe}_{B,C}\, \mathrm{coe}_{A,B}\, t| = |t| = |\mathrm{coe}_{A,C}\, t|$, hence by coherence the conversion $\Gamma \vdash_{\mathrm{coe}} \mathrm{coe}_{B,C}\, \mathrm{coe}_{A,B}\, t \cong \mathrm{coe}_{A,C}\, t \lhd$ must hold in $\mathrm{MLTT}_{\mathrm{coe}}$ as well.

## 6 RELATED AND FUTURE WORK

*Adding Definitional Equations to Dependent Type Theory.* Strub [53] endows a dependent type theory with additional equations from first order decidable theories, with further extensions to a universe hierarchy and large eliminations in Jouannaud and Strub [30] and Barras et al. [11]. Equational theories can sometimes be presented by a confluent set of rewrite rules, a case advocated by Cockx, Tabareau, and Winterhalter [18]. They show through counter-examples that ensuring type preservation in dependent type theory is a subtle matter and do not ensure normalization of the resulting theory. On the theoretical side, categorical tools are being developed to prove general conservativity and strictification results for type theories [12, 13] extending the seminal work of Hofmann [26] on conservativity of extensional type theory with respect to intensional type theory [60].

*Formalized Metatheory with Logical Relations.* Allais, McBride, and Boutillier [6] propose to add a variety of fusion laws for lists, including our functor laws, to a simply typed $\lambda$-calculus, only sketching an extension to dependent types. The three classes of normal forms (see Figures 9 and 13) is inspired from their work. While we depart from their normalization by evaluation approach to obtain fine-grained results on convergence of iterated weak-head reduction, we expect that the original strategy should extend to dependent types. Formalizing logical relations for MLTT is a difficult exercise, pioneered by Abel, Öhman, and Vezzosi [2] in Agda using inductive-recursive definitions, and Wieczorek and Biernacki [58] in Coq using impredicativity. We build upon and extend a Coq reimplementation of the former [3].

*Cast and Coercion Operators.* Pujet and Tabareau [47, 48] extend Abel, Öhman, and Vezzosi [2] to establish the metatheory of observational type theory [7]. Their work features a `cast` operator behaving similarly to coe, but guarded by an internal proof of equality instead of an external subtyping derivation. Their `cast` does not satisfy definitional transitivity, and we give evidence in Appendix A that such an extension would break metatheoretical properties. Another cast primitive with a similar operational behaviour appears in cast calculi for gradual typing [49], and indeed our proof that elaboration is type preserving in Section 5.3 is inspired by a similar one for GCIC, which combines gradual and dependent types [35]. In this case, casting is allowed between any two types, but the absence of guard is compensated by the possibility of runtime errors, making the type theory inconsistent.

*Functorial Maps for Inductive Type Schemes.* Luo and Adams [38] describe the construction of map for a class of strictly positive operators on paper, but do not implement it. Deriving map-like construction is a typical example of metaprogramming frameworks for proof assistants, *e.g.* Coq-Elpi [20, 55] in Coq, and the generics Agda library [22] derives a fold operation, from which map can be easily obtained. In a simply typed setting, Barral and Soloviev [10] employ rewriting techniques, in particular rewriting postponement, to show that an oriented variant of the functor laws are confluent and normalizing. These techniques rely on normalization, and could not be easily adapted to the dependent setting, however the idea of postponing the reduction step for identity appears in our logical relation as well. In a short abstract, McBride and Nordvall Forsberg [44] investigate a notion of functorial adapters that generalizes and unifies both the Check rule from bidirectional typing and the Coe rule from $\mathrm{MLTT}_{\mathrm{coe}}$.

*Subtyping, Dependent Types and Algorithmic Derivations.* Coherence of coercions in presence of structural subtyping is a challenging problem. To address the issue, Luo and Luo [39] introduce a notion of weak transitivity, weakening the coherence of the transitivity up to propositional equality. This solution does not interact well with dependency, forcing them to restrict structural subtyping to a class of non-dependent inductives, *e.g.* excluding (positive) $\Sigma$. Luo and Adams [38] show that the transitivity of coercions is admissible in presence of definitional compositions – called $\chi$-rules there – for inductive schemata. They rely on a conjecture that strong normalization and subject reduction hold in presence of these $\chi$-rules, explicitly mentioning that the metatheory with those additional equality rules is "largely unknown". We provide such results, and have formalized them for **List**. We use a completely different proof technique, that scales to a theory with universes and large elimination. Both aforementioned papers employ a strict order for subtyping and do not consider the functor law for the identity, nor tackle decidability of type-checking.

Aspinall and Compagnoni [9] investigate the relationship between subtyping and dependent types using algorithmic derivations to control the subtyping derivations for a variant of $\lambda P$, a type theory logically much weaker than MLTT. Lungu and Luo [37] study an elaboration of a subsumptive presentation into coercive one in presence of a coherent signature of subtyping relations between base types. Assuming normalization, they show that subtyping extends to Π types, setting aside other parametrized types. While they work over an abstract signature of coercions, the functor laws we study are needed to instantiate this signature with meaningful datatypes while respecting their assumptions. We explain the relation of these algorithmic system with bidirectional systems, notably the one of Abel, Öhman, and Vezzosi [2], contributing to a sharper picture.

*Integration with Other Forms of Subtyping.* As we mentioned in Section 5, our design of base subtyping was guided by simplicity. Our work on structural subtyping should integrate mostly seamlessly with other, more ambitious forms of subtyping. Coercions between dependent records form the foundation of hierarchical organizations of mathematical structures [4, 19, 59], and should be a simple extension of our framework. This could lead to vast simplification of the complex apparatus currently needed to deal with these hierarchies.

Refinement subtyping is heavily used in F* but also in Coq's Program [50] to specify the behaviour of programs. Relativizing any result of decidability of type-checking to that of the chosen fragment of refinements, an implementation of refinement subtyping using definitionally irrelevant propositions [24] to preserve coherence[14] should be within reach.

Our techniques for structural subtyping should also apply well in the context of algebraic approaches to cumulativity between universes [31, 52]. Cumulativity goes beyond mere subtyping, as it also involves definitional isomorphisms between two copies of the same type at different universe levels. Our definitional functor laws already allow these to interact well with map operations, but it would be interesting to investigate which extra definitional equations are needed – and can be realized – to make structural cumulativity work seamlessly, hopefully obtaining a translation from Russel-style to Tarski-style universes similar to our elaboration from $\mathrm{MLTT}_{\mathrm{sub}}$ to $\mathrm{MLTT}_{\mathrm{coe}}$.

*Data Availability Statement.* An archive containing the formalization presented in Section 4 is available at [33]. The Coq code is supplemented with a report and a Docker image.

*Acknowledgments.* The authors thank the anonymous reviewers for their feedback. The many useful discussions with Assia Mahboubi, Enzo Crance, Nicolas Tabareau and Loïc Pujet helped a lot to develop the material that led to this paper.

[14] With proof-relevant propositions, different proofs of $p \Rightarrow q$ induce different coercions between $\{x \mid p\}$ and $\{x \mid q\}$, breaking coherence.

## A INTERNAL SUBTYPING AND UNDECIDABILITY OF CONVERSION

The goal of the coercive approach is to reflect all the potential ambiguities present in a subtyping derivation. As such, wouldn't it be easier to just internalize the notion of subtype and let type theory deal with it? The following observation shows that there exists a big obstruction to any decidability result for conversion as long as we want to stay equivalent to the subsumptive presentation of subtyping.

Observation A.1 (No-go of internal subtyping). *Suppose that $\mathcal{T}$ is a type theory with a family* sub $A\,B$ *for any two types $A$ and $B$, equipped with reflexivity witnesses* $\texttt{refl}_A : \text{sub}\,A\,A$ *and transitivity witnesses* $\texttt{trans}\,w\,w' : \text{sub}\,A\,C$ *for* $w : \text{sub}\,A\,B$ *and* $w' : \text{sub}\,B\,C$*, as well as a coercion function* $\text{coe}_{A,B} : \text{sub}\,A\,B \to A \to B$*, such that* $\text{coe}_{A,A}\ \texttt{refl}_A \cong \text{id}_A$ *and* $\text{coe}_{B,C}\ w \circ \text{coe}_{A,B}\,w' \cong \text{coe}_{A,C}(\texttt{trans}\,w_{A,B}\,w_{B,C})$*. Then $\mathcal{T}$ embeds definitional models of the untyped $\lambda$-calculus, and in particular divergent terms.*

Indeed, whenever a context provides inhabitants of both sub $A\,B$ and sub $B\,A$, $\text{coe}_{A,B}$ and $\text{coe}_{B,A}$ provide a definitional isomorphism $A \cong B$. In particular any context inhabiting sub $A\,A \to A$ and sub $A \to A\,A$, for instance an inconsistent one, provides a definitional retraction of $A \to A$ onto $A$, hence a non-trivial model of the untyped $\lambda$-calculus with a divergent element $\Omega_A : A$. This observation motivates our external approach to subtyping with a specific judgement of subtyping that cannot be abstracted upon.

## B ARTEFACT DESCRIPTION

### B.1 Organization of the Artefact and Other Resources

This section describes in more details the Coq formalisation accompanying this paper, more specifically the content of section 4. To complement it, we also provide the following:

- the REQUIREMENTS.md and INSTALL.md file with installation instructions;
- the README.md file with a quick overview of the development with hyperlinks to the files of interest;
- a DOCKER.md file, with installation and usage instructions for the provided docker image;
- a Readme.v file, which gives a more in-depth overview of the development as a Coq file, using directly the main Coq definitions and theorems, and is roughly similar to the present PDF description;
- a doc/dependency_graph.png file, showing the structure of the development.

We utilize the logical relation proof technique presented in Abel, Öhman, and Vezzosi [2] and build upon its Coq implementation due to Adjedj et al. [3]. This artefact contributes an extension of the formalisation with lists and definitional functor laws for lists. We refer to both articles for further details on the proof technique and the general setup of the formalisation.

### B.2 Syntax

*Terms (AutoSubst/Ast).* The syntax of terms, along with the other files in the AutoSubst folder, are generated using the AutoSubst plugin. The definition of renaming and substitution are also automatically derived from the one of terms, and many boilerplate lemmas on them are too. Of particular interest are the constructors `tList`, `tNil`, `tCons`, `tElim` and `tMap`, respectively corresponding to the type constructor for lists, the empty list, list consing, the (dependent) eliminator for lists, and the definitionally functorial map operation.

*NormalForms.* Weak-head normal forms `whnf`, neutrals `whne` and compacted neutrals `whne_list` are defined as inductive predicate on terms, *i.e.* as function of type `term -> Prop`, corresponding to Fig. 4 and 10 from the paper. In particular, any compacted neutral is a normal form, and compacted neutrals can either consist of a map of a neutral, or simply of a neutral.

*Reduction (UntypedReduction).* Reduction, written `[l ⇒* l']`, is the transitive closure of one-step reduction `[l ⇒ *']`, defined as an inductive relation. In particular, we have the rules of Fig. 9, that is:

```
mapNil : forall {A A' B f : term}, [tMap A B f (tNil A') ⇒ tNil B]
mapCons : forall {A A' B f a l : term},
  [tMap A B f (tCons A' a l) ⇒ tCons B (tApp f a) (tMap A B f l)]
mapComp : forall {A B B' C f g l : term},
  whne l -> [tMap B C f (tMap A B' g l) ⇒ tMap A C (comp A f g) l]
```

### B.3 Typing and Conversion

*GenericTyping.* Following Abel, Öhman, and Vezzosi [2] and Adjedj et al. [3], the definition of the logical relation is parametrized by a notion of *generic typing*, a common interface to be instantiated with both the declarative and algorithmic notions of typing. This interface features a family of judgments for context well-formation, typing, conversion but also a conversion of neutrals and a (typed) reduction relation. These judgements should satisfy properties, listed for each predicate with a record (`TypingProperties`, `ConvProperties`, etc.), and grouped together in the `GenericTypingProperties` record. We use type-classes to automatically find these properties when needed, and attach generic notations (defined in Notations) to these type-classes too.

For lists, generic typing closely resembles declarative typing, as defined in Fig. 2. Generic conversion must contain reduction, which includes typed variants of the rules above. Moreover, we have congruence rules for constructors, for instance we have the following, where `ta` stands for an arbitrary generic conversion:

```
forall (Γ : context) (A A' : term),
  [Γ |-[ ta ] A ≅ A' : U] -> [Γ |-[ ta ] tList A ≅ tList A' : U]
```

Conversion is not constrained to be a congruence for destructors, but it must contain neutral conversion, which *is* a congruence for `tMap` and `tListElim`, provided its main argument is too. Functor laws are also specified at the level of neutral conversion.

*DeclarativeTyping.* The definition of the declarative judgments, as inductive predicates, corresponds to Fig. 2, 3, and 9 – the latter being restricted to the case of lists. The corresponding instance of generic typing is defined in DeclarativeTypingInstance. Neutral comparison is instantiated simply with conversion, *i.e.* the declarative instance does not distinguish between the two notions. Typed reduction is instantiated as the conjunction of declarative conversion and untyped reduction. All other judgments are directly instantiated with the corresponding declarative one.

*AlgorithmicTyping.* The raw algorithmic typing judgments, akin to Fig. 5 and 6, are again defined as inductive predicates. As we explain at the end of Section 2.3 in relation to Fig. 7, we must impose extra pre-conditions for these judgments to be well-behaved. The corresponding judgments, called *bundled*, are defined in BundledAlgorithmicTyping. In AlgorithmicConvProperties and AlgorithmicTypingProperties, we establish the properties of the conversion and typing judgments, to derive two new instances of generic typing. The first instance uses (bundled) algorithmic conversion, but declarative typing. It depends on consequences of the logical relation instantiated with the fully declarative instance. The second uses only bundled algorithmic judgments, but depends on consequences of the logical relation instantiated with the first, mixed instance.

## B.4 The Logical Relation

The logical relation is built from two layers, first the reducibility layer attaching witnesses of reducibility to weak-head normal form and second the validity layer that closes reducibility under substitution.

*Definition of reducibility (LogicalRelation).* The reducibility layer describes the types A that are reducible in a given context Γ and level l, noted [Γ ||-<l> A ]. Informally, a type is reducible when it weak-head reduces to a (weak-head) normal form, and the subterms of this normal form are themselves reducible. This weak-head normal form, when it exists, is unique by determinism of the weak-head reduction strategy. A witness of reducibility RA : [Γ ||-<l> A ] for the type A induce three subsequent predicates:

- reducible conversion of a type B to A, noted [Γ ||-<l> A ≅ B| RA],
- reducibility of terms t of type A, noted [Γ ||-<l> t : A | RA],
- reducible conversion of terms t,u of type A, noted [Γ ||-<l> t ≅ u : A | RA].

These three predicates are packed in a single record LRPack. Reducible types are characterized inductively together with their associated LRPack using an indexed inductive LR. This encoding of a seemingly inductive-recursive definition using the inductively generated graph of the functions is known as small-induction recursion. The actual content of the reducibility relation is defined independently for each type formers as well as the neutrals types. We focus here on the reducibility of lists and refer to [3, 2] for the other type formers.

A type A is reducible as a list if it weak-head reduces to a type of shape tList par where the parameter type par is itself reducible in any weakening of the context Γ. This Kripke-style quantification on all future (weakened) contexts Δ ≤ Γ is necessary for specifying reducibility in larger contexts.

Reducible terms of list type are defined inductively in two steps: ListProp holds of canonical forms of type list (nil, cons and neutrals) with reducible arguments ; ListRedTm holds of terms that weak-head reduce to a reducible canonical form. The two inductive definitions must be mutual since the tail of a reducible tCons need not to be in weak-head normal form. A neutral term of list type is reducible if it is a well-typed neutral and moreover, if it is of shape tMap A B f l with l necessary neutral itself, then the body f⟨wk1 Γ A⟩ (tRel 0) of f must be reducible in an extended context Γ,,A. In the latter case, the type B of the codomain of f cannot be required to be reducible since that would lead to non-well-founded definition for the logical relation, but it is reducibly convertible to the reducible parameter type par at which reducibility of lists is defined.

Reducible conversion between terms of list type follow a similar pattern. In order to account for the identity functor law, the additional reducibility datum needed to relate two neutral terms also depends on the shape of the terms:

- if both terms are respectively of the shape tMap A B f l and tMap A' B' f' l', then the bodies of f and f' must be reducibly convertible (congruence);
- if only one of the term is of shape tMap A B f l, then f must be reducibly convertible to the identity function, i.e. its body must be reducibly convertible to the first variable in context tRel 0.

*Properties of Reducibility.* In order to reason on reducibility, we derive the induction principle corresponding to the inductive-recursive definition of the logical relation in LogicalRelation/Induction. This induction principle is then employed to derive a variety of properties of reducibility in the LogicalRelation/ subdirectory: an inversion principle, irrelevance with respect to reducible conversion, reflexivity, symmetry and transitivity of reducible conversion, stability by weakening and by anti-reduction.

*Validity and the Fundamental Lemma.* Validity closes reducibility by reducible substitution using another encoding of an inductive-recursive schema. The fundamental lemma then states that all components of a derivable declarative judgement are valid, in particular, terms well-typed for the declarative presentation are valid. The proof of the fundamental lemma proceed by an induction on declarative typing derivations, using that each declarative derivation step is admissible for the validity logical relation. These admissibility results are shown independently for each type former in the Substitution/Introductions/ subdirectory. Most type and term formers related to lists are in List, while the eliminator for lists is in ListElim. The proofs follow the description of the logical relation: first, we show that each type, term or conversion equation is reducible using the definition and properties of reducibility, and then that it is valid. To show that the functor laws are valid, we use that composition of functions (e.g. morphisms for list) is definitionally associative and unital.

## B.5 Type-checker (Decidability folder)

*Open Recursion for Partial Functions.* To side-step issues with the complex termination argument of the conversion checker, we define it in an open recursion fashion, relying on a form of free monad. The functions for reduction, conversion and type checking are defined in Decidability/Functions. The main change compared to Adjedj et al. [3] is the addition of compaction to weak-head evaluation. Evaluation is implemented using a stack machine, on which elimination forms are pushed as they are encountered. When the machine hits a variable, for Adjedj et al. [3] it means the whole term – the variable against the stack of eliminations – is a neutral. However, this is not the case for us: we want to compute a compacted neutral. Thus, we add an extra compaction pass, implemented by the `compact` function, which merges successive map operations on the stack as we unpile them.

*Correctness of the Functions.* Correctness of the implementations is shown in three steps. First, we show Soundness, *i.e.* that a positive answer of the checker implies the corresponding (algorithmic) judgment. Next, we show Completeness, *i.e.* that whenever an algorithmic judgment holds, then the corresponding checker answers positively. Finally, we show Termination, *i.e.* that the checkers always terminates when run on well-typed inputs. Again, the main innovation has to do with compaction. To reason about it, we need to make explicit the invariant that the stack is always "well-typed", in a suitable sense, see `typed_stack` in Completeness.

## B.6 Main properties

The main properties we obtain from the logical relations and the certified checker are the following. First, every well-typed term and type are (weakly) normalising (proven in Normalisation):

```
Record WN (t : term) := {
  wn_val : term; wn_red : [ t ⇒* wn_val ]; wn_whnf : whnf wn_val; }.
Corollary normalisation {Γ A t} : [Γ |-[de] t : A] -> WN t.
Corollary type_normalisation {Γ A} : [Γ |-[de] A] -> WN A.
```

Conversion and typing are decidable (proven in Decidability):

```
Definition check_conv (Γ : context) (T t t' : term) (hΓ : [|- Γ])
  (hT : [Γ |- T]) (ht : [Γ |- t : T]) (ht' : [Γ |- t' : T]) :
  [Γ |- t ≅ t' : T] + ~[Γ |- t ≅ t' : T].
Definition check_full Γ (T t : term) : [Γ |- t : T] + ~[Γ |- t : T].
```

Finally, the type system seen as a logic is consistent, and canonicity holds at the type of natural numbers:

```
Lemma consistency {t} : [ε |- t : tEmpty] -> False.
Lemma nat_canonicity {t} : [ε |- t : tNat] ->
  ∑ n : nat, [ε |- t ≅ Nat.iter n tSucc tZero : tNat].
```

# C COMPLETE TYPING RULES

## C.1 Declarative MLTT

$\boxed{\vdash \Gamma}$ Context $\Gamma$ is well-formed

$$\frac{}{\vdash \cdot} \qquad \frac{\vdash \Gamma \qquad \Gamma \vdash A : \mathrm{Type}_i}{\vdash \Gamma, x : A}$$

$\boxed{\Gamma \vdash \sigma : \Delta}$ $\sigma$ is a well-typed substitution between contexts $\Gamma$ and $\Delta$

$$\frac{}{\Gamma \vdash \cdot : \cdot} \qquad \frac{\Gamma \vdash \sigma : \Delta \qquad \Gamma \vdash t : A[\sigma]}{\Gamma \vdash (\sigma, t) : \Delta, x : A}$$

$\boxed{\Gamma \vdash T}$ Type $T$ is well-formed in context $\Gamma$

$$\textsc{El}\ \frac{\Gamma \vdash A : \mathrm{Type}_i}{\Gamma \vdash A} \qquad \textsc{FunTy}\ \frac{\Gamma \vdash A \qquad \Gamma, x : A \vdash B}{\Gamma \vdash \Pi x : A.B} \qquad \textsc{ListTy}\ \frac{\Gamma \vdash A}{\Gamma \vdash \textbf{List}\ A}$$

$$\textsc{SigTy}\ \frac{\Gamma \vdash A \qquad \Gamma, x : A \vdash B}{\Gamma \vdash \Sigma x : A.B} \qquad \textsc{TreeTy}\ \frac{\Gamma \vdash A \qquad \Gamma, x : A \vdash B}{\Gamma \vdash \mathbf{W}\, x : A.B}$$

$$\textsc{IdTy}\ \frac{\Gamma \vdash A \qquad \Gamma \vdash a : A \qquad \Gamma \vdash a' : A}{\Gamma \vdash \mathbf{Id}_A\ a\, a'} \qquad \textsc{SumTy}\ \frac{\Gamma \vdash A \qquad \Gamma \vdash B}{\Gamma \vdash A + B}$$

$\boxed{\Gamma \vdash t : T}$ Term $t$ has type $T$ under context $\Gamma$

$$\textsc{Conv}\ \frac{\Gamma \vdash t : A \qquad \Gamma \vdash A \cong B}{\Gamma \vdash t : B} \qquad \textsc{Var}\ \frac{\vdash \Gamma \qquad (x : A) \in \Gamma}{\Gamma \vdash x : A} \qquad \textsc{Sort}\ \frac{\vdash \Gamma}{\Gamma \vdash \mathrm{Type}_i : \mathrm{Type}_{i+1}}$$

$$\textsc{FunUni}\ \frac{\Gamma \vdash A : \mathrm{Type}_i \quad \Gamma, x : A \vdash B : \mathrm{Type}_i}{\Gamma \vdash \Pi x : A.B : \mathrm{Type}_i} \qquad \textsc{Abs}\ \frac{\Gamma \vdash A \qquad \Gamma, x : A \vdash B \quad \Gamma, x : A \vdash t : B}{\Gamma \vdash \lambda x : A.t : \Pi x : A.B} \qquad \textsc{App}\ \frac{\Gamma \vdash t : \Pi x : A.B \quad \Gamma \vdash u : A}{\Gamma \vdash t\, u : B[u]}$$

$$\textsc{ListUni}\ \frac{\Gamma \vdash A : \mathrm{Type}_i}{\Gamma \vdash \textbf{List}\ A : \mathrm{Type}_i} \qquad \textsc{Nil}\ \frac{\Gamma \vdash A}{\Gamma \vdash \varepsilon_A : \textbf{List}\ A} \qquad \textsc{Cons}\ \frac{\Gamma \vdash A \qquad \Gamma \vdash a : A \quad \Gamma \vdash l : \textbf{List}\ A}{\Gamma \vdash a ::_A l : \textbf{List}\ A}$$

$$\textsc{ListInd}\ \frac{\Gamma, x : \textbf{List}\ A \vdash P \qquad \Gamma \vdash A \quad \Gamma \vdash b_\varepsilon : P[\varepsilon_A] \qquad \Gamma \vdash s : \textbf{List}\ A \quad \Gamma, x : A, y : \textbf{List}\ A, z : P[y] \vdash b_{::} : P[x ::_A y]}{\Gamma \vdash \mathrm{ind}_{\textbf{List}\ A}(s; z.P; b_\varepsilon, x.y.z.b_{::}) : P[s]}$$

$$\textsc{EmptyUni}\ \frac{}{\Gamma \vdash \mathbf{0} : \mathrm{Type}_0} \qquad \textsc{UnitUni}\ \frac{}{\Gamma \vdash \mathbf{1} : \mathrm{Type}_0} \qquad \textsc{UnitTm}\ \frac{}{\Gamma \vdash () : \mathbf{1}}$$

$$\textsc{EmptyInd}\ \frac{\Gamma \vdash s : \mathbf{0} \qquad \Gamma \vdash P}{\Gamma \vdash \mathrm{ind}_{\mathbf{0}}(s; P) : P} \qquad \textsc{UnitInd}\ \frac{\Gamma \vdash s : \mathbf{1} \qquad \Gamma, z : \mathbf{1} \vdash P \qquad \Gamma \vdash b_{()} : P[()]}{\Gamma \vdash \mathrm{ind}_{\mathbf{1}}(s; z.P; b_{()}) : P[s]}$$

$$\textsc{SigUni}\ \frac{\Gamma \vdash A : \mathrm{Type}_i \qquad \Gamma, x : A \vdash B : \mathrm{Type}_i}{\Gamma \vdash \Sigma\, x : A.B : \mathrm{Type}_i} \qquad \textsc{Pair}\ \frac{\Gamma \vdash t : A \qquad \Gamma \vdash u : B[t]}{\Gamma \vdash (t, u)_{x.B} : \Sigma\, x : A.B}$$

$$\textsc{Proj}_1\ \frac{\Gamma \vdash p : \Sigma\, x : A.B}{\Gamma \vdash \pi_1\, p : A} \qquad \textsc{Proj}_2\ \frac{\Gamma \vdash p : \Sigma\, x : A.B}{\Gamma \vdash \pi_2\, p : B[u]}$$

$$\textsc{TreeUni}\ \frac{\Gamma \vdash A : \mathrm{Type}_i \qquad \Gamma, x : A \vdash B : \mathrm{Type}_i}{\Gamma \vdash \mathbf{W}\, x : A.B : \mathrm{Type}_i}$$

$$\textsc{Sup}\ \frac{\begin{array}{c}\Gamma, x : A \vdash B \\ \Gamma \vdash a : A \qquad \Gamma \vdash k : B[a] \to \mathbf{W}\, x : A.B\end{array}}{\Gamma \vdash \mathrm{sup}_{x.B}\, a\, k : \mathbf{W}\, x : A.B}$$

$$\textsc{TreeInd}\ \frac{\begin{array}{c}\Gamma \vdash A \qquad \Gamma, x : A \vdash B \qquad \Gamma \vdash s : \mathbf{W}\, x : A.B \qquad \Gamma, z : \mathbf{W}\, x : A.B \vdash P \\ \Gamma, x : A, y : B[x] \to W x : A.B, h : \Pi\, z : B[x].P[y\, z] \vdash b : P\left[\mathrm{sup}_{x.B}\, x\, y\right]\end{array}}{\Gamma \vdash \mathrm{ind}_{\mathbf{W}\, x:A.B}(s; z.P; x.y.z.b) : P[s]}$$

$$\textsc{BoolUni}\ \frac{}{\Gamma \vdash \mathbf{B} : \mathrm{Type}_0} \qquad \textsc{True}\ \frac{}{\Gamma \vdash \mathrm{tt} : \mathbf{B}} \qquad \textsc{False}\ \frac{}{\Gamma \vdash \mathrm{ff} : \mathbf{B}}$$

$$\textsc{BoolInd}\ \frac{\Gamma \vdash s : \mathbf{B} \qquad \Gamma, z : \mathbf{B} \vdash P \qquad \Gamma \vdash b_{\mathrm{tt}} : P[\mathrm{tt}] \qquad \Gamma \vdash b_{\mathrm{ff}} : P[\mathrm{ff}]}{\Gamma \vdash \mathrm{ind}_{\mathbf{B}}(s; z.P; b_{\mathrm{tt}}, b_{\mathrm{ff}}) : P[s]}$$

$$\textsc{SumUni}\ \frac{\Gamma \vdash A : \mathrm{Type}_i \qquad \Gamma \vdash B : \mathrm{Type}_i}{\Gamma \vdash A + B : \mathrm{Type}_i} \qquad \textsc{SumInjLeft}\ \frac{\Gamma \vdash B \qquad \Gamma \vdash a : A}{\Gamma \vdash \mathrm{inj}^l_B\ a : A + B}$$

$$\textsc{SumInjRight}\ \frac{\Gamma \vdash A \qquad \Gamma \vdash b : B}{\Gamma \vdash \mathrm{inj}^r_A\ b : A + B}$$

$$\textsc{SumInd}\ \frac{\begin{array}{c}\Gamma \vdash s : A + B \qquad \Gamma, z : A + B \vdash P \\ \Gamma, x : A \vdash b_l : P\left[\mathrm{inj}^l_B\ x\right] \qquad \Gamma, x : B \vdash b_r : P\left[\mathrm{inj}^r_A\ x\right]\end{array}}{\Gamma \vdash \mathrm{ind}_{+}(s; z.P; x.b_l, x.b_r) : P[s]}$$

$$\textsc{IdTy}\ \frac{\Gamma \vdash A \qquad \Gamma \vdash a : A \qquad \Gamma \vdash a' : A}{\Gamma \vdash \mathbf{Id}_A\ a\, a'} \qquad \textsc{ReflTm}\ \frac{\Gamma \vdash A \qquad \Gamma \vdash a : A}{\Gamma \vdash \mathrm{refl}_{A,a} : \mathbf{Id}_A\ a\, a}$$

$$\textsc{IdInd}\ \frac{\begin{array}{c}\Gamma \vdash A \qquad \Gamma \vdash a : A \qquad \Gamma \vdash a' : A \\ \Gamma \vdash s : \mathbf{Id}_A\ a\, a' \qquad \Gamma, x : A, y : A, z : \mathbf{Id}_A\ x\, y \vdash P \qquad \Gamma, x : A \vdash b : P\left[\mathrm{id}, x, x, \mathrm{refl}_{A,x}\right]\end{array}}{\Gamma \vdash \mathrm{ind}_{\mathbf{Id}_A}(s; x.y.z.P; x.b) : P[a, a', s]}$$

$\boxed{\Gamma \vdash T \cong T'}$ Types $T$ and $T'$ are convertible in context $\Gamma$

$$\textsc{ReflTy}\ \frac{\Gamma \vdash A}{\Gamma \vdash A \cong A} \qquad \textsc{TransTy}\ \frac{\Gamma \vdash A \cong B \qquad \Gamma \vdash B \cong C}{\Gamma \vdash A \cong C} \qquad \textsc{ElC}\ \frac{\Gamma \vdash A \cong A' : \mathrm{Type}_i}{\Gamma \vdash A \cong A'}$$

$$\textsc{FunTyC}\ \frac{\Gamma \vdash A \cong A' \qquad \Gamma, x : A \vdash B \cong B'}{\Gamma \vdash \Pi\, x : A.B \cong \Pi\, x : A'.B'} \qquad \textsc{ListTyC}\ \frac{\Gamma \vdash A \cong A'}{\Gamma \vdash \mathbf{List}\, A \cong \mathbf{List}\, A'}$$

$$\textsc{SigTyC}\ \frac{\Gamma \vdash A \cong A' \qquad \Gamma, x : A \vdash B \cong B'}{\Gamma \vdash \Sigma\, x : A.B \cong \Sigma\, x : A'.B'} \qquad \textsc{TreeTyC}\ \frac{\Gamma \vdash A \cong A' \qquad \Gamma, x : A \vdash B \cong B'}{\Gamma \vdash \mathbf{W}\, x : A.B \cong \mathbf{W}\, x : A'.B'}$$

$$\textsc{IdTyC}\ \frac{\Gamma \vdash A \cong A' \qquad \Gamma \vdash t \cong t' : A \qquad \Gamma \vdash u \cong u' : A}{\Gamma \vdash \mathbf{Id}_A\ t\, u \cong \mathbf{Id}_{A'}\ t'\, u'} \qquad \textsc{SumTyC}\ \frac{\Gamma \vdash A \cong A' \qquad \Gamma \vdash B \cong B'}{\Gamma \vdash A + B \cong A' + B'}$$

$\boxed{\Gamma \vdash t \cong t' : T}$ Terms $t$ and $t'$ are convertible at type $T$ in context $\Gamma$

$$\textsc{Refl}\ \frac{\Gamma \vdash t : A}{\Gamma \vdash t \cong t : A} \qquad \textsc{Trans}\ \frac{\Gamma \vdash t \cong u : A \qquad \Gamma \vdash u \cong v : A}{\Gamma \vdash t \cong v : A} \qquad \textsc{Conv}\ \frac{\Gamma \vdash t \cong t' : A \qquad \Gamma \vdash A \cong B}{\Gamma \vdash t \cong t' : B}$$

$$\beta\textsc{Fun}\ \frac{\Gamma \vdash A \qquad \Gamma, x : A \vdash B \qquad \Gamma, x : A \vdash t : B \qquad \Gamma \vdash u : A}{\Gamma \vdash (\lambda\, x : A.t)\, u \cong t[u] : B[u]} \qquad \eta\textsc{Fun}\ \frac{\Gamma, x : A \vdash f\, x \cong g\, x : B}{\Gamma \vdash f \cong g : \Pi\, x : A.B}$$

$$\beta\textsc{Sig}_1\ \frac{\Gamma \vdash A \qquad \Gamma, x : A \vdash B \qquad \Gamma \vdash t : A \qquad \Gamma \vdash u : B[t]}{\Gamma \vdash \pi_1\, (t, u)_{x.B} \cong t : A} \qquad \beta\textsc{Sig}_2\ \frac{\Gamma \vdash A \qquad \Gamma, x : A \vdash B \qquad \Gamma \vdash t : A \qquad \Gamma \vdash u : B[t]}{\Gamma \vdash \pi_2\, (t, u)_{x.B} \cong u : B[t]}$$

$$\eta\textsc{Sig}\ \frac{\Gamma \vdash A \qquad \Gamma, x : A \vdash B \qquad \Gamma \vdash p : \Sigma\, x : A.B}{\Gamma \vdash p \cong (\pi_1\, p, \pi_2\, p)_{x.B} : \Sigma\, x : A.B}$$

$$\beta\textsc{Nil}\ \frac{\Gamma \vdash A \qquad \Gamma, x : \mathbf{List}\, A \vdash P \qquad \Gamma \vdash b_\varepsilon : P[\varepsilon_A] \qquad \Gamma, x : A, y : \mathbf{List}\, A, z : P[y] \vdash b_{::} : P[x ::_A y]}{\Gamma \vdash \mathrm{ind}_{\mathbf{List}\, A}(\varepsilon_A; z.P; b_\varepsilon, x.y.z.b_{::}) \cong b_\varepsilon : P[\varepsilon_A]}$$

$$\beta\textsc{Cons}\ \frac{\Gamma \vdash A \qquad \Gamma \vdash a : A \qquad \Gamma \vdash l : \mathbf{List}\, A \qquad \Gamma, x : \mathbf{List}\, A \vdash P \qquad \Gamma \vdash b_\varepsilon : P[\varepsilon_A] \qquad \Gamma, x : A, y : \mathbf{List}\, A, z : P[y] \vdash b_{::} : P[x ::_A y]}{\Gamma \vdash \mathrm{ind}_{\mathbf{List}\, A}(a ::_A l; z.P; b_\varepsilon, x.y.z.b_{::}) \cong b_{::}[a, l, \mathrm{ind}_{\mathbf{List}\, A}(l; z.P; b_\varepsilon, x.y.z.b_{::})] : P[a ::_A l]}$$

$$\beta\textsc{True}\ \frac{\Gamma, z : \mathbf{B} \vdash P \qquad \Gamma \vdash b_{\mathrm{tt}} : P[\mathrm{tt}] \qquad \Gamma \vdash b_{\mathrm{ff}} : P[\mathrm{ff}]}{\Gamma \vdash \mathrm{ind}_{\mathbf{B}}(\mathrm{tt}; z.P; b_{\mathrm{tt}}, b_{\mathrm{ff}}) \cong b_{\mathrm{tt}} : P[\mathrm{tt}]} \qquad \beta\textsc{False}\ \frac{\Gamma, z : \mathbf{B} \vdash P \qquad \Gamma \vdash b_{\mathrm{tt}} : P[\mathrm{tt}] \qquad \Gamma \vdash b_{\mathrm{ff}} : P[\mathrm{ff}]}{\Gamma \vdash \mathrm{ind}_{\mathbf{B}}(\mathrm{ff}; z.P; b_{\mathrm{tt}}, b_{\mathrm{ff}}) \cong b_{\mathrm{ff}} : P[\mathrm{ff}]}$$

$$\text{βInjLeft}\ \frac{\Gamma \vdash A+B \qquad \Gamma \vdash a : A \qquad \Gamma, z : A+B \vdash P \qquad \Gamma, x : A \vdash b_l : P\left[\mathsf{inj}^l_B\ x\right] \qquad \Gamma, x : B \vdash b_r : P\left[\mathsf{inj}^r_A\ x\right]}{\Gamma \vdash \mathsf{ind}_+(\mathsf{inj}^l_B\ a; z.P; x.b_l, x.b_r) \cong b_l[a] : P\left[\mathsf{inj}^l_B\ a\right]}$$

$$\text{βInjRiht}\ \frac{\Gamma \vdash A+B \qquad \Gamma \vdash b : B \qquad \Gamma, z : A+B \vdash P \qquad \Gamma, x : A \vdash b_l : P\left[\mathsf{inj}^l_B\ x\right] \qquad \Gamma, x : B \vdash b_r : P\left[\mathsf{inj}^r_A\ x\right]}{\Gamma \vdash \mathsf{ind}_+(\mathsf{inj}^l_A\ b; z.P; x.b_l, x.b_r) \cong b_r[b] : P\left[\mathsf{inj}^r_A\ b\right]}$$

$$\text{βTree}\ \frac{\Gamma, x : A \vdash B \qquad \Gamma \vdash a : A \qquad \Gamma \vdash k : B[a] \to \mathbf{W}\,x : A.B \qquad \Gamma, z : \mathbf{W}\,x : A.B \vdash P \qquad \Gamma, x : A, y : B[x] \to W x : A.B, h : \Pi\, z : B[x].P[y\ z] \vdash b : P\left[\mathsf{sup}_{x.B}\ x\ y\right]}{\Gamma \vdash \mathsf{ind}_{\mathbf{W}\,x:A.B}(\mathsf{sup}_{x.B}\ a\ k; z.P; x.y.z.b) \cong b[a, k, (\lambda\, z : B[x].\ \mathsf{ind}_{\mathbf{W}\,x:A.B}(k\ z; z.P; x.y.z.b))] : P\left[\mathsf{sup}_{x.B}\ a\ k\right]}$$

$$\text{βRefl}\ \frac{\Gamma \vdash A \qquad \Gamma \vdash a : A \qquad \Gamma, x : A, y : A, z : \mathbf{Id}_A\ x\ y \vdash P \qquad \Gamma, x : A \vdash b : P\left[x, x, \mathsf{refl}_{A,x}\right]}{\Gamma \vdash \mathsf{ind}_{\mathbf{Id}_A}(\mathsf{refl}_{A,a}; x.y.z.P; x.b) \cong b[a] : P\left[a, a, \mathsf{refl}_{A,a}\right]}$$

$$\text{FunCong}\ \frac{\Gamma \vdash A \cong A' : \mathsf{Type}_i \qquad \Gamma, x : A \vdash B \cong B' : \mathsf{Type}_i}{\Gamma \vdash \Pi\, x : A.B \cong \Pi\, x : A'.B' : \mathsf{Type}_i}$$

other congruences omitted

## C.2 Algorithmic MLTT

$\boxed{t \rightsquigarrow^1 t'}$ Term $t$ weak-head reduces in one step to term $t'$

$$\text{βFun}\ \frac{}{(\lambda\, x : A.t)\ u \rightsquigarrow^1 t[u]} \qquad \text{βSig}_1\ \frac{}{\pi_1\ (t, u)_{x.B} \rightsquigarrow^1 t} \qquad \text{βSig}_2\ \frac{}{\pi_2\ (t, u)_{x.B} \rightsquigarrow^1 u}$$

$$\text{βRedNil}\ \frac{}{\mathsf{ind}_{\mathbf{List}\,A}(\varepsilon_A; x.P; b_\varepsilon, x.y.z.b_{::}) \rightsquigarrow^1 b_\varepsilon}$$

$$\text{βRedCons}\ \frac{}{\mathsf{ind}_{\mathbf{List}\,A}(a ::_A l; x.P; b_\varepsilon, x.y.z.b_{::}) \rightsquigarrow^1 b_{::}[a, l, \mathsf{ind}_{\mathbf{List}\,A}(l; x.P; b_\varepsilon, x.y.z.b_{::})]}$$

$$\text{βTree}\ \frac{}{\mathsf{ind}_{\mathbf{W}\,x:A.B}(\mathsf{sup}_{x.B}\ a\ k; z.P; x.y.z.b) \rightsquigarrow^1 b[a, k, (\lambda\, z : B[x].\ \mathsf{ind}_{\mathbf{W}\,x:A.B}(k\ z; z.P; x.y.z.b))]}$$

$$\text{βTrue}\ \frac{}{\mathsf{ind}_{\mathbf{B}}(\mathsf{tt}; z.P; b_{\mathsf{tt}}, b_{\mathsf{ff}}) \rightsquigarrow^1 b_{\mathsf{tt}}} \qquad \text{βFalse}\ \frac{}{\mathsf{ind}_{\mathbf{B}}(\mathsf{ff}; z.P; b_{\mathsf{tt}}, b_{\mathsf{ff}}) \rightsquigarrow^1 b_{\mathsf{ff}}}$$

$$\text{βInjLeft}\ \frac{}{\mathsf{ind}_+(\mathsf{inj}^l\ a; z.P; x.b_l, x.b_r) \rightsquigarrow^1 b_l[a]}$$

$$\text{βInjRight}\ \frac{}{\mathsf{ind}_+(\mathsf{inj}^r\ b; z.P; x.b_l, x.b_r) \rightsquigarrow^1 b_r[b]} \qquad \text{βRefl}\ \frac{}{\mathsf{ind}_{\mathbf{Id}_A}(\mathsf{refl}_{A,a}; x.z.P; x.b) \rightsquigarrow^1 b[a]}$$

$$\text{RedApp}\ \frac{t \rightsquigarrow^1 t'}{t\ u \rightsquigarrow^1 t'\ u} \qquad \text{RedSig}_1\ \frac{t \rightsquigarrow^1 t'}{\pi_1\, t \rightsquigarrow^1 \pi_1\, t'} \qquad \text{RedSig}_2\ \frac{t \rightsquigarrow^1 t'}{\pi_2\, t \rightsquigarrow^1 \pi_2\, t'}$$

$$\textsc{RedInd}\ \frac{t \leadsto^1 t'}{\mathrm{ind}_T(t; P; \vec{b}) \leadsto^1 \mathrm{ind}_T(t'; P; \vec{b})}$$

$\boxed{t \leadsto^\star t'}$ Term $t$ weak-head reduces in multiple steps to term $t'$

$$\textsc{RedBase}\ \frac{}{t \leadsto^\star t} \qquad \textsc{RedStep}\ \frac{t \leadsto^1 t' \qquad t' \leadsto^\star t''}{t \leadsto^\star t''}$$

$\boxed{\mathrm{nf}\ f} \stackrel{\text{def}}{=} n \mid \mathrm{Type}_i \mid \Pi\, x : t.t \mid \lambda\, x : t.t \mid \mathbf{List}\, t \mid \varepsilon_t \mid t ::_t t \mid$ weak-head normal forms
$\Sigma\, x : t.t \mid (t, t) \mid \mathbf{W}\, x : t.t \mid \mathrm{sup}_t\, t\, t \mid \mathbf{0} \mid \mathbf{1} \mid () \mid$
$\mathbf{B} \mid \mathrm{tt} \mid \mathrm{ff} \mid \mathbf{Id}_t\ t\, t' \mid \mathrm{refl}_{t,t} \mid t + t \mid \mathrm{inj}^l_t\ t \mid \mathrm{inj}^r_t\ t$

$\boxed{\mathrm{ne}\ n} \stackrel{\text{def}}{=} x \mid n\, t \mid \mathrm{ind}_T(n; t; t) \mid \pi_1\, n \mid \pi_2\, n$ weak-head neutrals

$\boxed{\Gamma \vdash T \lhd}$ $T$ is a type in $\Gamma$

$$\textsc{FunTy}\ \frac{\Gamma \vdash A \lhd \qquad \Gamma, x : A \vdash B \lhd}{\Gamma \vdash \Pi\, x : A.B \lhd} \qquad \textsc{ListTy}\ \frac{\Gamma \vdash A \lhd}{\Gamma \vdash \mathbf{List}\, A \lhd}$$

$$\textsc{SigTy}\ \frac{\Gamma \vdash A \lhd \qquad \Gamma, x : A \vdash B \lhd}{\Gamma \vdash \Sigma\, x : A.B \lhd} \qquad \textsc{TreeTy}\ \frac{\Gamma \vdash A \lhd \qquad \Gamma, x : A \vdash B \lhd}{\Gamma \vdash \mathbf{W}\, x : A.B \lhd}$$

$$\textsc{EmptyTy}\ \frac{}{\Gamma \vdash \mathbf{0} \lhd} \qquad \textsc{UnitTy}\ \frac{}{\Gamma \vdash \mathbf{1} \lhd} \qquad \textsc{BoolTy}\ \frac{}{\Gamma \vdash \mathbf{B} \lhd}$$

$$\textsc{IdTy}\ \frac{\Gamma \vdash A \lhd \qquad \Gamma \vdash a \lhd A \qquad \Gamma \vdash a' \lhd A}{\Gamma \vdash \mathbf{Id}_A\ a\, a' \lhd}$$

$$\textsc{SumTy}\ \frac{\Gamma \vdash A \lhd \qquad \Gamma \vdash B \lhd}{\Gamma \vdash A + B \lhd} \qquad \textsc{El}\ \frac{\Gamma \vdash A \rhd_{\mathrm{h}} \mathrm{Type}_i \qquad A \text{ is not a canonical form}}{\Gamma \vdash A \lhd}$$

$\boxed{\Gamma \vdash t \rhd T}$ Term $t$ infers type $T$ in context $\Gamma$

$$\textsc{Sort}\ \frac{}{\Gamma \vdash \mathrm{Type}_i \rhd \mathrm{Type}_{i+1}} \qquad \textsc{Var}\ \frac{(x : T) \in \Gamma}{\Gamma \vdash x \rhd T} \qquad \textsc{Fun}\ \frac{\Gamma \vdash A \rhd_{\mathrm{h}} \mathrm{Type}_i \qquad \Gamma, x : A \vdash B \lhd \mathrm{Type}_i}{\Gamma \vdash \Pi\, x : A.B \rhd \mathrm{Type}_i}$$

$$\textsc{Abs}\ \frac{\Gamma \vdash A \lhd \qquad \Gamma, x : A \vdash t \rhd B}{\Gamma \vdash \lambda\, x : A.t \rhd \Pi\, x : A.B} \qquad \textsc{App}\ \frac{\Gamma \vdash t \rhd_{\mathrm{h}} \Pi\, x : A.B \qquad \Gamma \vdash u \lhd A}{\Gamma \vdash t\, u \rhd B[u]}$$

$$\textsc{List}\ \frac{\Gamma \vdash A \rhd_{\mathrm{h}} \mathrm{Type}_i}{\Gamma \vdash \mathbf{List}\, A \rhd \mathrm{Type}_i} \qquad \textsc{Nil}\ \frac{\Gamma \vdash A \lhd}{\Gamma \vdash \varepsilon_A \rhd \mathbf{List}\, A} \qquad \textsc{Cons}\ \frac{\Gamma \vdash A \lhd \qquad \Gamma \vdash a \lhd A \qquad \Gamma \vdash l \lhd \mathbf{List}\, A}{\Gamma \vdash a ::_A l \rhd \mathbf{List}\, A}$$

$$\textsc{ListInd}\ \frac{\Gamma \vdash A \lhd \qquad \Gamma \vdash s \lhd \mathbf{List}\,A \qquad \Gamma, x : \mathbf{List}\,A \vdash P \rhd \quad \Gamma \vdash b_\varepsilon \lhd P[\varepsilon_A] \qquad \Gamma, x : A, y : \mathbf{List}\,A, z : P[y] \vdash b_{::} \lhd P[x ::_A y]}{\Gamma \vdash \mathrm{ind}_{\mathbf{List}\,A}(s; x.P; b_\varepsilon, x.y.z.b_{::}) \rhd P[s]}$$

$$\textsc{Empty}\ \frac{}{\Gamma \vdash \mathbf{0} \rhd \mathrm{Type}_0} \qquad \textsc{EmptyInd}\ \frac{\Gamma \vdash s \lhd \mathbf{0} \qquad \Gamma \vdash P \lhd}{\Gamma \vdash \mathrm{ind}_{\mathbf{0}}(s; P) \rhd P} \qquad \textsc{UnitUni}\ \frac{}{\Gamma \vdash \mathbf{1} \rhd \mathrm{Type}_0}$$

$$\textsc{UnitTm}\ \frac{}{\Gamma \vdash () \rhd \mathbf{1}} \qquad \textsc{UnitInd}\ \frac{\Gamma \vdash s \lhd \mathbf{1} \qquad \Gamma, z : \mathbf{1} \vdash P \lhd \qquad \Gamma \vdash b_{()} \lhd P[()]}{\Gamma \vdash \mathrm{ind}_{\mathbf{1}}(s; z.P; b_{()}) \rhd P[s]}$$

$$\textsc{Sig}\ \frac{\Gamma \vdash A \rhd_{\mathrm{h}} \mathrm{Type}_i \qquad \Gamma, x : A \vdash B \lhd \mathrm{Type}_i}{\Gamma \vdash \Sigma\, x : A.B \rhd \mathrm{Type}_i} \qquad \textsc{Pair}\ \frac{\Gamma \vdash t \rhd A \qquad \Gamma, x : A \vdash B \lhd \quad \Gamma \vdash u \lhd B[t]}{\Gamma \vdash (t, u)_{x.B} \rhd \Sigma\, x : A.B}$$

$$\textsc{Proj}_1\ \frac{\Gamma \vdash p \rhd_{\mathrm{h}} \Sigma\, x : A.B}{\Gamma \vdash \pi_1\, p \rhd A} \qquad \textsc{Proj}_2\ \frac{\Gamma \vdash p \rhd \Sigma\, x : A.B}{\Gamma \vdash \pi_2\, p \rhd B[\pi_1\, p]}$$

$$\textsc{Tree}\ \frac{\Gamma \vdash A \rhd_{\mathrm{h}} \mathrm{Type}_i \qquad \Gamma, x : A \vdash B \lhd \mathrm{Type}_i}{\Gamma \vdash \mathbf{W}\, x : A.B \lhd \mathrm{Type}_i}$$

$$\textsc{Sup}\ \frac{\Gamma \vdash a \rhd A \qquad \Gamma, x : A \vdash B \lhd \qquad \Gamma \vdash k \lhd B[a] \to \mathbf{W}\, x : A.B}{\Gamma \vdash \mathrm{sup}_{x.B}\, a\, k \rhd \mathbf{W}\, x : A.B}$$

$$\textsc{TreeInd}\ \frac{\Gamma \vdash A \lhd \qquad \Gamma, x : A \vdash B \lhd \qquad \Gamma \vdash s \lhd \mathbf{W}\, x : A.B \qquad \Gamma, z : \mathbf{W}\, x : A.B \vdash P \lhd \quad \Gamma, x : A, y : B[x] \to \mathbf{W}\, x : A.B, h : \Pi\, z : B[x].P[y\, z] \vdash b \lhd P\left[\mathrm{sup}_{x.B}\, x\, y\right]}{\Gamma \vdash \mathrm{ind}_{\mathbf{W}\, x:A.B}(s; z.P; x.y.z.b) \rhd P[s]}$$

$$\textsc{BoolUni}\ \frac{}{\Gamma \vdash \mathbf{B} \rhd \mathrm{Type}_0} \qquad \textsc{True}\ \frac{}{\Gamma \vdash \mathrm{tt} \rhd \mathbf{B}} \qquad \textsc{False}\ \frac{}{\Gamma \vdash \mathrm{ff} \rhd \mathbf{B}}$$

$$\textsc{BoolInd}\ \frac{\Gamma \vdash s \lhd \mathbf{B} \qquad \Gamma, z : \mathbf{B} \vdash P \lhd \qquad \Gamma \vdash b_{\mathrm{tt}} \lhd P[\mathrm{tt}] \qquad \Gamma \vdash b_{\mathrm{ff}} \lhd P[\mathrm{ff}]}{\Gamma \vdash \mathrm{ind}_{\mathbf{B}}(s; z.P; b_{\mathrm{tt}}, b_{\mathrm{ff}}) \rhd P[s]}$$

$$\textsc{Sum}\ \frac{\Gamma \vdash A \rhd_{\mathrm{h}} \mathrm{Type}_i \qquad \Gamma \vdash B \lhd \mathrm{Type}_i}{\Gamma \vdash A + B \rhd \mathrm{Type}_i} \qquad \textsc{SumInjLeft}\ \frac{\Gamma \vdash B \lhd \qquad \Gamma \vdash a \rhd A}{\Gamma \vdash \mathrm{inj}^l_B\ a \rhd A + B}$$

$$\textsc{SumInjRight}\ \frac{\Gamma \vdash A \lhd \qquad \Gamma \vdash b \rhd B}{\Gamma \vdash \mathrm{inj}^r_A\ b \rhd A + B}$$

$$\textsc{SumInd}\ \frac{\Gamma \vdash A \lhd \qquad \Gamma \vdash B \lhd \qquad \Gamma \vdash s \lhd A + B \qquad \Gamma, z : A + B \vdash P \lhd \quad \Gamma, x : A \vdash b_l \lhd P\left[\mathrm{inj}^l_B\ x\right] \qquad \Gamma, x : B \vdash b_r \lhd P\left[\mathrm{inj}^r_A\ x\right]}{\Gamma \vdash \mathrm{ind}_{A+B}(s; z.P; x.b_l, x.b_r) \rhd P[s]}$$

$$\textsc{Id}\ \frac{\Gamma \vdash A \rhd_{\mathrm{h}} \mathsf{Type}_i \qquad \Gamma \vdash a \lhd A \qquad \Gamma \vdash a' \lhd A}{\Gamma \vdash \mathbf{Id}_A\ a\, a' \rhd \mathsf{Type}_i} \qquad \textsc{ReflTm}\ \frac{\Gamma \vdash A \lhd \qquad \Gamma \vdash a \lhd A}{\Gamma \vdash \mathsf{refl}_{A,a} \rhd \mathbf{Id}_A\ a\, a}$$

$$\textsc{IdInd}\ \frac{\begin{array}{c}\Gamma \vdash A \lhd \qquad \Gamma \vdash s \rhd_{\mathrm{h}} \mathbf{Id}_{A'}\ a\, a' \\ \Gamma, x : A, y : A, z : \mathbf{Id}_A\ x\, y \vdash P \lhd \qquad \Gamma, x : A \vdash b \lhd P[x, x, \mathsf{refl}_{A,x}]\end{array}}{\Gamma \vdash \mathsf{ind}_{\mathbf{Id}_A}(s; x.y.z.P; x.b) \rhd P[a, a', s]}$$

$\boxed{\Gamma \vdash t \lhd T}$ Term $t$ checks against type $T$

$$\textsc{Check}\ \frac{\Gamma \vdash t \rhd T' \qquad \Gamma \vdash T' \cong T \lhd}{\Gamma \vdash t \lhd T}$$

$\boxed{\Gamma \vdash t \rhd_{\mathrm{h}} T}$ Term $t$ infers the reduced type $T$

$$\textsc{InfRed}\ \frac{\Gamma \vdash t \rhd T \qquad \Gamma \vdash T \rightsquigarrow^\star T'}{\Gamma \vdash t \rhd_{\mathrm{h}} T'}$$

$\boxed{\Gamma \vdash T \cong T' \lhd}$ Types $T$ and $T'$ are convertible

$$\textsc{TyRed}\ \frac{T \rightsquigarrow^\star U \qquad T' \rightsquigarrow^\star U' \qquad \Gamma \vdash U \cong_{\mathrm{h}} U' \lhd}{\Gamma \vdash T \cong T' \lhd}$$

$\boxed{\Gamma \vdash t \cong t' \lhd A}$ Terms $t$ and $t'$ are convertible at type $T$

$$\textsc{TmRed}\ \frac{t \rightsquigarrow^\star u \qquad t' \rightsquigarrow^\star u' \qquad T \rightsquigarrow^\star U \qquad \Gamma \vdash u \cong_{\mathrm{h}} u' \lhd U}{\Gamma \vdash t \cong t' \lhd T}$$

$\boxed{\Gamma \vdash T \cong_{\mathrm{h}} T' \lhd}$ Reduced types $T$ and $T'$ are convertible

$$\textsc{CUniTy}\ \frac{}{\Gamma \vdash \mathsf{Type}_i \cong_{\mathrm{h}} \mathsf{Type}_i \lhd} \qquad \textsc{CProdTy}\ \frac{\Gamma \vdash A \cong A' \lhd \qquad \Gamma, x : A' \vdash B \cong B' \lhd}{\Gamma \vdash \Pi\, x : A.B \cong_{\mathrm{h}} \Pi\, x : A'.B' \lhd}$$

$$\textsc{CListTy}\ \frac{\Gamma \vdash A \cong A' \lhd}{\Gamma \vdash \mathbf{List}\, A \cong_{\mathrm{h}} \mathbf{List}\, A' \lhd} \qquad \textsc{CSigTy}\ \frac{\Gamma \vdash A \cong A' \lhd \qquad \Gamma, x : A \vdash B \cong B' \lhd}{\Gamma \vdash \Sigma\, x : A.B \cong_{\mathrm{h}} \Sigma\, x : A'.B' \lhd}$$

$$\textsc{CTreeTy}\ \frac{\begin{array}{c}\Gamma \vdash A \cong A' \lhd \\ \Gamma, x : A \vdash B \cong B' \lhd\end{array}}{\Gamma \vdash \mathbf{W}\, x : A.B \cong_{\mathrm{h}} \mathbf{W}\, x : A'.B' \lhd} \qquad \textsc{CIdTy}\ \frac{\begin{array}{c}\Gamma \vdash A \cong A' \lhd \qquad \Gamma \vdash t \cong t' \lhd A \\ \Gamma \vdash u \cong u' \lhd A\end{array}}{\Gamma \vdash \mathbf{Id}_A\ t\, u \cong_{\mathrm{h}} \mathbf{Id}_{A'}\ t'\, u' \lhd}$$

$$\textsc{CSumTy}\ \frac{\Gamma \vdash A \cong A' \lhd \qquad \Gamma \vdash B \cong B' \lhd}{\Gamma \vdash A + B \cong A' + B' \lhd} \qquad \textsc{CReflTy}\ \frac{T \text{ is } \mathbf{0}, \mathbf{1} \text{ or } \mathbf{B}}{\Gamma \vdash T \cong_{\mathrm{h}} T \lhd}$$

$$\textsc{NeuTy}\ \frac{\Gamma \vdash n \approx n' \rhd T}{\Gamma \vdash n \cong_{\mathrm{h}} n' \lhd}$$

$\boxed{\Gamma \vdash t \cong_{\mathrm{h}} t' \lhd A}$ Reduced terms $t$ and $t'$ are convertible at type $A$

$$\textsc{CUni}\ \frac{}{\Gamma \vdash \mathsf{Type}_i \cong_{\mathrm{h}} \mathsf{Type}_j \lhd \mathsf{Type}_k} \qquad \textsc{CFun}\ \frac{\Gamma \vdash A \cong A' \lhd \mathsf{Type}_i \quad \Gamma, x : A' \vdash B \cong B' \lhd \mathsf{Type}_i}{\Gamma \vdash \Pi\, x : A.B \cong_{\mathrm{h}} \Pi\, x : A'.B' \lhd \mathsf{Type}_i}$$

$$\textsc{CFunEta}\ \frac{\Gamma, x : A \vdash f\ x \cong f'\ x \lhd B}{\Gamma \vdash f \cong_{\mathrm{h}} f' \lhd \Pi\, x : A.B} \qquad \textsc{CSig}\ \frac{\Gamma \vdash A \cong A' \lhd \mathsf{Type}_i \quad \Gamma, x : A' \vdash B \cong B' \lhd \mathsf{Type}_i}{\Gamma \vdash \Sigma\, x : A.B \cong_{\mathrm{h}} \Sigma\, x : A'.B' \lhd \mathsf{Type}_i}$$

$$\textsc{CSigEta}\ \frac{\Gamma \vdash \pi_1\, p \cong \pi_1\, p' \lhd A \quad \Gamma \vdash \pi_2\, p \cong \pi_2\, p' \lhd B[\pi_1\, p]}{\Gamma \vdash p \cong_{\mathrm{h}} p' \lhd \Sigma\, x : A.B} \qquad \textsc{CList}\ \frac{\Gamma \vdash A \cong A' \lhd \mathsf{Type}_i}{\Gamma \vdash \textbf{List}\, A \cong_{\mathrm{h}} \textbf{List}\, A' \lhd \mathsf{Type}_i}$$

$$\textsc{CNil}\ \frac{}{\Gamma \vdash \varepsilon_A \cong_{\mathrm{h}} \varepsilon_{A'} \lhd \textbf{List}\, A''} \qquad \textsc{CCons}\ \frac{\Gamma \vdash a \cong a' \lhd A'' \qquad \Gamma \vdash l \cong l' \lhd \textbf{List}\, A''}{\Gamma \vdash a ::_A l \cong_{\mathrm{h}} a' ::_{A'} l' \lhd \textbf{List}\, A''}$$

$$\textsc{CReflUni}\ \frac{T \text{ is } \mathbf{0}, \mathbf{1} \text{ or } \mathbf{B}}{\Gamma \vdash T \cong_{\mathrm{h}} T \lhd \mathsf{Type}_0} \qquad \textsc{CUnitK}\ \frac{}{\Gamma \vdash () \cong_{\mathrm{h}} () \lhd \mathbf{1}} \qquad \textsc{CReflBool}\ \frac{t \text{ is } \mathsf{tt} \text{ or } \mathsf{ff}}{\Gamma \vdash t \cong_{\mathrm{h}} t \lhd \mathbf{B}}$$

$$\textsc{CTree}\ \frac{\Gamma \vdash A \cong A' \lhd \mathsf{Type}_i \quad \Gamma, x : A' \vdash B \cong B' \lhd \mathsf{Type}_i}{\Gamma \vdash \mathbf{W}\, x : A.B \cong_{\mathrm{h}} \mathbf{W}\, x : A'.B' \lhd \mathsf{Type}_i}$$

$$\textsc{CSup}\ \frac{\Gamma \vdash a \cong a' \lhd A'' \qquad \Gamma \vdash k \cong k' \lhd B''[a] \to \mathbf{W}\, x : A''.B''}{\Gamma \vdash \mathsf{sup}_{x.B}\, a\ k \cong_{\mathrm{h}} \mathsf{sup}_{x.B'}\, a'\ k' \lhd \mathbf{W}\, x : A''.B''}$$

$$\textsc{CSum}\ \frac{\Gamma \vdash A \cong A' \lhd \mathsf{Type}_i \qquad \Gamma \vdash B \cong B' \lhd \mathsf{Type}_i}{\Gamma \vdash A + B \cong A' + B' \lhd \mathsf{Type}_i}$$

$$\textsc{CInjLeft}\ \frac{\Gamma \vdash B \cong B' \lhd \qquad \Gamma \vdash a \cong a' \lhd A}{\Gamma \vdash \mathsf{inj}^l_B\ a \cong \mathsf{inj}^l_{B'}\ a' \lhd A + B} \qquad \textsc{CInjRight}\ \frac{\Gamma \vdash A \cong A' \lhd \qquad \Gamma \vdash b \cong b' \lhd B}{\Gamma \vdash \mathsf{inj}^r_A\ b \cong \mathsf{inj}^r_{A'}\ b' \lhd A + B}$$

$$\textsc{ReflRefl}\ \frac{}{\Gamma \vdash \mathsf{refl}_{A,a} \cong \mathsf{refl}_{A',a'} \lhd \mathbf{Id}_{A''}\ t\, u} \qquad \textsc{NeuNeu}\ \frac{\Gamma \vdash n \approx n' \rhd S \qquad \mathsf{ne}\, M}{\Gamma \vdash n \cong_{\mathrm{h}} n' \lhd M}$$

$$\textsc{NeuPos}\ \frac{\Gamma \vdash n \approx n' \rhd S \qquad T \text{ is } \mathsf{Type}_i, \mathbf{0}, \mathbf{1}, \mathbf{B}, \textbf{List}\, A, \mathbf{W}\, x : A.B \text{ or } \mathbf{Id}_A\ a\, a'}{\Gamma \vdash n \cong_{\mathrm{h}} n' \lhd T}$$

$\boxed{\Gamma \vdash t \approx_{\mathrm{h}} t' \rhd T}$ Neutrals $t$ and $t'$ are comparable, inferring the reduced type $T$

$$\text{NRed}\ \frac{\Gamma \vdash n \approx n' \rhd T \qquad T \rightsquigarrow^{\star} S}{\Gamma \vdash n \approx_{\mathrm{h}} n' \rhd S}$$

$\boxed{\Gamma \vdash t \approx t' \rhd T}$ Neutrals $t$ and $t'$ are comparable, inferring the type $T$

$$\text{NVar}\ \frac{(x : T) \in \Gamma}{\Gamma \vdash x \approx x \rhd T} \qquad \text{NApp}\ \frac{\Gamma \vdash n \approx_{\mathrm{h}} n' \rhd \Pi\, x : A.B \qquad \Gamma \vdash u \cong u' \lhd A}{\Gamma \vdash n\ u \approx n'\ u' \rhd B[u]}$$

$$\text{NListInd}\ \frac{\begin{array}{c}\Gamma \vdash A \cong A' \lhd \qquad \Gamma \vdash s \approx s' \rhd S \qquad \Gamma, z : \mathbf{List}\, A \vdash P \cong P' \lhd \\ \Gamma \vdash b_{\varepsilon} \cong b'_{\varepsilon} \lhd P[\varepsilon_A] \qquad \Gamma, x : A, y : \mathbf{List}\, A, z : P[y] \vdash b_{::} \cong b'_{::} \lhd P[x ::_A y]\end{array}}{\Gamma \vdash \mathrm{ind}_{\mathbf{List}\, A}(s; z.P; b_{\varepsilon}, x.y.z.b_{::}) \approx \mathrm{ind}_{\mathbf{List}\, A'}(s'; z.P'; b'_{\varepsilon}, x.y.z.b'_{::}) \rhd P[s]}$$

$$\text{NEmptyInd}\ \frac{\Gamma \vdash s \approx_{\mathrm{h}} s' \rhd \mathbf{0} \qquad \Gamma \vdash P \cong P' \lhd}{\Gamma \vdash \mathrm{ind}_{\mathbf{0}}(s; P) \approx \mathrm{ind}_{\mathbf{0}}(s'; P') \rhd P}$$

$$\text{NUnitInd}\ \frac{\Gamma \vdash s \approx_{\mathrm{h}} s' \rhd \mathbf{1} \qquad \Gamma, z : \mathbf{1} \vdash P \cong P' \lhd \qquad \Gamma \vdash b \cong b' \lhd P[()]}{\Gamma \vdash \mathrm{ind}_{\mathbf{1}}(s; z.P; b) \approx \mathrm{ind}_{\mathbf{0}}(s'; z.P'; b') \rhd P[s]}$$

$$\text{NSig}_1\ \frac{\Gamma \vdash n \approx_{\mathrm{h}} n' \rhd \Sigma\, x : A.B}{\Gamma \vdash \pi_1\, n \approx \pi_1\, n' \rhd A} \qquad \text{NSig}_2\ \frac{\Gamma \vdash n \approx_{\mathrm{h}} n' \rhd \Sigma\, x : A.B}{\Gamma \vdash \pi_2\, n \approx \pi_2\, n' \rhd B[\pi_1\, n]}$$

$$\text{NTreeInd}\ \frac{\begin{array}{c}\Gamma \vdash A \cong A' \lhd \\ \Gamma, x : A \vdash B \cong B' \lhd \qquad \Gamma \vdash s \approx s' \rhd S \qquad \Gamma, z : \mathbf{W}\, x : A.B \vdash P \cong P' \lhd \\ \Gamma, x : A, y : B[x] \to W x : A.B, h : \Pi\, z : B[x].P[y\, z] \vdash b \cong b' \lhd P\left[\mathrm{sup}_{x.B}\, x\, y\right]\end{array}}{\Gamma \vdash \mathrm{ind}_{\mathbf{W}\, x:A.B}(s; z.P; x.y.z.b) \approx \mathrm{ind}_{\mathbf{W}\, x:A'.B'}(s'; z.P'; x.y.z.b') \rhd P[s]}$$

$$\text{NBoolInd}\ \frac{\begin{array}{c}\Gamma \vdash s \approx_{\mathrm{h}} s' \rhd S \\ \Gamma, z : \mathbf{B} \vdash P \cong P' \lhd \qquad \Gamma \vdash b_{\mathrm{tt}} \cong b'_{\mathrm{tt}} \lhd P[\mathrm{tt}] \qquad \Gamma \vdash b_{\mathrm{ff}} \cong b'_{\mathrm{ff}} \lhd P[\mathrm{ff}]\end{array}}{\Gamma \vdash \mathrm{ind}_{\mathbf{B}}(s; z.P; b_{\mathrm{tt}}, b_{\mathrm{ff}}) \approx \mathrm{ind}_{\mathbf{B}}(s'; z.P'; b'_{\mathrm{tt}}, b'_{\mathrm{ff}}) \rhd P[s]}$$

$$\text{NSumInd}\ \frac{\begin{array}{c}\Gamma \vdash A \cong A' \lhd \qquad \Gamma \vdash B \cong B' \lhd \qquad \Gamma \vdash s \approx_{\mathrm{h}} s' \rhd S \qquad \Gamma, z : A + B \vdash P \cong P' \lhd \\ \Gamma, x : A \vdash b_l \cong b'_l \lhd P\left[\mathrm{inj}^l\ x\right] \qquad \Gamma, x : B \vdash b_r \cong b'_r \lhd P\left[\mathrm{inj}^r\ x\right]\end{array}}{\Gamma \vdash \mathrm{ind}_{A+B}(s; z.P; x.b_l, x.b_r) \approx \mathrm{ind}_{A'+B'}(s'; z.P'; x.b'_l, x.b'_r) \rhd P[s]}$$

$$\text{NIdInd}\ \frac{\begin{array}{c}\Gamma \vdash A \cong A' \lhd \qquad \Gamma \vdash s \approx_{\mathrm{h}} s' \rhd \mathbf{Id}_{A''}\ a\, a' \\ \Gamma, x : A, y : A, z : \mathbf{Id}_A\ x\, y \vdash P \cong P' \lhd \qquad \Gamma, x : A \vdash b \cong b' \lhd P\left[x, x, \mathrm{refl}_{A,x}\right]\end{array}}{\Gamma \vdash \mathrm{ind}_{\mathbf{Id}_A}(s; x.y.z.P; x.b) \approx \mathrm{ind}_{\mathbf{Id}_{A'}}(s'; x.y.z.P'; x.b') \rhd P[a, a', s]}$$

## C.3 Declarative $\mathbf{MLTT}_{\mathrm{map}}$

Extend the rules of Appendix C.1. In rule MapComp, we rely on conversion at domain and morphism types for a type former $F$. These are obtained from the judgements of Fig. 7 by replacing every typing judgement by a conversion judgement, and forgetting conversions. For instance, for $\Pi$ types, we have

$$\Delta \vdash_{\mathrm{map}} (A, B) \cong (A', B') : \mathrm{dom}(\Pi) \iff \Delta \vdash_{\mathrm{map}} A \cong A' \wedge \Delta, x : A \vdash_{\mathrm{map}} B \cong B'$$
$$\Delta \vdash_{\mathrm{map}} (f, g) \cong (f', g') : \mathrm{hom}_\Pi((A_1, B_1), (A_2, B_2)) \iff \Delta \vdash_{\mathrm{map}} f \cong f' : A_2 \to A_1 \quad \wedge \quad \Delta, x : A_2 \vdash_{\mathrm{map}} g \cong g' : B_1[f\,x] \to B_2$$

For each type former $F$ ($\Pi$, $\Sigma$, **List**, **W**, **Id**, $+$)

$$\textsc{Map}\ \frac{\Gamma \vdash_{\mathrm{map}} X, Y : \mathrm{dom}(F) \qquad \Gamma \vdash_{\mathrm{map}} f : \mathrm{hom}_F(X, Y)}{\Gamma \vdash_{\mathrm{map}} \mathrm{map}_F\, f : F\,X \to F\,Y}$$

$$\textsc{MapId}\ \frac{\Gamma \vdash_{\mathrm{map}} X : \mathrm{dom}(F) \qquad \Gamma \vdash_{\mathrm{map}} t : F\,X}{\Gamma \vdash_{\mathrm{map}} \mathrm{map}_F\, \mathrm{id}^F_X\ t \cong t : F\,X}$$

$$\textsc{MapComp}\ \frac{\Gamma \vdash_{\mathrm{map}} X, Y, Z : \mathrm{dom}(F) \qquad \Gamma \vdash_{\mathrm{map}} g : \mathrm{hom}_F(X, Y) \qquad \Gamma \vdash_{\mathrm{map}} f : \mathrm{hom}_F(Y, Z) \qquad \Gamma \vdash_{\mathrm{map}} t : F\,X}{\Gamma \vdash_{\mathrm{map}} \mathrm{map}_F\ f\,(\mathrm{map}_F\, g\, t) \cong \mathrm{map}_F (f \circ g)\, t : F\,Z}$$

$$\textsc{MapCong}\ \frac{\Gamma \vdash_{\mathrm{map}} X \cong X' : \mathrm{dom}(F) \qquad \Gamma \vdash_{\mathrm{map}} Y \cong Y' : \mathrm{dom}(F) \qquad \Gamma \vdash_{\mathrm{map}} f : \mathrm{hom}_F(X, Y) \qquad \Gamma \vdash_{\mathrm{map}} f' : \mathrm{hom}_F(X', Y') \qquad \Gamma \vdash_{\mathrm{map}} f \cong f' : \mathrm{hom}_F(X, Y)}{\Gamma \vdash_{\mathrm{map}} \mathrm{map}_F\ f \cong \mathrm{map}_F\ f' : F\,X \to F\,Y}$$

$\Gamma \vdash_{\mathrm{map}} t \cong u : A$

$$\textsc{mapFun}\ \frac{\Gamma \vdash_{\mathrm{map}} (f, g) : \mathrm{hom}_\Pi((A, B), (A', B')) \qquad \Gamma \vdash_{\mathrm{map}} h : \Pi\, x : A.B \qquad \Gamma \vdash_{\mathrm{map}} a' : A'}{\Gamma \vdash_{\mathrm{map}} \mathrm{map}_\Pi\, (f, g)\, h\, a' \cong g\,(h\,(f\,a')) : B'[a']}$$

$$\textsc{mapSig}_1\ \frac{\Gamma \vdash_{\mathrm{map}} (f, g) : \mathrm{hom}_\Sigma((A, B), (A', B')) \qquad \Gamma \vdash_{\mathrm{map}} p : \Sigma\, x : A.B}{\Gamma \vdash_{\mathrm{map}} \pi_1\, (\mathrm{map}_\Sigma\, (f, g)\, p) \cong f\,(\pi_1\, p) : A'}$$

$$\textsc{mapSig}_2\ \frac{\Gamma \vdash_{\mathrm{map}} (f, g) : \mathrm{hom}_\Sigma((A, B), (A', B')) \qquad \Gamma \vdash_{\mathrm{map}} p : \Sigma\, x : A.B}{\Gamma \vdash_{\mathrm{map}} \pi_2\, (\mathrm{map}_\Sigma\, (f, g)\, p) \cong g\,(\pi_2\, p) : B'[f\,(\pi_1\, p)]}$$

$$\textsc{mapListNil}\ \frac{\Gamma \vdash_{\mathrm{map}} f : \mathrm{hom}_{\mathbf{List}}(A, A')}{\Gamma \vdash_{\mathrm{map}} \mathrm{map}_{\mathbf{List}}\ f\, \varepsilon_A \cong \varepsilon_{A'} : \mathbf{List}\ A'}$$

$$\textsc{mapListCons}\ \frac{\Gamma \vdash_{\mathrm{map}} f : \mathrm{hom}_{\mathbf{List}}(A, A') \qquad \Gamma \vdash_{\mathrm{map}} hd : A \qquad \Gamma \vdash_{\mathrm{map}} tl : \mathbf{List}\ A}{\Gamma \vdash_{\mathrm{map}} \mathrm{map}_{\mathbf{List}}\ f\,(hd ::_A tl) \cong (f\,hd) ::_{A'} (\mathrm{map}_{\mathbf{List}}\ f\,tl) : \mathbf{List}\ A'}$$

$$\textsc{mapW}\ \frac{\Gamma \vdash_{\mathrm{map}} (f, g) : \mathrm{hom}_{\mathbf{W}}((A, B), (A', B')) \qquad \Gamma \vdash_{\mathrm{map}} a : A \qquad \Gamma \vdash_{\mathrm{map}} k : B\,a \to \mathbf{W}\, x : A.B}{\Gamma \vdash_{\mathrm{map}} \mathrm{map}_{\mathbf{W}}(f, g)\, (\mathrm{sup}_{x.B}\, a\, k) \cong \mathrm{sup}_{x.B'}(f\,a)\, (\lambda\, x : B'[f\,a].\, \mathrm{map}_{\mathbf{W}}(f, g)\, (k\,(g\,x))) : \mathbf{W}\, x : A'.B'}$$

$$\textsc{mapId}\ \frac{\Gamma \vdash_{\mathrm{map}} f : \mathrm{hom}_{\mathbf{Id}}(A, A') \qquad \Gamma \vdash_{\mathrm{map}} a : A}{\Gamma \vdash_{\mathrm{map}} \mathsf{map}_{\mathbf{Id}}\ f\ \mathsf{refl}_{A,a} \cong \mathsf{refl}_{A', f\,a} : \mathbf{Id}\ A'\,(f\,a)\,(f\,a)}$$

$$\textsc{mapSumLeft}\ \frac{\Gamma \vdash_{\mathrm{map}} (f, g) : \mathrm{hom}_{+}((A, B), (A', B')) \qquad \Gamma \vdash_{\mathrm{map}} a : A}{\Gamma \vdash_{\mathrm{map}} \mathsf{map}_{+}\ (f, g)\ (\mathsf{inj}^{l}\ a) \cong \mathsf{inj}^{l}\ (f\,a) : A' + B'}$$

$$\textsc{mapSumRight}\ \frac{\Gamma \vdash_{\mathrm{map}} (f, g) : \mathrm{hom}_{+}((A, B), (A', B')) \qquad \Gamma \vdash_{\mathrm{map}} b : B}{\Gamma \vdash_{\mathrm{map}} \mathsf{map}_{+}\ (f, g)\ (\mathsf{inj}^{r}\ b) \cong \mathsf{inj}^{r}\ (g\,b) : A' + B'}$$

## C.4 Algorithmic $\mathbf{MLTT}_{\mathrm{map}}$

Extends Appendix C.2. Replaces the rules already named with the same name in Appendix C.2.

$\boxed{\Gamma \vdash_{\mathrm{map}} t \cong_{\mathrm{h}} t' \lhd T}$

$$\textsc{NeuPosMap}\ \frac{\Gamma \vdash_{\mathrm{map}} n \approx_{\mathrm{map}} n' \lhd T \qquad T \text{ is } \mathsf{Type}_i, \mathbf{List}\ A, \mathbf{W}\, x : A.B, A + B \text{ or } \mathbf{Id}_A\ a\,a'}{\Gamma \vdash_{\mathrm{map}} n \cong_{\mathrm{h}} n' \lhd T}$$

$\boxed{\Gamma \vdash_{\mathrm{map}} n \approx n' \rhd T}$

$$\textsc{NListInd}\ \frac{\begin{array}{c}\Gamma \vdash_{\mathrm{map}} A \cong A' \lhd \qquad \Gamma \vdash_{\mathrm{map}} s \approx_{\mathrm{map}} s' \lhd \mathbf{List}\ A \qquad \Gamma, z : \mathbf{List}\ A \vdash_{\mathrm{map}} P \cong P' \lhd \\ \Gamma \vdash_{\mathrm{map}} b_{\varepsilon} \cong b'_{\varepsilon} \lhd P[\varepsilon_A] \qquad \Gamma, x : A, y : \mathbf{List}\ A, z : P[y] \vdash_{\mathrm{map}} b_{::} \cong b'_{::} \lhd P[x ::_A y]\end{array}}{\Gamma \vdash_{\mathrm{map}} \mathsf{ind}_{\mathbf{List}\,A}(s; z.P; b_{\varepsilon}, x.y.z.b_{::}) \approx \mathsf{ind}_{\mathbf{List}\,A'}(s'; z.P'; b'_{\varepsilon}, x.y.z.b'_{::}) \rhd P[s]}$$

$$\textsc{NTreeInd}\ \frac{\begin{array}{c}\Gamma \vdash_{\mathrm{map}} A \cong A' \lhd \qquad \Gamma, x : A \vdash_{\mathrm{map}} B \cong B' \lhd \\ \Gamma \vdash_{\mathrm{map}} s \approx_{\mathrm{map}} s' \lhd \mathbf{W}\, x : A.B \qquad \Gamma, z : \mathbf{W}\, x : A.B \vdash_{\mathrm{map}} P \cong P' \lhd \\ \Gamma, x : A, y : B[x] \to W x : A.B, h : \Pi\, z : B[x].P[y\,z] \vdash_{\mathrm{map}} b \cong b' \lhd P\left[\mathsf{sup}_{x.B}\, x\ y\right]\end{array}}{\Gamma \vdash_{\mathrm{map}} \mathsf{ind}_{\mathbf{W}\,x:A.B}(s; z.P; x.y.z.b) \approx \mathsf{ind}_{\mathbf{W}\,x:A'.B'}(s'; z.P'; x.y.z.b') \rhd P[s]}$$

$$\textsc{NSumInd}\ \frac{\begin{array}{c}\Gamma \vdash_{\mathrm{map}} A \cong A' \lhd \\ \Gamma \vdash_{\mathrm{map}} B \cong B' \lhd \qquad \Gamma \vdash_{\mathrm{map}} s \approx_{\mathrm{map}} s' \lhd A + B \qquad \Gamma, z : A + B \vdash_{\mathrm{map}} P \cong P' \lhd \\ \Gamma, x : A \vdash_{\mathrm{map}} b_l \cong b'_l \lhd P\left[\mathsf{inj}^{l}\ x\right] \qquad \Gamma, x : B \vdash_{\mathrm{map}} b_r \cong b'_r \lhd P\left[\mathsf{inj}^{r}\ x\right]\end{array}}{\Gamma \vdash_{\mathrm{map}} \mathsf{ind}_{A+B}(s; z.P; x.b_l, x.b_r) \approx \mathsf{ind}_{A'+B'}(s'; z.P'; x.b'_l, x.b'_r) \rhd P[s]}$$

$$\textsc{NIdInd}\ \frac{\begin{array}{c}\Gamma \vdash_{\mathrm{map}} A \cong A' \lhd \qquad \Gamma \vdash_{\mathrm{map}} s \approx_{\mathrm{map}} s' \lhd \mathbf{Id}_A \rhd a, a' \\ \Gamma, x, y : A, z : \mathbf{Id}_A\ x\,y \vdash_{\mathrm{map}} P \cong P' \lhd \qquad \Gamma, x : A \vdash_{\mathrm{map}} b \cong b' \lhd P\left[x, x, \mathsf{refl}_{A,x}\right]\end{array}}{\Gamma \vdash_{\mathrm{map}} \mathsf{ind}_{\mathbf{Id}_A}(s; x.z.P; b) \approx \mathsf{ind}_{\mathbf{Id}_{A'}}(s'; x.z.P'; b') \rhd P[a, a', s]}$$

$\boxed{\mathsf{unmap}, \mathsf{unmapfun}, \mathsf{unmapfun}_1, \mathsf{unmapfun}_2}$

$$\begin{aligned} \mathsf{unmap}(\mathsf{map}_F\, f\, t) &\stackrel{\mathrm{def}}{=} t & \mathsf{unmap}(t) &\stackrel{\mathrm{def}}{=} t \quad \text{otherwise} \\ \mathsf{unmapfun}(\mathsf{map}_F\, f\, t, x) &\stackrel{\mathrm{def}}{=} f\,x & \mathsf{unmapfun}(t, x) &\stackrel{\mathrm{def}}{=} x \quad \text{otherwise} \\ \mathsf{unmapfun}_1(\mathsf{map}_F\, f\, t, x) &\stackrel{\mathrm{def}}{=} \pi_1\, f\, x & \mathsf{unmapfun}_1(t, x) &\stackrel{\mathrm{def}}{=} x \quad \text{otherwise} \\ \mathsf{unmapfun}_2(\mathsf{map}_F\, f\, t, y) &\stackrel{\mathrm{def}}{=} \pi_2\, f\, y & \mathsf{unmapfun}_2(t, y) &\stackrel{\mathrm{def}}{=} y \quad \text{otherwise} \end{aligned}$$

$\boxed{\Gamma \vdash_{\text{map}} n \approx_{\text{map}} n' \lhd T}$

$$\textsc{UnmapList}\ \frac{\Gamma \vdash_{\text{map}} \texttt{unmap}(n) \approx_{\text{h}} \texttt{unmap}(n') \rhd \textbf{List}\, A \qquad \Gamma, x : A \vdash_{\text{map}} \texttt{unmapfun}(n, x) \cong \texttt{unmapfun}(n', x) \lhd B}{\Gamma \vdash_{\text{map}} n \approx_{\text{map}} n' \lhd \textbf{List}\, B}$$

$$\textsc{UnmapTree}\ \frac{\begin{array}{c}\Gamma \vdash_{\text{map}} \texttt{unmap}(n) \approx_{\text{h}} \texttt{unmap}(n') \rhd \mathbf{W}\, x : A.B \\ \Gamma, x : A \vdash_{\text{map}} \texttt{unmapfun}_1(n, x) \cong \texttt{unmapfun}_1(n', x) \lhd A' \\ \Gamma, x : A, y : B'[\texttt{unmapfun}_1(n, x)] \vdash_{\text{map}} \texttt{unmapfun}_2(n, y) \cong \texttt{unmapfun}_2(n', y) \lhd B\, x\end{array}}{\Gamma \vdash_{\text{map}} n \approx_{\text{map}} n' \lhd \mathbf{W}\, x : A'.B'}$$

$$\textsc{UnmapSum}\ \frac{\begin{array}{c}\Gamma \vdash_{\text{map}} \texttt{unmap}(n) \approx_{\text{h}} \texttt{unmap}(n') \rhd A + B \\ \Gamma, x : A \vdash_{\text{map}} \texttt{unmapfun}_1(n, x) \cong \texttt{unmapfun}_1(n', x) \lhd A' \\ \Gamma, x : B \vdash_{\text{map}} \texttt{unmapfun}_2(n, x) \cong \texttt{unmapfun}_2(n', x) \lhd B'\end{array}}{\Gamma \vdash_{\text{map}} n \approx_{\text{map}} n' \lhd A' + B'}$$

$$\textsc{UnmapId}\ \frac{\begin{array}{c}\Gamma \vdash_{\text{map}} \texttt{unmap}(n) \approx_{\text{h}} \texttt{unmap}(n') \rhd \mathbf{Id}_A\ a\, a' \\ \Gamma, x : A \vdash_{\text{map}} \texttt{unmapfun}(n, x) \cong \texttt{unmapfun}(n', x) \lhd A'\end{array}}{\Gamma \vdash_{\text{map}} n \approx_{\text{map}} n' \lhd \mathbf{Id}_{A'} \rhd \texttt{unmapfun}(n, a), \texttt{unmapfun}(n, a')}$$

$\boxed{t \rightsquigarrow^1 t'}$

$$\pi_1\,(\text{map}_\Sigma\ f\, p) \rightsquigarrow^1 \pi_1\, f\,(\pi_1\, p) \qquad \pi_2\,(\text{map}_\Sigma\ f\, p) \rightsquigarrow^1 \pi_2\, f\,(\pi_2\, p)$$

$$\text{map}_\Pi\ f\, h\, t \rightsquigarrow^1 (\pi_2\, f)\,(h\,((\pi_1\, f)\, t)) \qquad \text{map}_{\mathbf{Id}}\ f\ \text{refl}_{A,a} \rightsquigarrow^1 \text{refl}_{B, f\,a}$$

$$\text{map}_{\mathbf{List}}\, f\, \varepsilon \rightsquigarrow^1 \varepsilon \qquad \text{map}_{\mathbf{List}}\, f\,(hd :: tl) \rightsquigarrow^1 f\, hd :: \text{map}_{\mathbf{List}}\, f\, tl$$

$$\text{map}_{\mathbf{W}}\{T\}\{T'\} f\,(\sup a\ k) \rightsquigarrow^1 \sup_{x.\pi_2\, T'}(\pi_1\, f\, a)\ (\lambda\, x : (\pi_2\, T'\ (\pi_1\, f\, a)).\, \text{map}_{\mathbf{W}}\ f\,(k\,(\pi_2\, g\, x)))$$

$$\text{map}_+\,(f, g)\,(\text{inj}^l\ a) \rightsquigarrow^1 \text{inj}^l\,(f\, a) \qquad \text{map}_+\,(f, g)\,(\text{inj}^r\ b) \rightsquigarrow^1 \text{inj}^r\,(g\, b)$$

$$\textsc{RedMapComp}\ \frac{\text{ne}\ n \qquad F \in \{\mathbf{List}, \mathbf{Id}, +, \mathbf{W}\}}{\text{map}_F\, f\,(\text{map}_F\, g\, n) \rightsquigarrow^1 \text{map}_F(f \circ g)\, n}$$

### C.5 Declarative record types

Extends Appendix C.1.

$$\textsc{RecTy}\ \frac{\mathcal{L} \in \mathcal{P}_{\text{f}}(\text{Lbl}) \qquad \forall l \in \mathcal{L}.\quad \Gamma \vdash A_l}{\Gamma \vdash \{l : A_l\}_{l \in \mathcal{L}}}$$

$$\textsc{RecUni}\ \frac{\mathcal{L} \in \mathcal{P}_{\text{f}}(\text{Lbl}) \qquad \forall l \in \mathcal{L}.\quad \Gamma \vdash A_l : \text{Type}_i}{\Gamma \vdash \{l : A_l\}^i_{l \in \mathcal{L}} : \text{Type}_i}$$

$$\textsc{RecTm}\ \frac{\mathcal{L} \in \mathcal{P}_f(\mathrm{Lbl}) \qquad \forall l \in \mathcal{L}.\quad \Gamma \vdash u_l : A_l}{\Gamma \vdash \{l := u_l\}_{l\in\mathcal{L}} : \{l : A_l\}_{l\in\mathcal{L}}} \qquad \textsc{RecProj}\ \frac{\Gamma \vdash r : \{l : A_l\}_{l\in\mathcal{L}}}{\Gamma \vdash r.l : A_l}$$

$$\beta\textsc{Rec}\ \frac{\mathcal{L} \in \mathcal{P}_f(\mathrm{Lbl}) \qquad \forall l \in \mathcal{L}.\quad \Gamma \vdash u_l : A_l}{\Gamma \vdash \{l := u_l\}_{l\in\mathcal{L}}.l \cong u_l : A_l} \qquad \eta\textsc{Rec}\ \frac{\Gamma \vdash r : \{l : A_l\}_{l\in\mathcal{L}}}{\Gamma \vdash r \cong \{l := r.l\}_{l\in\mathcal{L}} : \{l : A_l\}_{l\in\mathcal{L}}}$$

### C.6 Algorithmic record types

Extends Appendix C.2. Record construction terms $\{l := u_l\}_{l\in\mathcal{L}}$ are normal forms, and $r.l$ is neutral whenever $r$ is.

$$\beta\textsc{Rec}\ \frac{}{\{l := u_l\}_{l\in\mathcal{L}}.l \rightsquigarrow^1 u_l} \qquad \textsc{RecUni}\ \frac{\mathcal{L} \in \mathcal{P}_f(\mathrm{Lbl}) \qquad \forall l \in \mathcal{L}.\quad \Gamma \vdash A_l \rhd_h \mathrm{Type}_i}{\Gamma \vdash \{l : A_l\}^i_{l\in\mathcal{L}} \rhd \mathrm{Type}_i}$$

$$\textsc{RecTm}\ \frac{\mathcal{L} \in \mathcal{P}_f(\mathrm{Lbl}) \qquad \forall l \in \mathcal{L}.\quad \Gamma \vdash u_l \rhd A_l}{\Gamma \vdash \{l := u_l\}_{l\in\mathcal{L}} \rhd \{l : A_l\}_{l\in\mathcal{L}}} \qquad \textsc{RecEta}\ \frac{\forall l \in \mathcal{L}.\quad \Gamma \vdash r.l \cong r'.l \lhd A_l}{\Gamma \vdash r \cong_h r' \lhd \{l : A_l\}_{l\in\mathcal{L}}}$$

$$\textsc{NRecProj}\ \frac{\Gamma \vdash n \approx_h n' \rhd \{l : A_l\}_{l\in\mathcal{L}}}{\Gamma \vdash n.l \approx n'.l \rhd A_l}$$

### C.7 Algorithmic MLTT$_{\mathrm{sub}}$

Extends Appendices C.2 and C.6, with rule CheckSub replacing Check.

$\boxed{\Gamma \vdash_{\mathrm{sub}} t \lhd T}$

$$\textsc{CheckSub}\ \frac{\Gamma \vdash_{\mathrm{sub}} t \rhd T' \qquad \Gamma \vdash_{\mathrm{sub}} T' \preccurlyeq T \lhd}{\Gamma \vdash_{\mathrm{sub}} t \lhd T}$$

$\boxed{\Gamma \vdash_{\mathrm{sub}} T \preccurlyeq T' \lhd}$ Type $T$ is a subtype of type $T'$

$$\textsc{TyRed}\ \frac{T \rightsquigarrow^\star U \qquad T' \rightsquigarrow^\star U' \qquad \Gamma \vdash_{\mathrm{sub}} U \preccurlyeq_h U' \lhd}{\Gamma \vdash_{\mathrm{sub}} T \preccurlyeq T' \lhd}$$

$\boxed{\Gamma \vdash_{\mathrm{sub}} T \preccurlyeq_h T' \lhd}$ Reduced type $T$ is a subtype of reduced type $T'$

$$\textsc{RecSub}\ \frac{\mathcal{K} \subseteq \mathcal{L} \qquad \forall k \in \mathcal{K}.\quad \Gamma \vdash_{\mathrm{sub}} A_k \preccurlyeq B_k \lhd}{\Gamma \vdash_{\mathrm{sub}} \{l : A_l\}_{l\in\mathcal{L}} \preccurlyeq_h \{k : A_k\}_{k\in\mathcal{K}} \lhd}$$

$$\textsc{ProdSub}\ \frac{\Gamma \vdash_{\mathrm{sub}} A' \preccurlyeq A \lhd \qquad \Gamma, x : A' \vdash_{\mathrm{sub}} B \preccurlyeq B' \lhd}{\Gamma \vdash_{\mathrm{sub}} \Pi x : A.B \preccurlyeq_h \Pi x : A'.B' \lhd}$$

$$\textsc{ListSub}\ \frac{\Gamma \vdash_{\mathrm{sub}} A \preccurlyeq A' \lhd}{\Gamma \vdash_{\mathrm{sub}} \mathbf{List}\ A \preccurlyeq_h \mathbf{List}\ A' \lhd} \qquad \textsc{SigSub}\ \frac{\Gamma \vdash_{\mathrm{sub}} A \preccurlyeq A' \lhd \qquad \Gamma, x : A \vdash_{\mathrm{sub}} B \preccurlyeq B' \lhd}{\Gamma \vdash_{\mathrm{sub}} \Sigma x : A.B \preccurlyeq_h \Sigma x : A'.B' \lhd}$$

$$\textsc{TreeSub}\ \frac{\Gamma \vdash_{\text{sub}} A \preccurlyeq A' \lhd \qquad \Gamma, x : A \vdash_{\text{sub}} B' \preccurlyeq B \lhd}{\Gamma \vdash_{\text{sub}} \mathbf{W}\, x : A.B \preccurlyeq_{\text{h}} \mathbf{W}\, x : A'.B' \lhd}$$

$$\textsc{IdSub}\ \frac{\Gamma \vdash_{\text{sub}} A \preccurlyeq A' \lhd \qquad \Gamma \vdash_{\text{sub}} t \cong t' \lhd A' \qquad \Gamma \vdash_{\text{sub}} u \cong u' \lhd A'}{\Gamma \vdash_{\text{sub}} \mathbf{Id}_A\ t\, u \preccurlyeq_{\text{h}} \mathbf{Id}_{A'}\ t'\, u' \lhd} \qquad \textsc{SubRefl}\ \frac{T \text{ is } \mathrm{Type}_i, \mathbf{0}, \mathbf{1} \text{ or } \mathbf{B}}{\Gamma \vdash_{\text{sub}} T \preccurlyeq_{\text{h}} T \lhd}$$

$$\textsc{SumSub}\ \frac{\Gamma \vdash_{\text{sub}} A \preccurlyeq A' \lhd \qquad \Gamma \vdash_{\text{sub}} B \preccurlyeq B' \lhd}{\Gamma \vdash_{\text{sub}} A + B \preccurlyeq_{\text{h}} A' + B' \lhd} \qquad \textsc{NeuSub}\ \frac{\Gamma \vdash_{\text{sub}} n \approx_{\text{h}} n' \rhd T}{\Gamma \vdash_{\text{sub}} n \preccurlyeq_{\text{h}} n' \lhd}$$

Admissible rules

$$\textsc{ConvSub}\ \frac{\Gamma \vdash_{\text{sub}} A \cong A' \lhd}{\Gamma \vdash_{\text{sub}} A \preccurlyeq A' \lhd} \qquad \textsc{SubAntiSym}\ \frac{\Gamma \vdash_{\text{sub}} A \preccurlyeq A' \lhd \qquad \Gamma \vdash_{\text{sub}} A' \preccurlyeq A \lhd}{\Gamma \vdash_{\text{sub}} A \cong A' \lhd}$$

$$\textsc{SubTrans}\ \frac{\Gamma \vdash_{\text{sub}} A \preccurlyeq A' \lhd \qquad \Gamma \vdash_{\text{sub}} A' \preccurlyeq A'' \lhd}{\Gamma \vdash_{\text{sub}} A \preccurlyeq A'' \lhd}$$

## C.8 Reduction rules and normal forms for $\mathsf{MLTT}_{\text{coe}}$

$t \leadsto^1 t'$

$$\textsc{RedCoeFun}\ \frac{\mathrm{nf}\, f}{(\mathrm{coe}_{\Pi\, x:A.B, \Pi\, x:A'.B'}\ f)\ a \leadsto^1 \mathrm{coe}_{B[\mathrm{coe}_{A',A}\ a], B'[a]} (f\ (\mathrm{coe}_{A',A}\ a))}$$

$$\textsc{RedCoeSig1}\ \frac{\mathrm{nf}\, p}{\pi_1\ (\mathrm{coe}_{\Sigma\, x:A.B, \Sigma\, x:A'.B'}\ p) \leadsto^1 \mathrm{coe}_{A,A'} (\pi_1\, p)}$$

$$\textsc{RedCoeSig2}\ \frac{\mathrm{nf}\, p}{\pi_2\ (\mathrm{coe}_{\Sigma\, x:A.B, \Sigma\, x:A'.B'}\ p) \leadsto^1 \mathrm{coe}_{B[\pi_1\, p], B'[\mathrm{coe}_{A,A'}(\pi_1\, p)]} (\pi_2\, p)}$$

$$\textsc{RedCoeRec}\ \frac{\mathrm{nf}\, r}{(\mathrm{coe}_{\{l : A_l\}_{l\in\mathcal{L}}, \{k : A_k\}_{k\in\mathcal{K}}}\ r).l \leadsto^1 \mathrm{coe}_{A_l, B_l}\ r.l} \qquad \textsc{CoeRedId}\ \frac{T \text{ is } \mathrm{Type}_i, \mathbf{0}, \mathbf{1} \text{ or } \mathbf{B}}{\mathrm{coe}_{T,T}\, t \leadsto^1 t}$$

$$\mathrm{coe}_{A+B, A'+B'} (\mathrm{inj}^l_B\ a) \leadsto^1 \mathrm{inj}^l_{B'}\ (\mathrm{coe}_{A,A'}\ a) \qquad \mathrm{coe}_{A+B, A'+B'} (\mathrm{inj}^r_A\ b) \leadsto^1 \mathrm{inj}^r_{A'}\ (\mathrm{coe}_{B,B'}\ b)$$

$$\mathrm{coe}_{\mathbf{List}\, A, \mathbf{List}\, A'}\ \varepsilon \leadsto^1 \varepsilon_{A'} \qquad \mathrm{coe}_{\mathbf{List}\, A, \mathbf{List}\, A'} (h :: t) \leadsto^1 \mathrm{coe}_{A,A'}\ h ::_{A'} \mathrm{coe}_{\mathbf{List}\, A, \mathbf{List}\, A'}\ t$$

$$\mathrm{coe}_{\mathbf{W}\, x:A.B, \mathbf{W}\, x:A.B'} (\sup a\ l) \leadsto^1$$
$$\sup_{x.B'} (\mathrm{coe}_{A,A'}\ a)\ (\lambda\, x : B'\left[\mathrm{coe}_{A,A'}\ a\right].\, \mathrm{coe}_{\mathbf{W}\, x:A.B, \mathbf{W}\, x:A.B'} (k\ (\mathrm{coe}_{B'[\mathrm{coe}_{A,A'}\ a], B[a]}\ x)))$$

$$\mathrm{coe}_{\mathbf{Id}_A\ a\, b, \mathbf{Id}_{A'}\ a'\, b'}\ \mathrm{refl}_{A,a} \leadsto^1 \mathrm{refl}_{A', (\mathrm{coe}_{A,A'}\ a)}$$

$$\textsc{CoeL}\ \frac{A \rightsquigarrow^1 A'}{\mathsf{coe}_{A,B}\, t \rightsquigarrow^1 \mathsf{coe}_{A',B}\, t} \qquad \textsc{CoeR}\ \frac{\mathsf{nf}^{\oplus} \text{ or ne } A \qquad B \rightsquigarrow^1 B'}{\mathsf{coe}_{A,B}\, t \rightsquigarrow^1 \mathsf{coe}_{A,B'}\, t}$$

$$\textsc{CoeTm}\ \frac{\mathsf{nf}^{\oplus} \text{ or ne } A, B \qquad t \rightsquigarrow^1 t'}{\mathsf{coe}_{A,B}\, t \rightsquigarrow^1 \mathsf{coe}_{A,B}\, t'} \qquad \textsc{CoeCoe}\ \frac{\mathsf{nf}^{\oplus} \text{ or ne } U, U', T, T' \qquad \text{ne } n}{\mathsf{coe}_{U,U'}\, \mathsf{coe}_{T,T'}\, n \rightsquigarrow^1 \mathsf{coe}_{T,U'}\, n}$$

| | | | |
|---|---|---|---|
| $\boxed{\mathsf{nf}\ f}$ | $\overset{\text{def}}{=}$ | $n \mid P \mid N \mid \lambda x : t.t \mid (t,t) \mid \{l := t_l\}_{l\in\mathcal{L}} \mid$ $\varepsilon_t \mid t ::_t t \mid \mathsf{sup}_t\, t\, t \mid () \mid \mathsf{tt} \mid \mathsf{ff} \mid \mathsf{refl}_{t,t} \mid \mathsf{inj}^l_t\, t \mid \mathsf{inj}^r_t\, t$ $\mathsf{coe}_{N,N}\, f$ | weak-head normal forms |
| $\boxed{\mathsf{nf}^{\ominus}\ N}$ | $\overset{\text{def}}{=}$ | $\Pi x : t.t \mid \Sigma x : t.t \mid \{l : t_l\}_{l\in\mathcal{L}}$ | negative whnf types |
| $\boxed{\mathsf{nf}^{\oplus}\ P}$ | $\overset{\text{def}}{=}$ | $\mathsf{Type}_i \mid \mathbf{0} \mid \mathbf{1} \mid \mathbf{B} \mid \mathbf{List}\, t \mid \mathbf{W}\, x : t.t \mid t + t \mid \mathbf{Id}_t\, t\, t'$ | other whnf types |
| $\boxed{\mathsf{ne}\ n}$ | $\overset{\text{def}}{=}$ | $x \mid n\, t \mid \pi_1\, n \mid \pi_2\, n \mid n.l \mid \mathsf{ind}_P(n; t; t)$ | weak-head neutrals |
| $\boxed{\mathsf{cne}\ c}$ | $\overset{\text{def}}{=}$ | $n \mid \mathsf{coe}_{P,P}\, n \mid \mathsf{coe}_{n,n}\, n$ | compacted neutrals |

## C.9 Algorithmic MLTT$_{\mathsf{coe}}$

Extends Appendix C.2 and Appendix C.6.

$\boxed{\Gamma \vdash_{\mathsf{coe}} t \rhd T}$

$$\textsc{Coe}\ \frac{\Gamma \vdash_{\mathsf{coe}} A \lhd \qquad \Gamma \vdash_{\mathsf{coe}} A' \lhd \qquad \Gamma \vdash_{\mathsf{coe}} t \lhd A \qquad \Gamma \vdash_{\mathsf{coe}} A \preccurlyeq A' \lhd}{\Gamma \vdash_{\mathsf{coe}} \mathsf{coe}_{A,A'}\, t \rhd A'}$$

$\boxed{\Gamma \vdash_{\mathsf{coe}} t \approx_{\mathsf{coe}} t' \lhd T}$ Compacted neutrals $t$ and $t'$ are comparable at type $T$

$$\textsc{NCoe}\ \frac{\Gamma \vdash_{\mathsf{coe}} n \approx n' \rhd S''}{\Gamma \vdash_{\mathsf{coe}} \mathsf{coe}_{S,T}\, n \approx_{\mathsf{coe}} \mathsf{coe}_{S',T'}\, n' \lhd T''} \qquad \textsc{NCoeL}\ \frac{\Gamma \vdash_{\mathsf{coe}} n \approx n' \rhd S''}{\Gamma \vdash_{\mathsf{coe}} \mathsf{coe}_{S,T}\, n \approx_{\mathsf{coe}} n' \lhd T''}$$

$$\textsc{NCoeR}\ \frac{\Gamma \vdash_{\mathsf{coe}} n \approx n' \rhd S''}{\Gamma \vdash_{\mathsf{coe}} n \approx_{\mathsf{coe}} \mathsf{coe}_{S',T'}\, n' \lhd T''} \qquad \textsc{NNoCoe}\ \frac{\Gamma \vdash_{\mathsf{coe}} n \approx n' \rhd S''}{\Gamma \vdash_{\mathsf{coe}} n \approx_{\mathsf{coe}} n' \lhd T''}$$

$\boxed{\Gamma \vdash_{\mathsf{coe}} t \cong_{\mathsf{h}} t' \lhd T}$

$$\textsc{NeuList}\ \frac{\Gamma \vdash_{\mathsf{coe}} n \approx_{\mathsf{coe}} n' \lhd \mathbf{List}\, A}{\Gamma \vdash_{\mathsf{coe}} n \cong_{\mathsf{h}} n' \lhd \mathbf{List}\, A} \qquad \textsc{NeuTree}\ \frac{\Gamma \vdash_{\mathsf{coe}} n \approx_{\mathsf{coe}} n' \lhd \mathbf{W}\, x : A.B}{\Gamma \vdash_{\mathsf{coe}} n \cong_{\mathsf{h}} n' \lhd \mathbf{W}\, x : A.B}$$

$$\textsc{NeuId}\ \frac{\Gamma \vdash_{\mathsf{coe}} n \approx_{\mathsf{coe}} n' \lhd \mathbf{Id}_A\, a\, a'}{\Gamma \vdash_{\mathsf{coe}} n \cong_{\mathsf{h}} n' \lhd \mathbf{Id}_A\, a\, a'} \qquad \textsc{NeuSum}\ \frac{\Gamma \vdash_{\mathsf{coe}} n \approx_{\mathsf{coe}} n' \lhd A + B}{\Gamma \vdash_{\mathsf{coe}} n \cong_{\mathsf{h}} n' \lhd A + B}$$

$$\textsc{NeuNeu}\ \frac{\Gamma \vdash_{\mathsf{coe}} n \approx_{\mathsf{coe}} n' \lhd M \qquad \text{ne } M}{\Gamma \vdash_{\mathsf{coe}} n \cong_{\mathsf{h}} n' \lhd M}$$

$\boxed{\Gamma \vdash_{\text{coe}} T \preccurlyeq T' \lhd}$

$$\textsc{TyRed}\ \frac{T \leadsto^{\star} U \qquad T' \leadsto^{\star} U' \qquad \Gamma \vdash_{\text{coe}} U \preccurlyeq_{\text{h}} U' \lhd}{\Gamma \vdash_{\text{coe}} T \preccurlyeq T' \lhd}$$

$\boxed{\Gamma \vdash_{\text{coe}} T \preccurlyeq_{\text{h}} T' \lhd}$

$$\textsc{RecSub}\ \frac{\mathcal{K} \subseteq \mathcal{L} \qquad \forall k \in \mathcal{K}.\quad \Gamma \vdash_{\text{coe}} A_k \preccurlyeq B_k \lhd}{\Gamma \vdash_{\text{coe}} \{l : A_l\}_{l \in \mathcal{L}} \preccurlyeq_{\text{h}} \{k : A_k\}_{k \in \mathcal{K}} \lhd}$$

$$\textsc{ProdSub}\ \frac{\Gamma \vdash_{\text{coe}} A' \preccurlyeq A \lhd \qquad \Gamma, x : A' \vdash_{\text{coe}} B\left[\text{coe}_{A',A}\, x\right] \preccurlyeq B' \lhd}{\Gamma \vdash_{\text{coe}} \Pi\, x : A.B \preccurlyeq_{\text{h}} \Pi\, x : A'.B' \lhd}$$

$$\textsc{ListSub}\ \frac{\Gamma \vdash_{\text{coe}} A \preccurlyeq A' \lhd}{\Gamma \vdash_{\text{coe}} \textbf{List}\ A \preccurlyeq_{\text{h}} \textbf{List}\ A' \lhd}$$

$$\textsc{SigSub}\ \frac{\Gamma \vdash_{\text{coe}} A \preccurlyeq A' \lhd \qquad \Gamma, x : A \vdash_{\text{coe}} B \preccurlyeq B'\left[\text{coe}_{A,A'}\, x\right] \lhd}{\Gamma \vdash_{\text{coe}} \Sigma\, x : A.B \preccurlyeq_{\text{h}} \Sigma\, x : A'.B' \lhd}$$

$$\textsc{TreeSub}\ \frac{\Gamma \vdash_{\text{coe}} A \preccurlyeq A' \lhd \qquad \Gamma, x : A \vdash_{\text{coe}} B'\left[\text{coe}_{A,A'}\, x\right] \preccurlyeq B \lhd}{\Gamma \vdash_{\text{coe}} \mathbf{W}\, x : A.B \preccurlyeq_{\text{h}} \mathbf{W}\, x : A'.B' \lhd}$$

$$\textsc{IdSub}\ \frac{\Gamma \vdash_{\text{coe}} A \preccurlyeq A' \lhd \qquad \Gamma \vdash_{\text{coe}} \text{coe}_{A,A'}\, t \cong t' \lhd A' \qquad \Gamma \vdash_{\text{coe}} \text{coe}_{A,A'}\, u \cong u' \lhd A'}{\Gamma \vdash_{\text{coe}} \textbf{Id}_A\ t\, u \preccurlyeq_{\text{h}} \textbf{Id}_{A'}\ t'\, u' \lhd}$$

$$\textsc{ListSub}\ \frac{\Gamma \vdash_{\text{coe}} A \preccurlyeq A' \lhd \qquad \Gamma \vdash_{\text{coe}} B \preccurlyeq B' \lhd}{\Gamma \vdash_{\text{coe}} A + B \preccurlyeq_{\text{h}} A' + B' \lhd}$$

$$\textsc{SubRefl}\ \frac{T \text{ is Type}_i,\ \mathbf{0},\ \mathbf{1} \text{ or } \mathbf{B}}{\Gamma \vdash_{\text{coe}} T \preccurlyeq T \lhd}$$

## C.10 Declarative $\text{MLTT}_{\text{coe}}$

Extends Appendix C.1 and Appendix C.5.

$\boxed{\Gamma \vdash_{\text{coe}} t : T}$

$$\textsc{Coe}\ \frac{\Gamma \vdash_{\text{coe}} A \qquad \Gamma \vdash_{\text{coe}} A' \qquad \Gamma \vdash_{\text{coe}} t : A \qquad \Gamma \vdash_{\text{coe}} A \preccurlyeq A'}{\Gamma \vdash_{\text{coe}} \text{coe}_{A,A'}\, t : A'}$$

$\boxed{\Gamma \vdash_{\text{coe}} t \cong t' : T}$

$$\textsc{CoeId}\ \frac{\Gamma \vdash_{\text{coe}} t : A}{\Gamma \vdash_{\text{coe}} \text{coe}_{A,A}\, t \cong t : A}$$

$$\textsc{CoeTrans}\ \frac{\Gamma \vdash_{\text{coe}} t : A \qquad \Gamma \vdash_{\text{coe}} A \preccurlyeq A' \qquad \Gamma \vdash_{\text{coe}} A' \preccurlyeq A''}{\Gamma \vdash_{\text{coe}} \text{coe}_{A',A''}\, \text{coe}_{A,A'}\, t \cong \text{coe}_{A,A''}\, t : A''}$$

$$\textsc{CoeCong}\ \frac{\Gamma \vdash_{\text{coe}} t \cong t' : A \qquad \Gamma \vdash_{\text{coe}} A \cong A' \qquad \Gamma \vdash_{\text{coe}} B \cong B'}{\Gamma \vdash_{\text{coe}} \text{coe}_{A,B}\, t \cong \text{coe}_{A',B'}\, t' : B}$$

$$\textsc{CoeFun}\ \frac{\Gamma \vdash_{\mathrm{coe}} A' \preccurlyeq A \qquad \Gamma, x : A' \vdash_{\mathrm{coe}} B[\mathrm{coe}_{A',A}\,x] \preccurlyeq B' \qquad \Gamma \vdash_{\mathrm{coe}} f : \Pi\, x : A.B \qquad \Gamma \vdash_{\mathrm{coe}} a : A'}{\Gamma \vdash_{\mathrm{coe}} (\mathrm{coe}_{\Pi\, x:A.B, \Pi\, x:A'.B'}\ f)\ a \cong \mathrm{coe}_{B[\mathrm{coe}_{A',A}\ a], B'[x]} (f\ (\mathrm{coe}_{A',A}\ a)) : \Pi\, x : A'.B'}$$

$$\textsc{CoeSig1}\ \frac{\Gamma \vdash_{\mathrm{coe}} A \preccurlyeq A' \qquad \Gamma, x : A \vdash_{\mathrm{coe}} B \preccurlyeq B'[\mathrm{coe}_{A,A'}\ x] \qquad \Gamma \vdash_{\mathrm{coe}} p : \Sigma\, x : A.B}{\Gamma \vdash_{\mathrm{coe}} \pi_1\ (\mathrm{coe}_{\Sigma\, x:A.B, \Sigma\, x:A'.B'}\ p) \cong \mathrm{coe}_{A,A'} (\pi_1\, p) : A'}$$

$$\textsc{CoeSig2}\ \frac{\Gamma \vdash_{\mathrm{coe}} A \preccurlyeq A' \qquad \Gamma, x : A \vdash_{\mathrm{coe}} B \preccurlyeq B'[\mathrm{coe}_{A,A'}\ x] \qquad \Gamma \vdash_{\mathrm{coe}} p : \Sigma\, x : A.B}{\Gamma \vdash_{\mathrm{coe}} \pi_2\ (\mathrm{coe}_{\Sigma\, x:A.B, \Sigma\, x:A'.B'}\ p) \cong \mathrm{coe}_{B[\pi_1\, p], B'[\mathrm{coe}_{A,A'} (\pi_1\, p)]} (\pi_2\, p) : B'[\mathrm{coe}_{A,A'} (\pi_1\, p)]}$$

$$\textsc{CoeRec}\ \frac{\mathcal{K} \subseteq \mathcal{L} \qquad \forall k \in \mathcal{K}.\quad \Gamma \vdash_{\mathrm{coe}} A_k \preccurlyeq B_k \qquad \Gamma \vdash r \rhd \{l : A_l\}_{l \in \mathcal{L}}}{\Gamma \vdash_{\mathrm{coe}} (\mathrm{coe}_{\{l : A_l\}_{l \in \mathcal{L}}, \{k : A_k\}_{k \in \mathcal{K}}}\ r).k \cong \mathrm{coe}_{A_k, B_k}\ r.k : B_k}$$

$$\textsc{CoeNil}\ \frac{\Gamma \vdash_{\mathrm{coe}} A \preccurlyeq A'}{\Gamma \vdash_{\mathrm{coe}} \mathrm{coe}_{\mathbf{List}\, A, \mathbf{List}\, A'}\ \varepsilon_A \cong \varepsilon_{A'} : \mathbf{List}\ A'}$$

$$\textsc{CoeCons}\ \frac{\Gamma \vdash_{\mathrm{coe}} A \preccurlyeq A' \qquad \Gamma \vdash_{\mathrm{coe}} a : A \qquad \Gamma \vdash_{\mathrm{coe}} l : \mathbf{List}\ A}{\Gamma \vdash_{\mathrm{coe}} \mathrm{coe}_{\mathbf{List}\, A, \mathbf{List}\, A'} (a ::_A l) \cong (\mathrm{coe}_{A,A'}\ a) ::_{A'} (\mathrm{coe}_{\mathbf{List}\, A, \mathbf{List}\, A'}\ l) : \mathbf{List}\ A'}$$

$$\textsc{CoeTree}\ \frac{\Gamma \vdash_{\mathrm{coe}} A \preccurlyeq A' \qquad \Gamma, x : A \vdash_{\mathrm{coe}} B'[\mathrm{coe}_{A,A'}\ x] \preccurlyeq B \qquad \Gamma \vdash_{\mathrm{coe}} a : A \qquad \Gamma \vdash_{\mathrm{coe}} k : B\, a \to \mathbf{W}\, x : A.B}{\begin{array}{l}\Gamma \vdash_{\mathrm{coe}} \mathrm{coe}_{\mathbf{W}\, x:A.B, \mathbf{W}\, x:A.B'} (\mathrm{sup}_{x.B}\ a\ l) \cong \\ \quad \mathrm{sup}_{x.B'} (\mathrm{coe}_{A,A'}\ a)\ (\lambda\, x : B'[\mathrm{coe}_{A,A'}\ a].\, \mathrm{coe}_{\mathbf{W}\, x:A.B, \mathbf{W}\, x:A.B'} (k\ (\mathrm{coe}_{B'[\mathrm{coe}_{A,A'}\ a], B[a]}\ x))) \\ \hfill : \mathbf{W}\, x : A'.B'\end{array}}$$

$$\textsc{CoeId}\ \frac{\Gamma \vdash_{\mathrm{coe}} A \preccurlyeq A' \qquad \Gamma \vdash_{\mathrm{coe}} a : A}{\Gamma \vdash_{\mathrm{coe}} \mathrm{coe}_{\mathbf{Id}_A\ a\, a, \mathbf{Id}_{A'}\ (\mathrm{coe}_{A,A'}\ a)\ (\mathrm{coe}_{A,A'}\ a)}\ \mathrm{refl}_{A,a} \cong \mathrm{refl}_{A', (\mathrm{coe}_{A,A'}\ a)} : \mathbf{Id}_A\ (\mathrm{coe}_{A,A'}\ a)\ (\mathrm{coe}_{A,A'}\ a)}$$

$$\textsc{CoeSumLeft}\ \frac{\Gamma \vdash_{\mathrm{coe}} A \preccurlyeq A' \qquad \Gamma \vdash_{\mathrm{coe}} B \preccurlyeq B' \qquad \Gamma \vdash_{\mathrm{coe}} a : A}{\Gamma \vdash_{\mathrm{coe}} \mathrm{coe}_{A+B, A'+B'} (\mathrm{inj}^l\ a) \cong \mathrm{inj}^l\ (\mathrm{coe}_{A,A'}\ a) : A' + B'}$$

$$\textsc{CoeSumRight}\ \frac{\Gamma \vdash_{\mathrm{coe}} A \preccurlyeq A' \qquad \Gamma \vdash_{\mathrm{coe}} B \preccurlyeq B' \qquad \Gamma \vdash_{\mathrm{coe}} b : B}{\Gamma \vdash_{\mathrm{coe}} \mathrm{coe}_{A+B, A'+B'} (\mathrm{inj}^r\ b) \cong \mathrm{inj}^r\ (\mathrm{coe}_{B,B'}\ b) : A' + B'}$$

$\boxed{\Gamma \vdash_{\mathrm{coe}} T \preccurlyeq T'}$ $T$ is a subtype of $T'$ in context $\Gamma$

$$\textsc{RecSub}\ \frac{\mathcal{K} \subseteq \mathcal{L} \qquad \forall k \in \mathcal{K}.\quad \Gamma \vdash_{\mathrm{coe}} A_k \preccurlyeq B_k}{\Gamma \vdash_{\mathrm{coe}} \{l : A_l\}_{l \in \mathcal{L}} \preccurlyeq \{k : A_k\}_{k \in \mathcal{K}}}$$

$$\textsc{ProdSub}\ \frac{\Gamma \vdash_{\mathrm{coe}} A' \preccurlyeq A \qquad \Gamma, x : A' \vdash_{\mathrm{coe}} B[\mathrm{coe}_{A',A}\, x] \preccurlyeq B'}{\Gamma \vdash_{\mathrm{coe}} \Pi\, x : A.B \preccurlyeq \Pi\, x : A'.B'}$$

$$\textsc{ListSub}\ \frac{\Gamma \vdash_{\text{coe}} A \preccurlyeq A'}{\Gamma \vdash_{\text{coe}} \textbf{List}\, A \preccurlyeq \textbf{List}\, A'} \qquad \textsc{SigSub}\ \frac{\Gamma \vdash_{\text{coe}} A \preccurlyeq A' \quad \Gamma, x : A \vdash_{\text{coe}} B \preccurlyeq B'\left[\text{coe}_{A,A'}\, x\right]}{\Gamma \vdash_{\text{coe}} \Sigma\, x : A.B \preccurlyeq \Sigma\, x : A'.B'}$$

$$\textsc{TreeSub}\ \frac{\Gamma \vdash_{\text{coe}} A \preccurlyeq A' \qquad \Gamma, x : A \vdash_{\text{coe}} B'\left[\text{coe}_{A,A'}\, x\right] \preccurlyeq B}{\Gamma \vdash_{\text{coe}} \mathbf{W}\, x : A.B \preccurlyeq \mathbf{W}\, x : A'.B'}$$

$$\textsc{IdSub}\ \frac{\Gamma \vdash_{\text{coe}} A \preccurlyeq A' \qquad \Gamma \vdash_{\text{coe}} \text{coe}_{A,A'}\, t \cong t' : A' \qquad \Gamma \vdash_{\text{coe}} \text{coe}_{A,A'}\, u \cong u' : A'}{\Gamma \vdash_{\text{coe}} \textbf{Id}_A\ t\, u \preccurlyeq \textbf{Id}_{A'}\ t'\, u'}$$

$$\textsc{SumSub}\ \frac{\Gamma \vdash_{\text{coe}} A \preccurlyeq A' \qquad \Gamma \vdash_{\text{coe}} B \preccurlyeq B'}{\Gamma \vdash_{\text{coe}} A + B \preccurlyeq A' + B'}$$

$$\textsc{SubRefl}\ \frac{\Gamma \vdash_{\text{coe}} A \cong A'}{\Gamma \vdash_{\text{coe}} A \preccurlyeq A'} \qquad \textsc{SubTrans}\ \frac{\Gamma \vdash_{\text{coe}} A \preccurlyeq A' \qquad \Gamma \vdash_{\text{coe}} A' \preccurlyeq A''}{\Gamma \vdash_{\text{coe}} A \preccurlyeq A''}$$

# D PROOFS OF LEMMAS

This section contains additional lemmas and proofs omitted from the body of the paper.

## D.1 From Section 3.2

$$\text{map}_\Pi\ ((g, f) : \text{hom}_\Pi((A, B), (A', B')))\ (h : \Pi(x : A)B) \stackrel{\text{def}}{=} \lambda\, x : A'.f\ (h\ (g\, x))$$
$$\text{map}_\Sigma\ ((g, f) : \text{hom}_\Pi((A, B), (A', B')))\ (p : \Sigma(x : A)B) \stackrel{\text{def}}{=} (g\ (\pi_1\, p), f\ (\pi_2\, p))$$

Lemma D.1. $\text{map}_\Pi$ *and* $\text{map}_\Sigma$ *satisfy the functor laws* MapId *and* MapComp*.*

Proof. For the preservation of identities, we have:

$$\text{map}_\Pi\ (\text{id}_A, \lambda\{x : A\}.\, \text{id}_{B\, x})\, h \cong \lambda\, x : A.\, \text{id}_{B\, x}\ (h\ (\text{id}_A\ \ x)) \cong \lambda\, x : A.g\, x \cong g$$
$$\text{map}_\Sigma\ (\text{id}_A, \lambda\{x : A\}.\, \text{id}_{B\, x})\, p \cong (\text{id}_A\ (\pi_1\, p), \text{id}_{B\,(\pi_1\, p)}\ (\pi_2\, p)) \cong (\pi_1\, p, \pi_2\, p) \cong p$$

For $(g, f) : \text{hom}_\Pi((A_2, B_2), (A_3, B_3)), (g', f') : \text{hom}_\Pi((A_1, B_1), (A_2, B_2))$ and $h : \Pi\, x : A_1.B_1\, x$ we compute

$$\begin{aligned} \text{map}_\Pi\ (g, f)\ (\text{map}_\Pi\ (g', f')\, h) &\cong \lambda\, x : A.f\ ((\lambda\, x' : A'.f'\ (h\ (g'\, x')))\ (g\, x)) \\ &\cong \lambda\, x : A, f\ (f'\ (h\ (g'\ (g\, x)))) \\ &\cong \lambda\, x : A, (f \circ f')\ (h\ ((g' \circ g)\, x)) \cong \text{map}_\Pi((g, f) \circ (g', f'))h \end{aligned}$$

Similarly, for $(g, f) : \text{hom}_\Sigma((A_2, B_2), (A_3, B_3)), (g', f') : \text{hom}_\Sigma((A_1, B_1), (A_2, B_2))$ and $p : \Sigma\, x : A_1.B_1\, x$ :

$$\begin{aligned} \text{map}_\Sigma\ (g, f)\ (\text{map}_\Sigma\ (g', f')\, p) &\cong (g\ (\pi_1\ \text{map}_\Sigma\ (g', f')\, p), f\ (\pi_2\ \text{map}_\Sigma\ (g', f')\, p)) \\ &\cong (g\ (g'\ (\pi_1\, p)), f\ (f'\ (\pi_2\, p))) \\ &\cong ((g \circ g')\ (\pi_1\, p)), (f \circ f')\ (\pi_2\, p))) \\ &\cong \text{map}_\Sigma\ ((g, f) \circ (g', f'))\ p \end{aligned}$$

□

## D.2 From Section 5.3

Lemma D.2 (Catch up, function type (Lemma 5.4)). *If* $\Gamma \vdash_{\text{coe}} f\ a \lhd B$ *and* $|f| = \lambda\, x : A'.\, t'$, *then there exists* $t$ *such that* $|t| = t'$ *and* $f\ a \rightsquigarrow^\star t[a]$.

Proof. We must have that $f = (\text{coe}_{T_1,\dots,T_n}(\lambda\, x : A.t_0))$[15] for some $A$, $t_0$ such that $|A| = A'$ and $|t_0| = t'$. Moreover, by well-typing we know that there exists some $B_0$ such that $\Gamma, x : A \vdash_{\text{coe}} t_0 \rhd B_0$, $\Gamma \vdash_{\text{coe}} \Pi\, x : A.B_0 \cong T_1 \preccurlyeq T_2 \cong \dots \preccurlyeq T_n \lhd$. By inversions, we must have $T_i \rightsquigarrow^\star \Pi\, x : A_i.\, B_i$, with the $A_i$ and $B_i$ again related. But now we can use the reduction rule of coe on product types, and get

$$f\ a \rightsquigarrow^\star \text{coe}_{B_0',B_1',\dots,B_n'}((\lambda\, x : A.t_0)\ (\text{coe}_{A_n,\dots,A_1,A}\ a))$$

where the $B_i'$ are obtained by adequately substituting coercions in the $B_i$. Now all the $B_i'$ are well-typed by subject reduction, so they must have weak-head normal forms $B_i''$, and once all of them have been reduced to weak-head normal form by a combination of CoeL, CoeR and CoeTm, we can finally reduce the inner β-redex, obtaining

$$f\ a \rightsquigarrow^\star \text{coe}_{B_0'',B_1'',\dots,B_n''}(t_0\left[\text{coe}_{A_n,\dots,A_1,A}\ a\right])$$

Now we can conclude, as indeed

$$\begin{aligned}\left|\text{coe}_{B_0'',B_1'',\dots,B_n''}(t_0\left[\text{coe}_{A_n,\dots,A_1,A}\ x\right])\right| &= \left|t_0\left[\text{coe}_{A_n,\dots,A_1,A}\ x\right]\right| \\ &= |t_0|\left[\left|\text{coe}_{A_n,\dots,A_1,A}\ x\right|\right] \\ &= |t_0|\,[|x|] = |t_0| = t'\end{aligned}$$

□

Lemma D.3 (Erasure is a backward simulation (Lemma 5.5)). *Assume that* $\Gamma \vdash_{\text{coe}} t \lhd T$. *If* $|t| \rightsquigarrow^\star u'$, *with* $u'$ *a weak-head normal form, then* $t \rightsquigarrow^\star u$, *with* $u$ *a weak-head normal form such that* $|u| = u'$.

Proof. First, if $|t| \rightsquigarrow^\star u'$, then there exists $u$ such that $t \rightsquigarrow^\star u$ and $|u| = u'$. Indeed, the previous catch-up lemmas ensure that redexes never get blocked by coercions. On function types, the lemma exactly says that a term erasing to a β-redex is able to simulate the β-reduction. On positive types, by the catch-up lemma again, coercions on a constructor reduce away until the constructor is exposed directly to the destructor, and so the reduction can kick in.

Second, if $|u|$ is a weak-head normal form, then there exists a weak-head normal form $v$ such that $u \rightsquigarrow^\star v$ and $|v| = |u|$. Indeed, if $|u|$ is a weak-head normal form but $u$ is not, it must be because either $|u|$ is a constructor of a positive type, or a neutral. In the first case, the catch-up lemmas let us conclude. In the second, we can iterate CoeCoe to fuse coercions until $u$ reduces to a compacted neutral, which is a weak-head normal form. □

Lemma D.4 (Elaboration preserves subtyping (Lemma 5.6)). *The following implications hold whenever the inputs of the conclusions are well-formed:*

(1) *if* $|\Gamma| \vdash_{\text{sub}} |T| \preccurlyeq^{\text{m}}_{\text{h}} |U| \lhd$, *then* $\Gamma \vdash_{\text{coe}} T \preccurlyeq^{\text{m}}_{\text{h}} U \lhd$;
(2) *if* $|\Gamma| \vdash_{\text{sub}} |T| \preccurlyeq^{\text{m}} |U| \lhd$, *then* $\Gamma \vdash_{\text{coe}} T \preccurlyeq U \lhd$;
(3) *if* $|\Gamma| \vdash_{\text{sub}} |t| \cong_{\text{h}} |u| \lhd |T|$, *then* $\Gamma \vdash_{\text{coe}} t \cong_{\text{h}} u \lhd T$;
(4) *if* $|\Gamma| \vdash_{\text{sub}} |t| \cong |u| \lhd |T|$, *then* $\Gamma \vdash_{\text{coe}} t \cong u \lhd T$;
(5) *if* $|\Gamma| \vdash_{\text{sub}} |t| \approx |u| \rhd T$, *then* $\Gamma \vdash_{\text{coe}} t \approx u \rhd T$;
(6) *if* $|\Gamma| \vdash_{\text{sub}} |t| \approx_{\text{h}} |u| \rhd T$, *then* $\Gamma \vdash_{\text{coe}} t \approx u \rhd T$.

[15] That is, a string of coercions $\text{coe}_{T_{n-1},T_n}(\dots \text{coe}_{T_1,T_2}(\lambda\, x : A.t_0))$.

PROOF. Lemma 5.5 ensures we can always match reductions to weak-head normal forms in $\text{MLTT}_{\text{sub}}$ with reductions to weak-head normal forms in $\text{MLTT}_{\text{coe}}$. As for conversion itself, the key cases are those where the term in $\text{MLTT}_{\text{coe}}$ is a coercion, that gets erased in $\text{MLTT}_{\text{sub}}$. Given the structure of normal forms from Figure 13, this can happen in three situations. If the coercions are between function types or record types, we do not inspect the terms, and instead eagerly η-expand in a type-directed fashion (which triggers further reduction of the now applied coercions). The third case is compacted neutrals. They can appear exactly in the places where $\text{MLTT}_{\text{coe}}$ uses the comparison of the compacted neutrals, which strips away the possibly present coercions, as expected. □

Finally, the main theorem states that we can elaborate terms using implicit subtyping to explicit coercions, in a type-preserving way.

THEOREM D.5 (ELABORATION – INDUCTION). *The following implications hold, whenever inputs to the conclusion are well-formed:*

1. *if* $|\Gamma| \vdash_{\text{sub}} t' \rhd T'$, *then there exists* $t$ *and* $T$ *such that* $t' = |t|, T' = |T|$, *and* $\Gamma \vdash_{\text{coe}} t \rhd T$;
2. *if* $|\Gamma| \vdash_{\text{sub}} t' \rhd_{\text{h}} T'$, *then there exists* $t$ *and* $T$ *such that* $t' = |t|, T' = |T|$, *and* $\Gamma \vdash_{\text{coe}} t \rhd_{\text{h}} T$;
3. *if* $|\Gamma| \vdash_{\text{sub}} t' \lhd |T|$, *then there exists* $t$ *such that* $t' = |t|$ *and* $\Gamma \vdash_{\text{coe}} t \lhd T$.

PROOF. Once again, by mutual induction. Each rule is mapped to its counterpart, but for CHECK-SUB, where we need to insert a coercion in the elaborated term. This coercion is well-typed by Lemma 5.6. □

### D.3 Translation from $\text{MLTT}_{\text{coe}}$ to $\text{MLTT}_{\text{map}}$

$\text{MLTT}_{\text{coe}}$ terms contain enough information to entirely capture the subtyping derivations. We exploit this information to define a relation $[\![t]\!] \simeq t'$ between a $\text{MLTT}_{\text{coe}}$ term $t$ and a $\text{MLTT}_{\text{map}}$ term $t'$, that makes explicit the functorial nature of coercions. The definition of $[\![t]\!] \simeq t'$ employs an auxiliary relation $[\![A \rightsquigarrow B]\!] \simeq x$ to translate coercions from $A$ to $B$, where $x$ is either the special value $\star$ or a $\text{MLTT}_{\text{map}}$ term $f$. The value $\star$ arises in the case of an identity coercion that should be erased by the translation. In order to translate records, we assume that we have access to an (effective, decidable) total order on the countable set Lbl of labels, so that we can order in a canonical fashion every finite subsets $\mathcal{L} \subseteq \text{Lbl}$ as $\mathcal{L} = \{l_1 < \ldots < l_n\}$.

$$\text{TslTy}\ \frac{}{[\![\text{Type}_i]\!] \simeq \text{Type}_i} \qquad \text{TslList}\ \frac{[\![A]\!] \simeq A'}{[\![\textbf{List}\ A]\!] \simeq \textbf{List}\ A'} \qquad \text{TslPi}\ \frac{[\![A]\!] \simeq A' \qquad [\![B]\!] \simeq B'}{[\![\Pi\, x : A.B]\!] \simeq \Pi\, x : A'.B'}$$

$$\text{TslSig}\ \frac{[\![A]\!] \simeq A' \qquad [\![B]\!] \simeq B'}{[\![\Sigma\, x : A.B]\!] \simeq \Sigma\, x : A'.B'} \qquad \text{TslRec}\ \frac{\forall l \in \mathcal{L}.[\![A_l]\!] \simeq A'_l \qquad \mathcal{L} = \{l_1 < \ldots < l_n\}}{[\![\{l : A_l\}_{l \in \mathcal{L}}]\!] \simeq \Sigma\, x_{l_1} : A'_{l_1}. \ldots A'_{l_n}}$$

$$\text{TslVar}\ \frac{}{[\![x]\!] \simeq x} \qquad \text{TslLam}\ \frac{[\![A]\!] \simeq A' \qquad [\![t]\!] \simeq t'}{[\![\lambda\, x : A.t]\!] \simeq \lambda\, x : A'.t'} \qquad \text{TslApp}\ \frac{[\![u]\!] \simeq u' \qquad [\![v]\!] \simeq v'}{[\![u\ v]\!] \simeq u'\ v'}$$

$$\text{TslPair}\ \frac{[\![u]\!] \simeq u' \qquad [\![v]\!] \simeq v'}{[\![(u, v)]\!] \simeq (u', v')} \qquad \text{TslFst}\ \frac{[\![p]\!] \simeq p'}{[\![\pi_1\, p]\!] \simeq \pi_1\, p'} \qquad \text{TslSnd}\ \frac{[\![p]\!] \simeq p'}{[\![\pi_2\, p]\!] \simeq \pi_2\, p'}$$

$$\text{TslRecTm}\ \frac{\forall l \in \mathcal{L}.[\![u_l]\!] \simeq u'_l \qquad \mathcal{L} = \{l_1 < \ldots < l_n\}}{[\![\{l := u_l\}]\!] \simeq (u'_{l_1}, \ldots u'_{l_n})}$$

$$\textsc{TslProj}\ \frac{[\![p]\!] \simeq p' \qquad \Gamma \vdash_{\mathrm{coe}} p\{l : A_l\}_{l\in\mathcal{L}} \qquad \mathcal{L} = \{l_1 < \ldots < l_n\}}{[\![p.l_i]\!] \simeq \pi_1 \circ \pi_2^{i-1}(p')}$$

$$\textsc{TslCoeId}\ \frac{[\![A \rightsquigarrow B]\!] \simeq \star \qquad [\![t]\!] \simeq t'}{[\![\mathrm{coe}_{A,B}\, t]\!] \simeq t'} \qquad\qquad \textsc{TslCoe}\ \frac{[\![A \rightsquigarrow B]\!] \simeq f \qquad [\![t]\!] \simeq t'}{[\![\mathrm{coe}_{A,B}\, t]\!] \simeq f\, t'}$$

$$\textsc{TslCoeNf}\ \frac{A \rightsquigarrow^\star A' \text{ nf} \qquad B \rightsquigarrow^\star B' \text{ nf} \qquad [\![A' \rightsquigarrow B']\!] \simeq x \qquad A \neq A' \text{ or } B \neq B'}{[\![A \rightsquigarrow B]\!] \simeq x}$$

$$\textsc{TslCoeListId}\ \frac{[\![A \rightsquigarrow B]\!] \simeq \star}{[\![\mathbf{List}\, A \rightsquigarrow \mathbf{List}\, B]\!] \simeq \star} \qquad\qquad \textsc{TslCoeList}\ \frac{[\![A \rightsquigarrow B]\!] \simeq f}{[\![\mathbf{List}\, A \rightsquigarrow \mathbf{List}\, B]\!] \simeq \mathrm{map}_{\mathbf{List}}\, f}$$

$$\textsc{TslCoePiIdBoth}\ \frac{[\![A_2 \rightsquigarrow A_1]\!] \simeq \star \qquad [\![B_1 \rightsquigarrow B_2]\!] \simeq \star}{[\![\Pi\, x : A_1.B_1 \rightsquigarrow \Pi\, x : A_2.B_2]\!] \simeq \star}$$

$$\textsc{TslCoePiIdDom}\ \frac{[\![A_2 \rightsquigarrow A_1]\!] \simeq \star \qquad [\![B_1 \rightsquigarrow B_2]\!] \simeq g \qquad [\![A_1]\!] \simeq A_1'}{[\![\Pi\, x : A_1.B_1 \rightsquigarrow \Pi\, x : A_2.B_2]\!] \simeq \mathrm{map}_\Pi\, (\mathrm{id}_{A_1'}, g)}$$

$$\textsc{TslCoePiIdCod}\ \frac{[\![A_2 \rightsquigarrow A_1]\!] \simeq f \qquad [\![B_1 \left[\mathrm{coe}_{A_2,A_1}\, x\right] \rightsquigarrow B_2]\!] \simeq \star \qquad [\![B_2]\!] \simeq B_2'}{[\![\Pi\, x : A_1.B_1 \rightsquigarrow \Pi\, x : A_2.B_2]\!] \simeq \mathrm{map}_\Pi(f, \mathrm{id}_{B_2'})}$$

$$\textsc{TslCoePi}\ \frac{[\![A_2 \rightsquigarrow A_1]\!] \simeq f \qquad [\![B_1 \left[\mathrm{coe}_{A_2,A_1}\, x\right] \rightsquigarrow B_2]\!] \simeq g}{[\![\Pi\, x : A_1.B_1 \rightsquigarrow \Pi\, x : A_2.B_2]\!] \simeq \mathrm{map}_\Pi(f, g)} \qquad \text{and similarly for } \Sigma$$

$$\textsc{TslCoeRecId}\ \frac{\mathcal{K} = \mathcal{L} \qquad \forall k \in \mathcal{K}.[\![A_k \rightsquigarrow B_k]\!] \simeq \star}{[\![\{l : A_l\}_{l\in\mathcal{L}} \rightsquigarrow \{k : B_k\}_{k\in\mathcal{K}}]\!] \simeq \star}$$

$$\textsc{TslCoeRec}\ \frac{\mathcal{K} \subseteq \mathcal{L} \qquad \forall k \in \mathcal{K}.[\![A_k \rightsquigarrow B_k]\!] \simeq f_k \qquad \mathcal{K} = \{k_1 < \ldots < k_n\}}{[\![\{l : A_l\}_{l\in\mathcal{L}} \rightsquigarrow \{k : B_k\}_{k\in\mathcal{K}}]\!] \simeq \lambda p.(f_{k_1}\, (\pi_1\, p), \ldots f_{k_n}\, (\pi_2^{n-1}\, p))}$$

$$\textsc{TslCoeTy}\ [\![\mathrm{Type}_i \rightsquigarrow \mathrm{Type}_i]\!] \simeq \star \qquad\qquad \textsc{TslCoeNe}\ \frac{\mathrm{ne}\, N \qquad \mathrm{ne}\, M}{[\![N \rightsquigarrow M]\!] \simeq \star}$$

The translation is extended to contexts pointwise.

$$\frac{}{[\![\cdot]\!] \simeq \cdot} \qquad\qquad \frac{[\![\Gamma]\!] \simeq \Gamma' \qquad [\![A]\!] \simeq A'}{[\![\Gamma, x : A]\!] \simeq \Gamma', x : A'}$$

We note $[\![t]\!]\downarrow$ when $t$ is in the domain of the relation and $[\![t]\!]$ for the image of $t$ when it is defined.

Lemma D.6 (Determinism of translation). *The translation relation $[\![t]\!] \simeq t'$ is a partial function, i.e. it is deterministic: for any $t\, t_1'\, t_2'$, if $[\![t]\!] \simeq t_1'$ and $[\![t]\!] \simeq t_2'$ then $t_1' = t_2'$.*

Proof. We show by mutual induction on a derivation that $[\![A \rightsquigarrow B]\!] \simeq x$ is a partial function as well from pairs of $\mathrm{MLTT}_{\mathrm{coe}}$ types to either $\star$ or a $\mathrm{MLTT}_{\mathrm{map}}$ term. In the key case TslCoeNf, note that the reduction relation $\rightsquigarrow^\star$ is deterministic as well, so we can conclude by induction hypothesis. All other cases are immediate or simple applications of the inductive hypothesis, using the fact that at each step, at most one rule apply. ◻

Lemma D.7 (Stability of translation by weakening). *If $\rho$ is a substitution that maps variables to variables then $[\![t]\!][\rho] = [\![t[\rho]]\!]$.*

Proof. Immediate by induction on $t$, the only case interesting case being the translation of variables, with a similar lemma for $[\![A \rightsquigarrow B]\!] \simeq x$ using that neutrals are preserved. □

Lemma D.8 (Well-typed terms translate). *If $\Gamma \vdash_{\text{coe}} t : A$ then $[\![\Gamma]\!]\downarrow$, $[\![A]\!]\downarrow$ and $[\![t]\!]\downarrow$.*

Proof. We prove by a straightforward mutual induction on an algorithmic typing derivation that:

- If $\vdash_{\text{coe}} \Gamma$ then $[\![\Gamma]\!]\downarrow$;
- If $\Gamma \vdash_{\text{coe}} A \lhd$ and $[\![\Gamma]\!]\downarrow$ then $[\![A]\!]\downarrow$;
- If $\Gamma \vdash_{\text{coe}} t \lhd A$ and $[\![\Gamma]\!]\downarrow$ then $[\![t]\!]\downarrow$;
- If $\Gamma \vdash_{\text{coe}} t \rhd A$ and $[\![\Gamma]\!]\downarrow$ then $[\![t]\!]\downarrow$;
- If $\Gamma \vdash_{\text{coe}} A \preccurlyeq B \lhd$ or $\Gamma \vdash_{\text{coe}} A \preccurlyeq_{\text{h}} B \lhd$ then there exists $x$ such that $[\![A \rightsquigarrow B]\!] \simeq x$.

□

Lemma D.9 (Identity coercions). *If $\Gamma \vdash_{\text{coe}} A \cong B \lhd$ or $\Gamma \vdash_{\text{coe}} A \cong_{\text{h}} B \lhd$ then $[\![A \rightsquigarrow B]\!] \simeq \star$.*

Proof. Straightforward mutual induction on the bidirectional conversion derivation. □

Lemma D.10 (Stability of translation by substitution). *If $\Gamma \vdash_{\text{coe}} t : A$ and $\Delta \vdash_{\text{coe}} \sigma : \Gamma$ then $[\![t]\!][[\![\sigma]\!]] = [\![t[\sigma]]\!]$ and similarly for typing.*

*If $\Gamma \vdash_{\text{coe}} A \preccurlyeq B \lhd$, $\Delta \vdash_{\text{coe}} \sigma : \Gamma$ and*

- $[\![A \rightsquigarrow B]\!] \simeq \star$ *then* $[\![A[\sigma] \rightsquigarrow B[\sigma]]\!] \simeq \star$;
- $[\![A \rightsquigarrow B]\!] \simeq f$ *then* $[\![A[\sigma] \rightsquigarrow B[\sigma]]\!] \simeq f[\sigma]$.

Proof. Straightforward mutual induction on the bidirectional derivation. □

*Forward simulation.* Following the proof strategy employed for the equivalence between subsumptive and coercive subtyping, the nest step would require to prove that the translation is a forward simulate, i.e. if $\Gamma \vdash_{\text{coe}} t : A$ and $t \rightsquigarrow^1 t'$ then $[\![t]\!] \rightsquigarrow^\star [\![t']\!]$. As stated, this lemma does not hold. Indeed, the rule CoeCoe leads to reductions of coercions with type annotations which may be convertible but not reduce correctly. We conjecture that a weaker version of the simulation with respect to conversion in $\text{MLTT}_{\text{map}}$ should hold, that is if $\Gamma \vdash_{\text{coe}} t : A$ and $t \rightsquigarrow^1 t'$ then $[\![\Gamma]\!] \vdash_{\text{map}} [\![t]\!] \cong [\![t']\!] : [\![A]\!]$. Such statement should be proved mutually with other properties stating that the translation preserves typing, as follows.

Conjecture D.11 (Translation preserves typing).

1. *If* $\vdash_{\text{coe}} \Gamma$ *then* $\vdash_{\text{map}} [\![\Gamma]\!]$
2. *If* $\Gamma \vdash_{\text{coe}} A$ *then* $[\![\Gamma]\!] \vdash_{\text{map}} [\![A]\!]$
3. *If* $\Gamma \vdash_{\text{coe}} t : A$ *then* $[\![\Gamma]\!] \vdash_{\text{map}} [\![t]\!] : [\![A]\!]$
4. *If* $\Gamma \vdash_{\text{coe}} A \cong B$ *then* $[\![\Gamma]\!] \vdash_{\text{map}} [\![A]\!] \cong [\![B]\!]$
5. *If* $\Gamma \vdash_{\text{coe}} t \cong u : A$ *then* $[\![\Gamma]\!] \vdash_{\text{map}} [\![t]\!] \cong [\![u]\!] : [\![A]\!]$
6. *If* $\Gamma \vdash_{\text{coe}} A \preccurlyeq B$ *then either*
   (a) $[\![A \rightsquigarrow B]\!] \simeq \star$ *and* $[\![\Gamma]\!] \vdash_{\text{map}} [\![A]\!] \cong [\![B]\!]$
   (b) $[\![A \rightsquigarrow B]\!] \simeq f$ *and* $[\![\Gamma]\!] \vdash_{\text{map}} f : [\![A]\!] \to [\![B]\!]$

Preservation of typing, together with catch up lemmas, and a backward simulation lemma, would then allow to lift bidirectional conversion derivations in $\text{MLTT}_{\text{map}}$ between the translation of terms from $\text{MLTT}_{\text{coe}}$. The use of bidirectional conversion is essential here to remain at each step within the translation of $\text{MLTT}_{\text{coe}}$ terms.

Conjecture D.12 (Embedding). $[\![-]\!]$ *embeds* $\mathit{MLTT}_{\mathrm{coe}}$ *into* $\mathit{MLTT}_{\mathrm{map}}$*: well-typed* $\mathit{MLTT}_{\mathrm{coe}}$ *terms translate to well-typed* $\mathit{MLTT}_{\mathrm{map}}$ *terms, preserving and reflecting conversion.*